\documentclass[english,aps,pra,reprint,twocolumn,footinbib,utf8,superscriptaddress,floatfix]{revtex4-2}

\usepackage[T1]{fontenc}
\usepackage[utf8]{inputenc}

\usepackage{mathtools}
\usepackage{amsmath}
\usepackage{amssymb}

\usepackage[dvipdfmx]{graphicx}
\usepackage[dvipsnames]{xcolor}
\definecolor{pdarkblue}{rgb}{0.1797, 0.1875, 0.5703}
\usepackage{hyperref}
\hypersetup{
    colorlinks = true,
    citecolor = pdarkblue,
    linkcolor = pdarkblue,
    urlcolor = pdarkblue,
}

\allowdisplaybreaks
\usepackage[normalem]{ulem}
\usepackage[varg]{txfonts}
\usepackage{newtxtext}
\usepackage{bm}

\usepackage{titlesec}
\usepackage{indentfirst}

\titleformat*{\section}{\centering\fontsize{10.5pt}{\baselineskip}\selectfont\bfseries}
\titleformat*{\subsection}{\centering\fontsize{10.5pt}{\baselineskip}\selectfont\bfseries}
\titleformat*{\subsubsection}{\centering\fontsize{10.5pt}{\baselineskip}\selectfont\itshape}

\titlespacing*{\section}{\linewidth}{3em}{0.4em}
\titlespacing*{\subsection}{\linewidth}{1.5em}{0.4em}
\titlespacing*{\subsubsection}{\linewidth}{1.5em}{0.4em}

\AtBeginDocument{
\fontsize{10.5pt}{12.36pt}\selectfont
\fontdimen2\font = 2.9pt 
}

\newcommand{\ket}[1]{\vert#1\rangle}
\newcommand{\bra}[1]{\langle#1\vert}
\newcommand{\braket}[2]{\langle#1\vert#2\rangle}
\newcommand{\ketbra}[2]{\vert#1\rangle\langle#2\vert}
\newcommand{\tr}{\mathop{\mathrm{tr}}}
\newcommand{\argmin}{\mathop{\mathrm{argmin}}}

\newcommand{\td}{t_{\mathrm{d}}}

\begin{document}

\title{Improving variational counterdiabatic driving with weighted actions and computer algebra}
\author{Naruo Ohga}
\email{naruo.ohga@ubi.s.u-tokyo.ac.jp}
\affiliation{Department of Physics, Graduate School of Science, The University of Tokyo, 7-3-1 Hongo, Bunkyo-ku, Tokyo 113-0033, Japan}
\affiliation{Basic Research Laboratories, NTT, Inc., Kanagawa 243-0198, Japan}

\author{Takuya Hatomura}
\affiliation{Basic Research Laboratories, NTT, Inc., Kanagawa 243-0198, Japan}
\affiliation{NTT Research Center for Theoretical Quantum Information, NTT, Inc., Kanagawa 243-0198, Japan}

\date{\today}

\begin{abstract}
Variational counterdiabatic (CD) driving is a disciplined and widely used method to robustly control quantum many-body systems by mimicking adiabatic processes with high fidelity and reduced duration. Central to this technique is a universal structure of the adiabatic gauge potential (AGP) over a parameterized Hamiltonian.
Here, we reveal that introducing a new degree of freedom into the theory of the AGP can significantly improve variational CD driving. Specifically, we find that the algebraic characterization of the AGP is not unique, and we exploit this nonuniqueness to develop the \textit{weighted variational method} for deriving a refined driving protocol. This approach extends the conventional method in two aspects: it assigns customized weights to matrix elements relevant to specific problems, and it effectively incorporates nonlocal information into local driving coefficients. We also develop an efficient numerical algorithm to compute the refined driving protocol using computer algebra. Our framework is broadly applicable and, in principle, it can replace any previous use of variational CD driving. We demonstrate its practicality by applying it to adiabatic evolution along the ground state of a parameterized Hamiltonian. This proposal outperforms the conventional method in terms of fidelity, as confirmed by extensive numerical simulations on quantum Ising models. 
\end{abstract}

\maketitle

\onecolumngrid
\vspace{-.5em}
\twocolumngrid

\section{Introduction 
\label{sec:introduction}}

Fast, accurate, and stable control of quantum systems is of paramount importance to realize quantum state preparation, state processing, and information processing for advancing quantum technologies~\cite{Acin2018TheQuantumTechnologies}. 
High-quality quantum control is realized by combining a theoretical scheme to compute a driving protocol on classical computers with experimental techniques to faithfully implement the protocol, both forming broad research areas~\cite{Glaser2015TrainingSchrodingers}. While the early development of quantum control focused on few-body systems, recent advances in quantum technology highlight the need for methods tailored to many-body systems, such as systems with tens to hundreds of qubits~\cite{Arute2019QuantumSupremacy,Wu2021StrongQuantum,Bluvstein2023LogicalQuantum}.

Controlling many-body systems poses a huge theoretical challenge, especially due to the exponential computational cost of simulating quantum dynamics on classical computers.
Control methods such as quantum optimal control~\cite{Glaser2015TrainingSchrodingers} and reinforcement learning~\cite{Bukov2018ReinforcementLearning} rely on time-evolution simulations to maximize the final fidelity, but direct simulation requires exponentially large memory and time for many-body systems. Compression methods such as tensor networks have been incorporated to reduce this cost~\cite{Orus2019TensorNetworks}, but they involve tradeoffs between accuracy and computational cost. In particular, this reliance on classical simulation is inappropriate for quantum computation, which should exploit quantum dynamics beyond classical simulations.

An alternative approach that avoids time-evolution simulations is to determine driving protocols from fundamental and universal structures of quantum systems. One such structure is represented by an operator called the adiabatic gauge potential (AGP), which endows a geometric structure with a parameterized Hamiltonian~\cite{Kolodrubetz2017GeometryAndNonAdiabatic}. By implementing the AGP as a driving Hamiltonian, we can perform counterdiabatic (CD) driving~\cite{Demirplak2003AdiabaticPopulation, Demirplak2005AssistedAdiabatic, Demirplak2008OnTheConsistencyExtremal, Berry2009Transitionless}. CD driving mimics adiabatic processes with---in principle---an arbitrarily short duration, with the increased energy cost manageable by the quantum speed limit~\cite{Campbell2017TradeOffBetween, Santos2015SuperadiabaticControlled, Funo2017UniversalWork, Abah2019EnergeticCost}. CD driving inherits the robustness of adiabatic processes~\cite{Ruschhaupt2012OptimallyRobust, Takahashi2013HowFastAndRobust} and is widely used to stably guide a quantum state along an eigenstate of a time-dependent Hamiltonian. Explicit expressions of the AGP have been derived for a variety of simple systems, including two- and three-level systems~\cite{Chen2010ShortcutToAdiabatic}, harmonic oscillators~\cite{Muga2010TransitionlessQuantum}, many-body quantum spin systems associated with free fermions~\cite{del2012AssistedFinite}, classical particles in potentials~\cite{Jarzynski2013GeneratingShortcuts}, many-body and/or nonlinear systems with scale-invariant dynamics~\cite{del2013ShortcutsToAdiabaticity,Deffner2014ClassicalAndQuantum}, quantum systems associated with classical nonlinear integrable systems~\cite{Okuyama2016FromClassical}, and classical spin systems~\cite{Hatomura2018ShortcutsToAdiabatic}.

In recent years, \textit{the variational AGP}, a systematic approximation of the AGP~\cite{Sels2017Minimizing,Kolodrubetz2017GeometryAndNonAdiabatic}, has attracted attention for its ability to overcome the difficulty of computing and implementing the exact AGP in many-body systems. The variational AGP is defined by minimizing an action (a functional of operators) over experimentally feasible operators. The action is derived from an algebraic characterization of the exact AGP~\cite{Jarzynski2013GeneratingShortcuts}, and it could yield the exact AGP if minimized over arbitrary Hermitian operators. The minimization can be easily computed even for large systems, making the variational AGP practical for many-body systems. Using the variational AGP as a driving Hamiltonian, we can realize approximated CD driving, called \textit{variational CD driving} or \textit{local CD driving}.

Variational CD driving has been tested and applied to a wide range of systems, leading to significantly improved quantum control. Numerical tests have been performed on spin systems with various lattice structures and interactions~\cite{Hartmann2019RapidCounter, Hartmann2022PolynomialScaling, Passarelli2020CounterdiabaticDriving, Prielinger2021TwoParameterCounter, Kumar2021CounterdiabaticRoute, Barone2024CounterdiabaticOptimized,Andras2024FightingExponentially} and on fermion lattice systems~\cite{Xie2022VariationalCounterdiabatic,Sels2017Minimizing}. It has been experimentally realized with nuclear magnetic resonance~\cite{Zhou2020ExperimentalRealization}, a tight-binding lattice of ultracold atoms~\cite{Meier2020CounterdiabaticControl}, and superconducting quantum computers~\cite{Hegade2021ShortcutsToAdiabaticity}. Practical applications have been considered and tested, such as heat engines~\cite{Hartmann2020ManyBody,Hartmann2020MultiSpin}, quantum state transfer~\cite{Villazon2021ShortcutsToDynamic,Ji2022CounterdiabaticTransfer}, quantum annealing and its varieties~\cite{Hartmann2019RapidCounter, Hartmann2022PolynomialScaling, Passarelli2020CounterdiabaticDriving, Prielinger2021TwoParameterCounter, Kumar2021CounterdiabaticRoute, Barone2024CounterdiabaticOptimized,Passarelli2023CounterdiabaticReverse}, and optimization problems including integer factorization, portfolio optimization, protein folding, and logistics~\cite{Hegade2021DigitizedAdiabatic,Hegade2022DigitizedCounterdiabatic, Hegade2022PortfolioOptimization, Hegade2023DigitizedCounterdiabatic, Guan2024SingleLayer, Romero2025BiasField}.
In addition, variational CD driving has been combined with other techniques and ideas to derive a number of practical methods, including Floquet engineering~\cite{Petiziol2024QuantumControl,Petiziol2018FastAdiabatic,Petiziol2019AcceleratingAdiabatic,Claeys2019FloquetEngineering, Schindler2024CounterdiabaticDriving}, tensor networks~\cite{Kim2024PRXQuantum, McKeever2024PRXQuantum}, digital quantum simulation~\cite{Hegade2021DigitizedAdiabatic,Hegade2022DigitizedCounterdiabatic}, Lie algebra~\cite{Hatomura2021ControllingAndExploring,Petiziol2018FastAdiabatic}, the Lanczos algorithm~\cite{Bhattacharjee2023ALanczosApproach, Takahashi2024ShortcutsToAdiabaticity}, and optimizations of the intermediate path~\cite{Prielinger2021TwoParameterCounter, Mbeng2022RotatedAnsatz, Cepaite2023CounterdiabaticOptimized}. 
From a broader perspective, variational AGPs have been extended to variational quantum algorithms~\cite{Yao2021ReinforcementLearning,Wurtz2022CounterdiabaticityAndTheQuantum,Chandarana2022DigitizedCounterdiabatic, Chandarana2023MetaLearning}, approximate diagonalization~\cite{Wurtz2020VariationalSchrieffer, Wurtz2020EmergentConservation}, open quantum systems \cite{Santos2021GeneralizedTransitionless,Passarelli2022OptimalQuantum}, and classical Hamiltonian systems \cite{Gjonbalaj2022CounterdiabaticDriving}, and they have been used as a sensitive probe for quantum phase transitions~\cite{Hatomura2021ControllingAndExploring,Takahashi2024ShortcutsToAdiabaticity}, quantum chaos~\cite{Bhattacharjee2023ALanczosApproach}, and singular macroscopic degeneracies~\cite{Sugiura2021AdiabaticLandscape}.

In most applications, variational CD driving significantly improves the driving quality, such as the fidelity to the target state, compared to the bare time evolution without adding the variational AGP\@. Nevertheless, the fidelity often remains far from unity, particularly for many-body and random systems (e.g., Ref.~\cite{Hartmann2022PolynomialScaling}). This is presumably due to two key insufficiencies in the driving protocols determined by variational CD driving. First, the driving protocol is ignorant of which eigenstate is targeted, for example, whether it is the ground state or the highest energy state. Second, the local driving coefficient of the protocol is often determined by the system's local parameters~\cite{Sels2017Minimizing, Hartmann2019RapidCounter} and fails to globally integrate nonlocal information, which is unfavorable for controlling highly entangled eigenstates of many-body systems.

To overcome these two difficulties, this paper focuses on a fundamental yet unaddressed aspect of variational AGP\@. The variational method has always employed a specific form of action. However, any functionals that attain their minimum at the exact AGP could, in principle, play the role of action for determining a variational AGP\@. There should be an infinite number of such functionals. Different actions should yield different approximated AGPs and may help overcome the key issues of the existing method.
Although some literature mentions an alternative action involving a Gibbs state~\cite{Sels2017Minimizing,Takahashi2024ShortcutsToAdiabaticity}, this action is difficult to compute in many-body systems without compromising practicality.

In this paper, we materialize this observation by both advancing the fundamental theory of the AGP and developing a computational technique, revealing that the conventional approach did not fully exploit the potential of variational CD driving. We find that the algebraic characterization of the AGP is not unique, and we introduce an infinite variety of characterizations. By exploiting this freedom, we develop a systematic framework for constructing actions tailored to specific purposes, termed \textit{weighted actions}. The weighted action overcomes the above two insufficiencies of the conventional method. First, it allows us to assign weights to relevant energy eigenstates to reduce the approximation error of important matrix elements.
Second, the weighted action contains nonlocal terms, allowing the resulting variational AGPs to reflect nonlocal information of the system effectively.

We show that this theoretical advance can be practically implemented by using computer algebra in general spin and fermionic systems. Computer algebra is the computation performed with algebraic relations between operators rather than matrix representations~\cite{Steeb2010QuantumMechanics}, and it can compute variational AGPs from weighted actions in polynomial time with respect to the system size. In particular, we construct a concrete algorithm for general spin-$1/2$ systems and provide an explicit computational time analysis, extending similar algebraic methods implicitly used in some literature to compute AGPs~\cite{Xie2022VariationalCounterdiabatic, Barone2024CounterdiabaticOptimized}. An efficient C++ implementation is available in Ref.~\cite{GitHub}.

This proposal, which we term the \textit{weighted variational method}, is practical and widely applicable, as it can---in principle---generalize any previous application of variational AGPs by simply replacing the conventional action with weighted actions. This generalization is experimentally implementable without substantial new difficulties, such as new driving fields/interactions or faster control of the driving fields. To demonstrate the power of our proposal, we apply it to quantum control along the ground state of an arbitrary time-dependent Hamiltonian by designing a weighted action that emphasizes lower-energy eigenstates. We numerically test this application in quantum Ising models with random couplings. The final ground-state fidelity increases in almost all samples, and the median increase reaches, for example, 19 times for spin-glass systems. This enhancement is achieved even in the short-duration limit and remains valid for larger system sizes.

The remainder of this paper is organized as follows. 
In Secs.~\ref{sec:general-framework}--\ref{sec:results}, we present our proposal at three different levels of generality. 
In Sec.~\ref{sec:general-framework}, we develop the most general framework, including the extended algebraic characterizations of the AGP, weighted actions, and the numerical algorithm with computer algebra.
Section~\ref{sec:ground_state} is at the second level of generality, where we specialize the framework to ground-state evolution. This specialization is still valid for arbitrary quantum systems. 
Section~\ref{sec:results} presents the most concrete results obtained from numerical simulations of quantum Ising models.
Section~\ref{sec:analysis} analyzes the possible mechanisms behind the improvement in fidelity using numerical analysis and the quantum speed limit formula. 
Section~\ref{sec:conclusion} concludes the paper.

\section{General framework 
\label{sec:general-framework}}

\subsection{Setup}

We start with a general setup: consider a general quantum system---few-body or many-body---with a parameterized Hamiltonian $H(\lambda)$. Without loss of generality, we assume that $\lambda$ is scalar and takes values $0\leq \lambda\leq1$. We aim to realize an adiabatic process from $\lambda = 0$ to $\lambda = 1$, keeping the population of each eigenstate intact except at degeneracies. If we are allowed to take a very long time, this goal is achieved simply by driving the system with the Hamiltonian $H(\lambda)$ with slowly modulating $\lambda$ from 0 to 1, as ensured by the adiabatic theorem~\cite{Kato1950OnTheAdiabaticTheorem}. However, practical driving protocols must achieve the process in a short time to combat decoherence and dissipation.

\begin{figure}
    \includegraphics{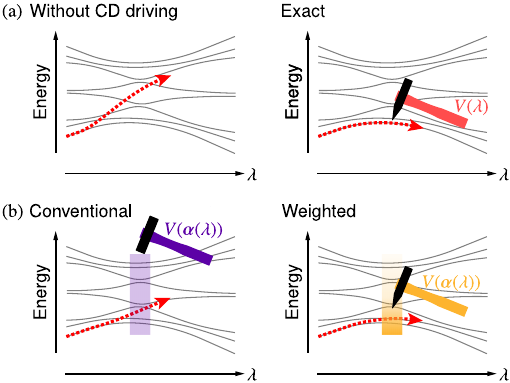}
    \caption{Schematics of the methods of CD driving. (a) Concept of exact CD driving, depicted in the energy eigenbasis of the original Hamiltonian $H(\lambda)$. Gray curves represent the energy of the eigenstates of $H(\lambda)$. Without CD driving, the population escapes from the target eigenstates (e.g., the ground state in this figure) during time evolution, as depicted by the red curve (left). CD driving uses a driving force $V(\lambda) =\Phi(\lambda)$, depicted as a ``hammer'' hitting the system, to ensure that the system correctly tracks the target eigenstate during time evolution (right). Exact CD driving keeps the system perfectly on the target, but it is impractical for many-body systems.
    (b) Comparison between the conventional and weighted variational methods of CD driving. The conventional variational method tries to design a hammer $V(\bm\alpha(\lambda))$ that hits all eigenstates equally well (left). However, this goal may not be fully achieved with a limited number of driving fields. In contrast, our weighted variational method designs a specialized hammer for important eigenstates (e.g., low-energy states in this figure), providing a tailored approach to a specific problem (right). As a result, our method can more effectively suppress the escape of the population from the target eigenstates, as shown by the red curves.%
    \label{fig:concept}}
\end{figure}

A quick adiabatic process with an arbitrary schedule $\lambda_t$ can be achieved by CD driving.
Consider adding a driving Hamiltonian of the form $\dot{\lambda}_{t}V(\lambda_{t})$, where the dot denotes the time derivative, and $V(\lambda)$ is an operator depending on $\lambda$. We evolve the system according to the Schr\"{o}dinger equation,
\begin{equation}
i\hbar\partial_{t}\ket{\psi(t)} = [H(\lambda_{t})+\dot{\lambda}_{t}V(\lambda_{t})]\ket{\psi(t)}.
\label{eq:Schroedinger}
\end{equation}
By introducing the spectral decomposition, 
\begin{equation}
H(\lambda) = \sum_{n = 1}^{D}\epsilon_{n}(\lambda)\ketbra{\phi_{n}(\lambda)}{\phi_{n}(\lambda)}
\label{eq:eigen_decomposition}
\end{equation}
with $\epsilon_{1}(\lambda) \leq \cdots \leq \epsilon_{D}(\lambda)$, we can rewrite the Schr\"{o}dinger equation with the probability amplitudes of the eigenstates, $c_{n}(t) \coloneqq \braket{\phi_{n}(\lambda_{t})}{\psi(t)}$:
\begin{equation}
i\hbar\dot{c}_{n}(t) = \kappa_{n}(t)c_{n}(t)+\dot{\lambda}_{t}\sum_{\substack{m\\
(m \neq  n)
}
}[V(\lambda_{t})-\Phi(\lambda_{t})]_{nm}c_{m}(t).
\label{eq:Schroedinger_coeff_driving}
\end{equation}
Here, $\kappa_{n}(t) \coloneqq \epsilon_{n}(\lambda_{t})+\dot{\lambda}_{t}[V(\lambda_{t})]_{nn} - i\hbar\dot{\lambda}_{t}\braket{\phi_{n}(\lambda_{t})}{\partial_{\lambda}\phi_{n}(\lambda_{t})}$ is a real-valued function, $[\cdots]_{nm}$ is the shorthand for $\bra{\phi_{n}(\lambda_{t})}\cdots\ket{\phi_{m}(\lambda_{t})}$, and $\Phi(\lambda)$ is a Hermitian operator called the AGP,
\begin{equation}
\Phi(\lambda) \coloneqq i\hbar \sum_{\substack{nm\\(n \neq  m)}} \ket{\phi_{n}(\lambda)}\braket{\phi_{n}(\lambda)}{\partial_{\lambda}\phi_{m}(\lambda)}\bra{\phi_{m}(\lambda)}.
\label{eq:Hcd}
\end{equation}
In the right-hand side of Eq.~\eqref{eq:Schroedinger_coeff_driving}, the first term $\kappa_{n}(t)c_{n}(t)$ only changes the complex phase of $c_{n}(t)$, keeping the population $\vert c_{n}(t)\vert^{2}$ intact, but the second term causes undesired nonadiabatic transitions, i.e., transitions from one eigenstate to another. The nonadiabatic transitions can be perfectly suppressed by choosing $V(\lambda) = \Phi(\lambda)$, which is called CD driving~\cite{Demirplak2003AdiabaticPopulation,Demirplak2005AssistedAdiabatic,Demirplak2008OnTheConsistencyExtremal,Berry2009Transitionless}. In the energy eigenbasis of $H(\lambda)$, the CD driving term $V(\lambda)$ is intuitively depicted as a ``hammer'' that exerts an external force on the system to make it follow an eigenstate faithfully [Fig.~\ref{fig:concept}(a)]. 

The exact CD driving, however, has two crucial drawbacks when applied to many-body systems.
First, the calculation of the AGP involves the diagonalization of the Hamiltonian $H(\lambda)$, which is usually impossible for many-body systems. 
Second, the nonlocal many-body interactions in the AGP hinder experimental implementation. Implementable control fields are usually local, such as one- and two-body. It has been proposed to use Floquet engineering to effectively expand the implementable set of control fields, including nonlocal interactions~\cite{Petiziol2024QuantumControl,Petiziol2018FastAdiabatic,Petiziol2019AcceleratingAdiabatic,Claeys2019FloquetEngineering}, but the number of independently controllable fields is still practically limited due to the constraint on the highest achievable control frequency. It thus remains difficult to implement the full AGP in many-body systems.

\subsection{Conventional variational method}

{Given these practical difficulties in implementing the full AGP, it is desirable to establish an optimal approximation of CD driving within the limited control fields at hand. Reference~\cite{Sels2017Minimizing} addressed this problem by proposing a variational method to determine approximated AGPs.}

The variational method starts with an algebraic characterization of the exact AGP~\cite{Jarzynski2013GeneratingShortcuts},
\begin{equation}
[H(\lambda),\partial_{\lambda}H(\lambda)-i\hbar^{-1}[H(\lambda),\Phi(\lambda)]] = 0.
\label{eq:algebraic_char_1}
\end{equation}
This characterization is equivalent to the following variational formula that minimizes an action $\mathcal{S}_{\lambda}^{(1)}[V]$~\cite{Sels2017Minimizing}:%
\begin{subequations}
\label{eq:variational_char-1}
\begin{align}
\Phi(\lambda) &  \in \argmin_{V}\mathcal{S}_{\lambda}^{(1)}[V],
\label{eq:S_min_1-1}\\
\mathcal{S}_{\lambda}^{(1)}[V]\, & \! \coloneqq \bigl\Vert\partial_{\lambda}H(\lambda)-i\hbar^{-1}[H(\lambda),V]\bigr\Vert^{2},
\label{eq:S_min_2-1}
\end{align}
\end{subequations}
where $\Vert A\Vert \coloneqq \sqrt{\tr(A^{\dagger}A)}$ denotes the Hilbert--Schmidt norm. The minimization is performed over any Hermitian operator $V$.
Note that this variational formula only fixes the part of $V$ noncommuting with $H(\lambda)$, as the action $\mathcal{S}_{\lambda}^{(1)}[V]$ is independent of the part of $V$ that commutes with $H(\lambda)$. Thus, the minimizer of $\mathcal{S}_{\lambda}^{(1)}[V]$ is not unique, as indicated by the symbol ``$\in $'', and $\Phi(\lambda)$ is one of them. The meaning of the superscript ``$(1)$'' will be clarified later.
Since the action $\mathcal{S}_{\lambda}^{(1)}[V]$ is minimized by the true AGP, the difference $\mathcal{S}_{\lambda}^{(1)}[V] - \mathcal{S}_{\lambda}^{(1)}[\Phi(\lambda)]$ can be interpreted as quantifying how far $V$ is apart from $\Phi(\lambda)$ without explicitly using the matrix elements of $V$ and $\Phi(\lambda)$.

Motivated by this interpretation, the variational method determines the variational AGP by minimizing the action $\mathcal{S}^{(1)}_{\lambda}[V]$ over an ansatz consisting of experimentally implementable Hermitian operators. For concreteness, we consider an ansatz of the form
\begin{equation}
 V(\bm{\alpha}(\lambda)) = \sum_{\mu = 1}^{M}\alpha_{\mu}(\lambda)A_{\mu},
 \label{eq:ansatz}
\end{equation}
where $\bm{\alpha}(\lambda) \equiv (\alpha_{1}(\lambda),\ldots,\alpha_{M}(\lambda))$ are $\lambda$-dependent scalar coefficients to be optimized, and $A_{1},\dots,A_{M}$ are $\lambda$-independent Hermitian operators. For practical applications, $A_{\mu}$'s are taken to be experimentally feasible operators. For each $\lambda$, the coefficient $\bm{\alpha}(\lambda)$ is determined by the minimization,
\begin{align}
\bm{\alpha}^{(1)}(\lambda) &  \coloneqq \argmin_{\bm{\alpha} \in \mathbb{R}^{M}}\mathcal{S}_{\lambda}^{(1)}[V(\bm{\alpha})].
\label{eq:alpha-1}
\end{align}
Notably, the action $\mathcal{S}_{\lambda}^{(1)}[V(\bm{\alpha})]$ can be calculated using algebraic relations between the operators appearing in $H(\lambda)$ and $A_{\mu}$, such as the spin operators for spin systems and the creation/annihilation operators for fermionic systems~\cite{Sels2017Minimizing, Hatomura2021ControllingAndExploring}, without diagonalizing the Hamiltonian. This property makes the variational method practical for many-body systems.

We can obtain further insight into this variational method by expanding the action with the eigenbasis of $H(\lambda)$:
\begin{equation}
\mathcal{S}_{\lambda}^{(1)}[V] = \frac{1}{\hbar^{2}}\!\!\sum_{nm}[\epsilon_{m}(\lambda)-\epsilon_{n}(\lambda)]^{2}\bigl\vert[V-\Phi(\lambda)]_{mn}\bigr\vert^{2}+\mathrm{const.},
\label{eq:action_explicit}
\end{equation}
where ``const.''~refers to a term independent of $V$. This expression shows that the minimization of $\mathcal{S}_{\lambda}^{(1)}[V]$ makes $V_{mn}$ as close to $[\Phi(\lambda)]_{mn}$ as possible within a given ansatz.

\vspace{.25em}

\subsection{Weighted variational method 
\label{subsec:weighted_variational_general}}

Although the driving coefficient $\bm{\alpha}^{(1)}(\lambda)$ is determined through an optimization, it is not guaranteed to provide the best possible driving protocol because the optimization depends on the choice of variational functional, i.e., how we measure the ``distance'' between $V$ and the exact AGP\@. In fact, the conventional variational functional $\mathcal{S}^{(1)}_\lambda$ has two significant limitations, despite its widespread applications. 

First, the conventional action does not reflect which eigenstate is the target of the driving. Equation~\eqref{eq:action_explicit} shows that the action $\mathcal{S}_{\lambda}^{(1)}[V]$ is the sum over the matrix elements $\vert[V-\Phi(\lambda)]_{mn}\vert^{2}$ for all $(m,n)$, and the minimization of $\mathcal{S}_{\lambda}^{(1)}[V]$ tries to suppress all of these $O(D^{2})$ terms. On the other hand, minimization is carried out with only $M$ degrees of freedom, which is usually much smaller than $O(D^{2})$. Thus, the minimization problem is severely underparameterized, and suppressing one matrix element may conflict with suppressing another. This fact suggests that we can improve variational CD driving by introducing an alternative action that assigns weights to relevant energy eigenstates, prioritizing the suppression of errors in important elements over unimportant ones. 

Second, the action $\mathcal{S}_{\lambda}^{(1)}[V]$ often fails to incorporate nonlocal information into variational AGPs. In simple systems, the coefficient $\alpha_{\mu}^{(1)}(\lambda)$ from Eq.~\eqref{eq:alpha-1} often depends only on local parameters. For example, in the Ising model with $A_{\mu}$ acting on one spin, the coefficient $\alpha_{\mu}(\lambda)$ is determined solely by the local magnetic field and the nearest-neighbor coupling constants around the spin~\cite{Sels2017Minimizing}. More generally, the action $\mathcal{S}_{\lambda}^{(1)}[V(\bm{\alpha})]$ is always a sum of local contributions whenever $H(\lambda)$ and $A_{\mu}$'s are sums of local operators (see Appendix~\ref{subsec:apdx_nonlocal}). 
This locality is unfavorable for approximating the exact AGP, which is usually nonlocal and sensitive to small perturbations in complex many-body systems~\cite{Lychkovskiy2017TimeScale}. Therefore, variational CD driving may be improved by using an alternative action that incorporates nonlocal information. 

Motivated by these arguments, we introduce a systematic framework for constructing alternative actions, termed \textit{weighted actions}. Alternative actions should solve the above two issues of the conventional action. At the same time, they should be easy to compute without diagonalizing the Hamiltonian.

We start with developing alternative algebraic characterizations of the AGP $\Phi(\lambda)$. Let us introduce a fictitious Hamiltonian $\mathcal{P}_{\lambda}(H(\lambda)$), where
\begin{equation}
\mathcal{P}_{\lambda}(x) = \sum_{k = 0}^{K}p_{k}(\lambda)x^{k}
\label{eq:polynomial}
\end{equation}
is an arbitrary polynomial with an arbitrary degree $K$ and possibly $\lambda$-dependent coefficients $p_{k}(\lambda)$. The key observation is that, regardless of the choice of $\mathcal{P}_{\lambda}(x)$, the AGP $\Phi(\lambda)$ of the original Hamiltonian in Eq.~\eqref{eq:Hcd} satisfies the following algebraic characterization (see Appendix~\ref{subsec:apdx_derivations} for derivation):
\begin{equation}
[H(\lambda),\partial'_{\lambda}\mathcal{P}_{\lambda}(H(\lambda))-i\hbar^{-1}[\mathcal{P}_{\lambda}(H(\lambda)),\Phi(\lambda)]] = 0.
\label{eq:algebraic_char-general}
\end{equation}
Here, $\partial'_{\lambda}$ is the $\lambda$-derivative that ignores the $\lambda$-dependence of the coefficients $p_k(\lambda)$. More precisely, it is defined by $\partial'_{\lambda}\mathcal{P}_{\lambda}(x(\lambda)) = \sum_{k\geq1}p_{k}(\lambda)\partial_{\lambda}[x(\lambda)^{k}]$ for any $\lambda$-dependent quantity $x(\lambda)$. This algebraic characterization extends Eq.~\eqref{eq:algebraic_char_1} and introduces an extra degree of freedom to the theory of the AGP.

This algebraic characterization can be converted into the following variational formula over a weighted action $\mathcal{S}_{\lambda}^{(\mathcal{P})}[V]$ (see Appendix~\ref{subsec:apdx_derivations} for derivation):%
\begin{subequations}
\label{eq:variational_char-general}
\begin{align}
\Phi(\lambda) &  \in \argmin_{V}\mathcal{S}_{\lambda}^{(\mathcal{P})}[V],
\label{eq:S_min_1-general}\\
\mathcal{S}_{\lambda}^{(\mathcal{P})}[V]\, & \! \coloneqq \bigl\Vert\partial'_{\lambda}\mathcal{P}_{\lambda}(H(\lambda))-i\hbar^{-1}[\mathcal{P}_{\lambda}(H(\lambda)),V]\bigr\Vert^{2}.
\label{eq:S_min_2-general}
\end{align}
\end{subequations}
This expression extends Eq.~\eqref{eq:variational_char-1}, and the minimizer is unique up to the part of $V$ that commutes with $\mathcal{P}_\lambda(H(\lambda))$. When $\mathcal{P}_\lambda(x)$ is a degree-one polynomial ($K = 1$), the action $\mathcal{S}_{\lambda}^{(\mathcal{P})}[V]$ is equivalent to the conventional action $\mathcal{S}_{\lambda}^{(1)}[V]$ except for an overall multiplicative factor, which follows from $\partial'_{\lambda}[p_{1}(\lambda)H(\lambda)] = p_{1}(\lambda)\partial_{\lambda}H(\lambda)$ and $\partial'_{\lambda}p_{0}(\lambda) = 0$. The difference $\mathcal{S}_{\lambda}^{(\mathcal{P})}[V]-\mathcal{S}_{\lambda}^{(\mathcal{P})}[\Phi(\lambda)]$ of this action can be interpreted as measuring the ``distance'' between $V$ and $\Phi(\lambda)$ in a way different from the original action and dependent on the choice of the polynomial. This distance measure is still written without using the matrix elements of $V$ and $\Phi(\lambda)$. 

Our proposal, the \textit{weighted variational method}, determines the driving coefficients $\bm{\alpha}(\lambda)$ by the constrained minimization of the weighted action:
\begin{equation}
\bm{\alpha}^{(\mathcal{P})}(\lambda) \coloneqq \argmin_{\bm{\alpha} \in \mathbb{R}^{M}}\mathcal{S}_{\lambda}^{(\mathcal{P})}[V(\bm{\alpha})],
\label{eq:alpha-general}
\end{equation}
where $V(\bm\alpha)$ denotes the ansatz in Eq.~\eqref{eq:ansatz}.
While the minimization of $\mathcal{S}_{\lambda}^{(\mathcal{P})}[V]$ over all Hermitian operators gives the exact AGP irrespectively of $\mathcal{P}_\lambda(x)$, the constrained minimization generally gives different results depending on $\mathcal{P}_\lambda(x)$.

The weighted variational method extends the conventional method in the two aforementioned aspects. First, it enables us to assign weights to some of the matrix elements $\vert[V-\Phi(\lambda)]_{mn}\vert^{2}$. This property is clearly seen by expanding $\mathcal{S}_{\lambda}^{(\mathcal{P})}[V]$ in terms of the Hamiltonian eigenbasis (see Appendix~\ref{subsec:apdx_derivations} for derivation):
\begin{align}
\mathcal{S}_{\lambda}^{(\mathcal{P})}[V] &  = \frac{1}{\hbar^{2}}\sum_{nm}\bigl[\mathcal{P}_{\lambda}(\epsilon_{m}(\lambda))-\mathcal{P}_{\lambda}(\epsilon_{n}(\lambda))\bigr]^{2}\bigl\vert[V-\Phi(\lambda)]_{mn}\bigr\vert^{2}\nonumber \\[-0.7em]
 & \hspace{14em}+\mathrm{const}.
\label{eq:actionK_explicit-general}
\end{align}
Compared with the conventional action in Eq.~\eqref{eq:action_explicit}, the deviation between $V_{mn}$ and $[\Phi(\lambda)]_{mn}$ receives an additional weight,
\begin{equation}
w_{mn}^{(\mathcal{P})}(\lambda) \coloneqq \frac{\bigl[\mathcal{P}_{\lambda}(\epsilon_{m}(\lambda))-\mathcal{P}_{\lambda}(\epsilon_{n}(\lambda))\bigr]^{2}}{[\epsilon_{m}(\lambda)-\epsilon_{n}(\lambda)]^{2}}.
\label{eq:weight_def-general}
\end{equation}
In a practical application, the polynomial $\mathcal{P}_{\lambda}(x)$ should be designed to assign weights to important elements for a specific purpose. One example is to assign weights to low-energy eigenstates to assist the adiabatic process along the ground state, which will be explored in later sections.

Second, the weighted action contains nonlocal terms. Even if the original Hamiltonian contains only local couplings, the fictitious Hamiltonian for $K\geq2$ has nonlocal interaction terms due to $H(\lambda)^{2}, H(\lambda)^{3}, \dots$ appearing in $\mathcal{P}_\lambda(H(\lambda))$. Due to these nonlocal interactions, the weighted actions contain nonlocal terms, enabling us to effectively incorporate nonlocal information into the driving coefficients $\bm{\alpha}^{(\mathcal{P})}(\lambda)$.

The comparison between the conventional and weighted variational methods is summarized more intuitively in Fig.~\ref{fig:concept}(b). 
The ``hammer,'' i.e., the driving term $V$, from the weighted variational method is tuned to work best for a specific part of the energy spectrum, such as the low-energy states. This is in contrast to the hammer from the conventional method, which is not optimized for the specific part of the spectrum. 
By this tuning, the hammer is expected to hit the target state more accurately, thereby suppressing deviations from the target dynamics more effectively.

In general, a higher-degree polynomial $\mathcal{P}_{\lambda}(x)$ is favorable for designing a better action because we have more freedom to choose the coefficients $p_k(\lambda)$. However, a higher-degree polynomial requires more computation time to minimize the action (Sec.~\ref{subsec:algorithm_general}). Therefore, the degree $K$ of the polynomial $\mathcal{P}_{\lambda}(x)$ should be set to the highest possible number under a practical limitation on computational time.

We remark that if the Hamiltonian has a conserved charge $J(\lambda)$ satisfying $[H(\lambda), J(\lambda)]=0$, we can make use of $J(\lambda)$ to construct more general variational characterizations and, hence, more fine-tuned weighted actions. This could be used to suppress nonadiabatic transitions more selectively, as discussed in more detail in Appendix~\ref{subsec:apdx_conserved_charge}\@.

\subsection{Computational algorithm using computer algebra 
\label{subsec:algorithm_general}}

The minimization in Eq.~\eqref{eq:alpha-general} is efficiently solved by an algebraic method, avoiding the costly manipulation of matrix expressions of the Hamiltonian and driving operators. This is enabled because the algebraic expression of $\mathcal{S}^{(\mathcal{P})}_\lambda[V]$ in Eq.~\eqref{eq:S_min_2-general} involves no explicit matrix elements.

We first fix $\lambda$ and discuss the calculation for a single $\lambda$. Inserting the ansatz in Eq.~\eqref{eq:ansatz} into the definition of $\mathcal{S}_{\lambda}^{(\mathcal{P})}[V]$ in Eq.~\eqref{eq:S_min_2-general}, we obtain
\begin{align}
 \mathcal{S}_{\lambda}^{(\mathcal{P})}[V(\bm{\alpha})] 
 &= -\frac{2}{\hbar^{2}}\sum_{\mu=1}^{M} r_{\mu}^{(\mathcal{P})}(\lambda)\alpha_{\mu}
 +\frac{1}{\hbar^{2}}\sum_{\mu,\nu=1}^{M} Q_{\mu\nu}^{(\mathcal{P})}(\lambda)\alpha_{\mu}\alpha_{\nu}
 \nonumber \\
 &\hspace{12em} {\vphantom{A}+\mathrm{const.}},
\label{eq:actionK_ansatz-general}
\end{align}
where we define%
\begin{subequations}
\label{eq:def_Qr-general}
\begin{align}
Q_{\mu\nu}^{(\mathcal{P})}(\lambda) &  \coloneqq -\tr\left\{ [\mathcal{P}_{\lambda}(H(\lambda)),A_{\mu}][\mathcal{P}_{\lambda}(H(\lambda)),A_{\nu}]\right\} ,
\label{eq:def_Qr_Q-general}\\
r_{\mu}^{(\mathcal{P})}(\lambda) &  \coloneqq  i\hbar\,\tr\left\{ \partial'_{\lambda}\mathcal{P}_{\lambda}(H(\lambda))[\mathcal{P}_{\lambda}(H(\lambda)),A_{\mu}]\right\} ,
\label{eq:def_Qr_r-general}
\end{align}
\end{subequations}
for $\mu,\nu = 1,\dots,M$. The optimal $\bm{\alpha}$ is found by solving $\partial\mathcal{S}_{\lambda}^{(\mathcal{P})}[V(\bm{\alpha})]/\partial\alpha_{\mu} = 0$, which turns into a coupled linear equation,
\begin{equation}
\sum_{\nu = 1}^{M}Q_{\mu\nu}^{(\mathcal{P})}(\lambda)\alpha_{\nu} = r_{\mu}^{(\mathcal{P})}(\lambda)\qquad(\mu = 1,\dots,M).
\label{eq:linear_equation-general}
\end{equation}
Thus, once we find $Q_{\mu\nu}^{(\mathcal{P})}(\lambda)$ and $r_{\mu}^{(\mathcal{P})}(\lambda)$, we can easily get the optimal driving coefficients $\bm{\alpha}^{(\mathcal{P})}(\lambda)$ by solving this $M$-variable linear equation with a standard numerical method.

We can compute $Q_{\mu\nu}^{(\mathcal{P})}(\lambda)$ and $r_{\mu}^{(\mathcal{P})}(\lambda)$ efficiently using computer algebra for spin systems with an arbitrary spin quantum number and fermionic systems. Computer algebra stores operators, such as a Hamiltonian, by their algebraic representations composed of elementary operators, such as spin operators for spin systems and creation/annihilation operators for fermionic systems~\cite{Steeb2010QuantumMechanics}. 
It performs computations such as addition, multiplication, and trace by directly manipulating the algebraic representations, just as humans do by hand.

In physically natural setups, the computational time of $Q_{\mu\nu}^{(\mathcal{P})}(\lambda)$ and $r_{\mu}^{(\mathcal{P})}(\lambda)$ with computer algebra scales polynomially in system size, which is much more efficient than the computation with matrix representations. Consider a system on a lattice with $N$ sites. The matrix representation of an operator usually contains an exponential number of nonzero elements with respect to $N$. Thus, the computation time with matrix representations scales exponentially with the system size. In contrast, the computation time with computer algebra is roughly proportional to the number of elementary operators involved in the computation. Assuming that the number of elementary operators in $H(\lambda)$ and $\{A_\mu\}$ is a polynomial of $N$, which is valid in almost all physically natural setups, the number of elementary operators appearing in $Q_{\mu\nu}^{(\mathcal{P})}(\lambda)$ and $r_{\mu}^{(\mathcal{P})}(\lambda)$ is also a polynomial of $N$. Thus, the computational time with algebraic representations scales polynomially with the system size.

For concreteness, we below focus on a general spin-$1/2$ system consisting of $N$ spins, labeled by $i = 1,\dots,N$. We use $X_i$, $Y_i$, and $Z_i$ to denote the Pauli operators acting on the $i$th spin. The algebraic representation of an operator has the form of a sum of terms, and each term consists of a scalar coefficient and a tensor product of Pauli operators, e.g., $1.5X_1Y_2$, where we omit the identity operators acting on the remaining spins. Computer algebra manipulates this representation using algebraic relations such as $X_{j}Y_{j} = iZ_{j}$, $X_{j}X_{j} = I$, $\tr X_{j} = 0$, $\tr I = 2^{N}$, and so on. We develop a detailed algorithm for basic operations such as addition, multiplication, trace, and commutators in Appendices~\ref{subsec:apdx_elementary_operation} and \ref{subsec:apdx_trace_commutator}\@. 
These algorithms inherit basic ideas from existing computer algebra frameworks~\cite{Baylis1996PauliAlgebra,Filip2010SdCasSpinDynamics,YiZhuangMathematicaPackages,Steeb2010QuantumMechanics,Barone2024CounterdiabaticOptimized,Loizeau2025QuantumMany} and make them more efficient and suitable for our purposes. Using these basic algorithms, we present a concrete algorithm for computing $\bm{\alpha}^{(\mathcal{P})}(\lambda)$ for a fixed $\lambda$ in Appendix~\ref{subsec:apdx_algorithm-general}.

Our computer algebra algorithm for general spin-$1/2$ systems can compute $Q_{\mu\nu}^{(\mathcal{P})}(\lambda)$ and $r_{\mu}^{(\mathcal{P})}(\lambda)$ in $O(N^{K})$ time, as shown in Appendix~\ref{subsec:apdx_algorithm-general}. This computation time is analyzed under the following four assumptions. First, $H(\lambda)$ and $A_{1},\dots,A_{M}$ are $k$-local operators with $k$ small and independent of $N$. Second, the Hamiltonian consists of $O(N)$ terms. Third, the driving operators $\{A_{\mu}\}$ satisfy either one of the following conditions: (i) the number of driving operators is $M = O(N)$, and each $A_{\mu}$ consists of $O(N^{0})$ terms; (ii) the number of driving operators is $M = O(N^{0})$, and each $A_{\mu}$ consists of $O(N)$ terms. In either case, the driving operators $\{A_{\mu}\}$ have $O(N)$ terms in total. Fourth, these $O(N)$ terms of $\{A_{\mu}\}$ are evenly distributed over $O(N)$ spins so that each spin is acted on only by $O(N^{0})$ terms. These four assumptions are satisfied by most physically natural setups, such as local control of the Ising and Heisenberg models on a regular lattice of any dimension. Under these assumptions, the computational time scales as $O(N^{K})$, where we neglect a multiplicative factor that mildly depends on $K$ because we will focus on a small $K$ such as $K\leq5$ below. Due to this polynomial scaling, our algorithm can treat from tens to hundreds of spins (see Sec.~\ref{subsec:numerical_setup} for an example of the computational time).

This $O(N^{K})$-time algebraic computation is followed by a standard numerical method to solve the linear equation in Eq.~\eqref{eq:linear_equation-general}. This step takes $O(M^3)$ time if we use exact methods, such as the Householder QR factorization~\cite{Trefethen1997NumericalLinear}. Alternatively, iterative (approximate) methods, such as the conjugate gradient method, can solve Eq.~\eqref{eq:linear_equation-general} in $O(R M^2)$ time, where $R$ is the number of iterations required for convergence within a specified tolerance~\cite{Trefethen1997NumericalLinear}. Either way, the $M\times M$ matrix $Q_{\mu\nu}^{(\mathcal{P})}(\lambda)$ is efficiently handled in polynomial time of $N$ since $M$ is of $O(N)$ or less.

So far, we have considered the calculation for a single value of $\lambda$. To calculate the driving coefficients for the entire range of $\lambda$, we can, of course, repeat the calculation independently for many values of $\lambda$. However, we can perform the calculation more efficiently if the Hamiltonian has the form 
\begin{equation}
H(\lambda) = \sum_{\gamma = 1}^{\Gamma}f_{\gamma}(\lambda)F_{\gamma},
\label{eq:Hamiltonian_fF}
\end{equation}
where $F_{1},\dots,F_{\Gamma}$ are operators independent of $\lambda$, $f_{1}(\lambda),\dots,f_{\Gamma}(\lambda)$ are scalar coefficients, and $\Gamma$ is a small number such as 2 or 3. This condition on the Hamiltonian is satisfied in a large portion of the applications of variational CD driving reviewed in Sec.~\ref{sec:introduction}\@. By inserting Eq.~\eqref{eq:Hamiltonian_fF} into the expressions of $Q_{\mu\nu}^{(\mathcal{P})}(\lambda)$ and $r_{\mu}^{(\mathcal{P})}(\lambda)$ in Eq.~\eqref{eq:def_Qr-general}, we can separate the scalar coefficients dependent on $\lambda$ from the traces of operators independent of $\lambda$. Then, the evaluation of the traces, which is the most time consuming step, is common for all $\lambda$ and needs to be done only once (see Appendix \ref{subsec:apdx_time-dependent-general} for details). With our concrete algorithm for spin-$1/2$ systems, the computational complexity for these traces scales as $O(N^{K})$, similarly to the single-$\lambda$ case (see Appendix~\ref{subsec:apdx_algorithm-general}).

\section{Weighted variational method applied to ground-state evolution 
\label{sec:ground_state}}

\subsection{Theoretical framework}
\label{subsec:ground_state_theory}

We proceed to a more concrete analysis by specializing our general framework to ground-state evolution. We assume that $H(\lambda)$ has a nondegenerate ground state for all $\lambda$, and we consider guiding the quantum state from $\ket{\phi_{1}(0)}$ to $\ket{\phi_{1}(1)}$ along the ground state $\ket{\phi_{1}(\lambda)}$ as closely as possible. For simplicity, we assume that the system is initially prepared in $\ket{\phi_{1}(0)}$ without error.

To assist ground-state evolution, we choose a polynomial $\mathcal{P}_{\lambda}(x)$ that assigns more weights to lower-energy eigenstates. A simple reasoning behind this choice is that we do not need to suppress nonadiabatic transitions between high-energy levels. We provide a further justification in Sec.~\ref{subsec:partial_action}\@. As discussed in Sec.~\ref{subsec:weighted_variational_general}, a larger-degree polynomial $\mathcal{P}_{\lambda}(x)$ can give a better driving protocol but requires more computational time. Therefore, we specify $\mathcal{P}_{\lambda}(x)$ for each degree $K\geq 1$, allowing users to choose $K$ based on realistic limitations on computational time. For every $K\geq 1$, we heuristically take a polynomial of the form
\begin{equation}
\mathcal{P}_{\lambda}^{\mathrm{GS},K}(x) \coloneqq (x-E_{\lambda}^{(K)})^{K},
\label{eq:polynomial-GS}
\end{equation}
where GS stands for ``ground state,'' and $E_{\lambda}^{(K)}$ is a constant optimized later so that the weights are concentrated on low-energy states as much as possible. This polynomial $\mathcal{P}_{\lambda}^{\mathrm{GS},K}(x)$ corresponds to choosing the coefficients $p_{k}(\lambda) = \binom{K}{k} (-E_{\lambda}^{(K)})^{K-k}$ in Eq.~\eqref{eq:polynomial}, where $\binom{x}{y}$ denotes the binomial coefficient. Below, we use the notation $\mathcal{S}_{\lambda}^{(K)}[V] \equiv \mathcal{S}_{\lambda}^{(\mathcal{P}^{\mathrm{GS},K})}[V]$, $\bm{\alpha}^{(K)}(\lambda) \equiv \bm{\alpha}^{(\mathcal{P}^{\mathrm{GS},K})}(\lambda)$, $Q_{\mu\nu}^{(K)}(\lambda) \equiv  Q_{\mu\nu}^{(\mathcal{P}^{\mathrm{GS},K})}(\lambda)$, $r_{\mu}^{(K)}(\lambda) \equiv  r_{\mu}^{(\mathcal{P}^{\mathrm{GS},K})}(\lambda)$, and $w_{mn}^{(K)} \equiv  w_{mn}^{(\mathcal{P}^{\mathrm{GS},K})}(\lambda)$ for conciseness. In this notation, the minimization problem for determining the coefficient reads
\begin{equation}
    \bm{\alpha}^{(K)}(\lambda) = \argmin_{\bm{\alpha}\in \mathbb{R}^M} \mathcal{S}^{(K)}_\lambda [V(\bm{\alpha})].
    \label{eq:minimum-GS}
\end{equation}
The degree-one ($K = 1$) action equals the conventional action in Eq.~\eqref{eq:S_min_2-1}, consistent with the notation $\mathcal{S}_{\lambda}^{(1)}[V]$ for the conventional action.

We analyze the weighted action $\mathcal{S}_\lambda^{(K)}[V]$ by examining the additional weight $w_{mn}^{(K)}(\lambda)$ onto the $(m,n)$th element, introduced in Eq.~\eqref{eq:weight_def-general}. The weight is explicitly calculated as (see Appendix~\ref{subsec:apdx_deriv-weight} for derivation)
\begin{align}
    & w_{mn}^{(K)}(\lambda) 
    \nonumber \\
    &=  \frac{\bigl\{[\epsilon_m(\lambda) - E_\lambda^{(K)}]^K - [\epsilon_n(\lambda) - E_\lambda^{(K)}]^K\bigr\}^2}{[\epsilon_m(\lambda) - \epsilon_n (\lambda)]^2}
    \nonumber \\
    &=\! \sum_{\smash{s = -(K-1)}}^{K-1} \! (K-\vert s\vert)[\epsilon_{n}(\lambda) -E_\lambda^{(K)}]^{K-1-s}[\epsilon_{m}(\lambda) -E_\lambda^{(K)}]^{K-1+s}.
    \label{eq:weight-GS}
\end{align}
This weight is better understood by introducing a state-dependent weight, $w_n^{(K)}(\lambda)\coloneqq K^2[\epsilon_{n}(\lambda) -E_\lambda^{(K)}]^{2K-2} \geq 0$. We can then prove
\begin{equation}
0 \leq  w_{mn}^{(K)}(\lambda) \leq \max\bigl\{w_n^{(K)}(\lambda),w_m^{(K)}(\lambda)\bigr\}.
\label{eq:weight_upper}
\end{equation}
Moreover, when $[\epsilon_n(\lambda)-E_\lambda^{(K)}]$ and $[\epsilon_m(\lambda)-E_\lambda^{(K)}]$ have the same sign, $w_{mn}^{(K)}(\lambda)$ lies between $w_n^{(K)}(\lambda)$ and $w_m^{(K)}(\lambda)$. This fact implies that the weight is approximated as $w_{mn}^{(K)}(\lambda) \simeq w_n^{(K)}(\lambda) \simeq w_m^{(K)}(\lambda)$ when the state-dependent weights further satisfy $w_n^{(K)}(\lambda) \simeq w_m^{(K)}(\lambda)$ (see Appendix~\ref{subsec:apdx_deriv-weight} for the derivation of these properties). Due to these properties, we regard $w_n^{(K)}(\lambda)$ as the weight associated with the $n$th eigenstate for heuristic discussions in the following.

The constant $E_{\lambda}^{(K)}$, which we call the \textit{energy shift}, is determined so that the heuristic weight $w_n^{(K)}(\lambda)$ is concentrated on low-energy states as much as possible. This is achieved by the following optimization:%
\begin{subequations}
\label{eq:E_min}
\begin{align}
E_{\lambda}^{(K)} &  = \argmin_{E \in \mathbb{R}}\Omega_{\lambda}^{(K)}(E),
\label{eq:E_min_1}\\
\Omega_{\lambda}^{(K)}(E)\, & \! \coloneqq \frac{\sum_{n}[\epsilon_{n}(\lambda)-E]^{2K-2}\epsilon_{n}(\lambda)}{\sum_{n}[\epsilon_{n}(\lambda)-E]^{2K-2}}.
\label{eq:E_min_2}
\end{align}
\end{subequations}
The function $\Omega_{\lambda}^{(K)}(E)$ is the average energy over the weight $[\epsilon_{n}(\lambda)-E]^{2K-2}$, and a smaller $\Omega_{\lambda}^{(K)}(E)$ implies that the weight $[\epsilon_{n}(\lambda)-E]^{2K-2}$ is more concentrated on low-energy states. Therefore, by choosing $E_\lambda^{(K)}$ as the minimizer of $\Omega_{\lambda}^{(K)}(E)$, we can make the resulting weight $w_n^{(K)}(\lambda) = K^2 [\epsilon_{n}(\lambda)-E_\lambda^{(K)}]^{2K-2}$ emphasize low-energy states as much as possible. We discuss the typical behavior of the function $\Omega_{\lambda}^{(K)}(E)$ and the energy shift $E_{\lambda}^{(K)}$ in Appendix~\ref{subsec:apdx_Omega}\@.

We remark that the particular polynomial form in Eq.~\eqref{eq:polynomial-GS} is not the only possible choice. For example, we could directly use the general polynomial form in Eq.~\eqref{eq:polynomial} and determine the coefficients $p_1(\lambda),\dots,p_K(\lambda)$ by a multivariate numerical optimization similar to Eq.~\eqref{eq:E_min} so that the corresponding weight concentrates the most in the lower-energy state. Nevertheless, the particular form in Eq.~\eqref{eq:polynomial-GS} is simpler and performs sufficiently well, as demonstrated in the following. Furthermore, the particular form is expected to be robust, as its behavior can be theoretically analyzed and supported (Sec.~\ref{subsec:partial_action}).

\subsection{Computational algorithm}

The numerical implementation of the weighted variational method for ground-state evolution involves two optimization problems, one for the action $\mathcal{S}_{\lambda}^{(K)}[V(\bm\alpha)]$ and the other for $\Omega_{\lambda}^{(K)}(E)$. The minimization of $\mathcal{S}_{\lambda}^{(K)}[V(\bm\alpha)]$ is efficiently performed with computer algebra, as discussed in the general framework in Sec.~\ref{subsec:algorithm_general}\@. The minimization of $\Omega_{\lambda}^{(K)}(E)$ is also efficiently computed using computer algebra. To do so, we rewrite the function $\Omega_{\lambda}^{(K)}(E)$ without explicitly using the energy eigenvalues as
\begin{align}
\Omega_{\lambda}^{(K)}(E) &  = \frac{\tr\left\{ [H(\lambda)-E]^{2K-2}H(\lambda)\right\} }{\tr\left\{ [H(\lambda)-E]^{2K-2}\right\} } \nonumber \\
 &  = \frac{\sum_{k = 0}^{2K-2}(-1)^{k}\binom{2K-2}{k}  \omega_{2K-1-k}(\lambda)E^{k}} {\sum_{k = 0}^{2K-2}(-1)^{k}\binom{2K-2}{k} \omega_{2K-2-k}(\lambda)E^{k}},
\label{eq:E_cost_function}
\end{align}
where we define $\omega_{k}(\lambda) \coloneqq \tr[H(\lambda){}^{k}]$ for $k = 0,\dots,2K-1$. To minimize this function, we first compute the values of $\omega_{k}(\lambda)$ using computer algebra, which is completed in polynomial time in system size. Then, Eq.~\eqref{eq:E_cost_function} becomes a rational function of $E$ with known coefficients, and the minimization of $\Omega_{\lambda}^{(K)}(E)$ is easily conducted by standard numerical methods such as the Newton method. See Appendix \ref{subsec:apdx_Omega} for further discussion on this optimization. 

For general spin-$1/2$ systems, we summarize the entire algorithm for the ground-state evolution in Appendix~\ref{subsec:apdx_algorithm-GS}, where we show that the time for the algebraic computation scales as $O(N^{K})$, similarly to the general framework. Furthermore, if the $\lambda$ dependence of the Hamiltonian is in the form of Eq.~\eqref{eq:Hamiltonian_fF}, we can reduce the cost of evaluating the trace $\omega_{k}(\lambda)$ by separating the $\lambda$-dependent coefficients from the $\lambda$-independent traces, as has been done for $Q_{\mu\nu}^{(\mathcal{P})}(\lambda)$ and $r_{\mu}^{(\mathcal{P})}(\lambda)$ in the general framework, without changing the overall $O(N^K)$ scaling (see Appendices~\ref{subsec:apdx_time-dependent-GS} and \ref{subsec:apdx_algorithm-GS} for details).

\section{Numerical tests 
\label{sec:results}}

\subsection{System and setup
\label{subsec:numerical_setup}}

We test the weighted variational method with a quantum annealing protocol on a transverse-field Ising model with inhomogeneous parameters. Quantum annealing solves optimization problems via ground-state evolution, with a variety of real-world applications proposed, ranging from chemistry to finance~\cite{Yarkoni2022QuantumAnnealing}. An optimization problem is encoded in the final Hamiltonian so that its ground state corresponds to the optimal solution. The solution is found by adiabatically reaching the ground state and performing a measurement on it. Since the target ground state is unknown, the control protocol must be determined without the target information. To meet this requirement, variational CD driving has been used to enhance quantum annealing protocols~\cite{Hartmann2019RapidCounter, Hartmann2022PolynomialScaling, Passarelli2020CounterdiabaticDriving, Prielinger2021TwoParameterCounter, Kumar2021CounterdiabaticRoute, Barone2024CounterdiabaticOptimized,Passarelli2023CounterdiabaticReverse,Hegade2021DigitizedAdiabatic,Hegade2022DigitizedCounterdiabatic, Hegade2022PortfolioOptimization, Hegade2023DigitizedCounterdiabatic, Guan2024SingleLayer, Romero2025BiasField}. The conventional method increases the final fidelity, i.e., the probability of obtaining the correct optimal solution, but the fidelity drops considerably for large systems~\cite{Hartmann2022PolynomialScaling}.

Quantum annealing is precisely one of the situations that suffer from the two shortcomings of the conventional variational method (Sec.~\ref{subsec:weighted_variational_general}). First, the desired time evolution is along the unknown ground state. Thus, suppressing nonadiabatic transitions between high-energy states is irrelevant, and a control method targeted at the ground state is desired.
Second, the local magnetic fields and coupling constants between spins are spatially inhomogeneous to encode a specific problem, and the solution to the problem is determined by globally integrating all information from these system parameters. Thus, local driving coefficients determined solely from nearby parameters are unfavorable. Our proposal, which overcomes these two shortcomings, can indeed enhance the fidelity of quantum annealing processes, as we demonstrate below.

Consider the ground-state evolution of the following transverse-field Ising model with random couplings,
\begin{equation}
H(\lambda) = (1-\lambda)\sum_{i = 1}^{N}X_{i}+\lambda\left(\sum_{i = 1}^{N}h_{i}Z_{i}\pm\!\!\sum_{(i,j) \in \Lambda_{\mathrm{NN}}}\!\! J_{ij}Z_{i}Z_{j}\right)
\label{eq:Ising_Hamiltonian}
\end{equation}
for $0 \leq \lambda\leq1$, where $\Lambda_{\mathrm{NN}}$ is the set of the nearest-neighbor pairs of spins ($(i,j)$ and $(j,i)$ are counted only once). We set $\hbar = 1$ in all numerical results. We use the two-dimensional square lattice of width $N_{\mathrm{w}}$ and height $N_{\mathrm{h}}$ with open boundary condition. The number of spins is $N = N_{\mathrm{w}}N_{\mathrm{h}}$, and the number of nearest-neighbor pairs is $\vert\Lambda_{\mathrm{NN}}\vert = N_{\mathrm{w}}(N_{\mathrm{h}}-1)+(N_{\mathrm{w}}-1)N_{\mathrm{h}}$. The longitudinal magnetic field $h_{i}$ is sampled from an i.i.d.~gamma distribution with mean $1.0$ and standard deviation $0.5$, which ensures $h_{i}\geq0$. By changing the sign and coefficients of the coupling term, we realize ferromagnetic, antiferromagnetic, and spin-glass systems. For the ferromagnetic system, we choose the minus sign from the $\pm$ sign and sample $J_{ij}$ from an i.i.d.~gamma distribution with mean $1.0$ and standard deviation $0.5$. For the antiferromagnetic system, we choose the plus sign and draw $J_{ij}$ from the same gamma distribution. For the spin-glass system, we sample $J_{ij}$ from an i.i.d.~Gaussian distribution with mean $0$ and standard deviation $1.0$ (the choice of the $\pm$ sign is irrelevant). 

We employ two types of ansatz for driving operators: the \textit{one-body driving} ansatz,
\begin{equation}
V(\bm{\alpha}(\lambda)) = \sum_{i = 1}^{N}\alpha_{i}(\lambda)Y_{i},
\label{eq:one-body}
\end{equation}
and the \textit{two-body driving} ansatz,
\begin{equation}
V(\bm{\alpha}(\lambda)) = \sum_{i = 1}^{N}\alpha_{i}(\lambda)Y_{i}+\!\!\sum_{(i,j) \in \Lambda_{\mathrm{NN}}}\!\!\alpha_{ij}(\lambda)(Y_{i}Z_{j}+Z_{i}Y_{j}).
\label{eq:two-body}
\end{equation}
The number of basis operators $M$ is $M = N$ for the one-body driving and $M = N+\vert\Lambda_{\mathrm{NN}}\vert$ for the two-body driving. In the latter case, we identify the set of coefficients $\{\alpha_{N+1}(\lambda),\dots,\alpha_{N+\vert\Lambda_{\mathrm{NN}}\vert}(\lambda)\}$ with the set $\{\alpha_{ij}(\lambda) \mid(i,j) \in \Lambda_{\mathrm{NN}}\}$ in an arbitrary order for notational convenience. These driving operators are taken from the first order of the nested commutator ansatz for variational AGPs~\cite{Claeys2019FloquetEngineering}, and they can be experimentally realized by using gauge transformations~\cite{Sels2017Minimizing} or digitized driving~\cite{Hegade2021ShortcutsToAdiabaticity,Romero2025BiasField}.

\begin{figure}[!tp]
\includegraphics{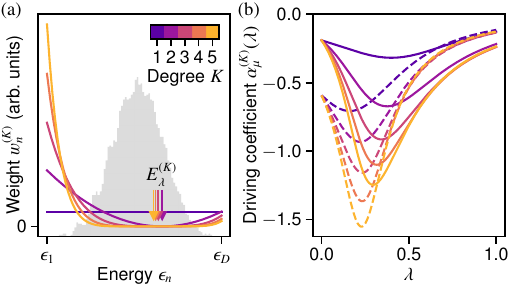}
\caption{Determination of the driving coefficients using the weighted variational method with different values of $K$, exemplified with a ferromagnetic Ising model with $N = 12$. (a) The weight on the eigenstates $w_n^{(K)}(\lambda)$ plotted against the energy eigenvalues $\epsilon_n(\lambda)$ for $K = 1,\dots,5$ (purple to yellow). We use $\lambda = 0.25$ as an example. Each weight is normalized so that the area under the curve is constant for better visualization, as the overall normalization is unimportant. 
The arrows show the energy shift $E_{\lambda}^{(K)}$ for $K = 2,\dots,5$, which corresponds to the center of the weight function. The energies $\epsilon_{1}$ and $\epsilon_{D}$ are the lowest and the highest energy eigenvalues, and the gray histogram shows the density of states.
(b) The driving coefficients $\alpha^{(K)}_{\mu}(\lambda)$ at two representative sites of the one-body driving obtained from the weighted variational method with $K = 1,\dots,5$. Solid curves represent $\alpha^{(K)}_{\mu}(\lambda)$ of one of the sites, and dashed curves are for another site. We plot the coefficients for all sites in Fig.~S1 in Supplemental Material~\cite{Suppli}.
\label{fig:ferromagnetic_weight_coeff}}
\end{figure}

We determine the driving coefficients $\bm{\alpha}^{(K)}(\lambda)$ by the weighted variational method in Eq.~\eqref{eq:minimum-GS} with degrees $K = 1,\dots,5$, and we numerically simulate the time evolution. We take a simple linear schedule, $\lambda_{t} = t/\td$ for $0 \leq  t \leq \td$, where $\td$ is the protocol duration, and solve the Schr\"{o}dinger equation in Eq.~\eqref{eq:Schroedinger} with Hamiltonian $H(\lambda_{t})+\dot{\lambda}_{t}V(\bm{\alpha}^{(K)}(\lambda_{t}))$ and the initial state $\ket{\phi_{1}(0)}$ using the QuTiP package in Python~\cite{Johansson2012QuTiP,Johansson2013QuTiP}. We write the resulting time evolution as $\ket{\psi^{(K)}(t)}$. For comparison, we also compute the time evolution without the driving term by setting $V(\lambda_{t}) = 0$ in Eq.~\eqref{eq:Schroedinger}, whose result is written as $\ket{\psi^{(\varnothing)}(t)}$.

We measure the performance of variational CD driving by the fidelity to the instantaneous ground state,
\begin{equation}
\mathcal{F}^{(K)}(t) \coloneqq \vert\braket{\phi_{1}(\lambda_{t})}{\psi^{(K)}(t)}\vert^{2},
\label{eq:fidelity_def}
\end{equation}
for $K \in \{\varnothing,1,\dots,5\}$, and in particular, the final fidelity $\mathcal{F}_{\mathrm{f}}^{(K)} \coloneqq \mathcal{F}^{(K)}(\td) = \vert\braket{\phi_{1}(1)}{\psi^{(K)}(\td)}\vert^{2}$ at the end of the protocol. We also introduce the gain $\mathcal{G}_{\mathrm{f}}^{(K)}$ as 
\begin{equation}
\mathcal{G}_{\mathrm{f}}^{(K)} \coloneqq \frac{\mathcal{F}_{\mathrm{f}}^{(K)}}{\mathcal{F}_{\mathrm{f}}^{(1)}} = \frac{\vert\braket{\phi_{1}(1)}{\psi^{(K)}(\td)}\vert^{2}}{\vert\braket{\phi_{1}(1)}{\psi^{(1)}(\td)}\vert^{2}}.
\label{eq:gain_def}
\end{equation}
The gain measures the relative increase of the fidelity compared to the conventional method, and it trivially satisfies $\mathcal{G}_{\mathrm{f}}^{(1)} = 1$. These performance measures depend not only on degree $K$ but also on the realization of the random parameters $\{h_{i}\}$ and $\{J_{ij}\}$, the choice of driving terms (one-body or two-body), the system size $N$, and the protocol duration $\td$, but we make these dependencies implicit for brevity.

Our algorithm for computing the driving coefficients $\bm{\alpha}^{(K)}(\lambda)$ is efficient, and it can be applied to systems with a large number of spins. For example, ten minutes of calculation can deal with $N = 4330$ for $K = 2$, $N = 360$ for $K = 3$, $N = 84$ for $K = 4$, and $N = 37$ for $K = 5$ with our C++ implementation available in Ref.~\cite{GitHub}. These values are obtained from the calculations of the driving coefficients for a one-dimensional Ising model ($N_{\mathrm{h}} = 1$ and $N_{\mathrm{w}} = N$ in the above setup) with the one-body driving at 100 values of $\lambda\in [0,1]$. We used Algorithm 4 in Appendix~\ref{subsec:apdx_algorithm-GS} for the algebraic calculation and the conjugate gradient method to solve Eq.~\eqref{eq:linear_equation-general}. We used an ordinary laptop computer (MacBook Pro with Apple M3 Pro chip and 18 GB memory) and the Clang compiler with the level-O2 compiler optimization and CPU parallelization.

Despite this efficiency, we will focus entirely on smaller systems, $N_{\mathrm{h}} = 3$ and $N_{\mathrm{w}} = 3,4,5$, and hence $N = 9,12,15$ in the following demonstration. The system size is constrained by the cost of the numerical simulation for our demonstrative purposes, which would not be necessary when testing our method on real quantum systems.

\subsection{Ferromagnetic systems}

We first demonstrate how the weighted variational method with different degrees results in different driving coefficients using a ferromagnetic system with $N = 12$ as an example. Figure~\ref{fig:ferromagnetic_weight_coeff}(a) shows the weight associated with the $n$th eigenstate, $w_n^{(K)}(\lambda)$, for $K = 1,\dots,5$. The weight is more concentrated on low-energy states as $K$ increases. In Fig.~\ref{fig:ferromagnetic_weight_coeff}(b), we plot the resulting driving coefficients $\alpha_\mu^{(K)}(\lambda)$ for $K = 1,\dots,5$ for the one-body driving ansatz. As seen from the figure, both the shape and the magnitude of the driving coefficients vary with the degree $K$. A higher degree method tends to result in a larger magnitude in this specific system, while this tendency is not always true for other systems (see Fig.~S1 in Supplemental Material~\cite{Suppli}). The increase in the maximum magnitude is at most three times for almost all cases. Thus, the difficulty of experimental implementation does not qualitatively increase. 

\begin{figure}
\includegraphics{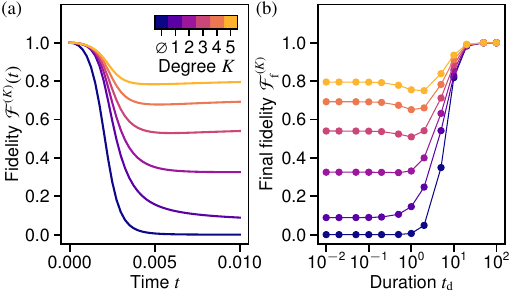}
\caption{Performance of the weighted variational method in a ferromagnetic Ising model of $N=12$ with the one-body driving. We simulate the time evolution with the driving protocol obtained by the weighted variational method with $K = 1,\dots,5$, where $K = 1$ is the conventional variational method, and $K \geq 2$ are our proposal. We also simulate the time evolution without CD driving, represented by $K = \varnothing$. (a) The time evolution of the fidelity to the ground state $\mathcal{F}^{(K)}(t)$. 
(b) The final fidelity $\mathcal{F}_{\mathrm{f}}^{(K)}$ plotted over different protocol durations $\td$. 
\label{fig:ferromagnetic_single}}
\end{figure}

We test the weighted variational method in the same ferromagnetic system of $N = 12$ with the one-body driving, which results in significantly improved fidelity. Figure~\ref{fig:ferromagnetic_single}(a) shows the time evolution of the fidelity $\mathcal{F}^{(K)}(t)$ with a short duration $\td=0.01$. The fidelity drops from unity to almost zero ($\simeq2.4\times10^{-4}$) in the absence of the driving term ($K = \varnothing$). The final fidelity remains around $0.089$ with the driving protocol calculated by the conventional method ($K = 1$). With our weighted variational method ($K\geq2$), the final fidelity is significantly improved up to around $0.80$ at $K = 5$, which is $8.9$ times larger than the conventional method. 

The final fidelity $\mathcal{F}_{\mathrm{f}}^{(K)}$ depends on the protocol duration $\td$, as shown in Fig.~\ref{fig:ferromagnetic_single}(b). When the duration is sufficiently long, the time evolution with and without CD driving both achieve fidelities close to unity. In this regime, $\dot{\lambda}_{t} = 1/\td$ is small, and the CD driving term becomes negligible in the Schr\"{o}dinger equation in Eq.~\eqref{eq:Schroedinger_coeff_driving}. Then, the unit fidelity is understood from the adiabatic theorem~\cite{Kato1950OnTheAdiabaticTheorem}. As the duration $\td$ becomes shorter, the fidelity decreases and converges to a constant value that depends on $K$. In this regime, $\dot{\lambda}_{t} = 1/\td$ is large, and the first term $\kappa_{n}(t)c_{n}(t)$ becomes negligible in Eq.~\eqref{eq:Schroedinger_coeff_driving}. We can then recast Eq.~\eqref{eq:Schroedinger_coeff_driving} into a form independent of the duration $\td$ by using the chain rule $(\dot{\lambda}_{t})^{-1}\partial_{t}c_{n}(t) = \partial_{\lambda}c_{n}(t_{\lambda})$, where $t_{\lambda}$ is the inverse function of $\lambda_{t}$. This makes the final fidelity independent of $\td$~\cite{Sels2017Minimizing,Sugiura2021AdiabaticLandscape}. In other words, the time evolution in this short-duration limit is determined solely by the discrepancy between the exact and approximated AGPs. In the following, we always focus on this regime and set $\td = 0.01$ to investigate the impact of the discrepancy.

\begin{figure}
\includegraphics{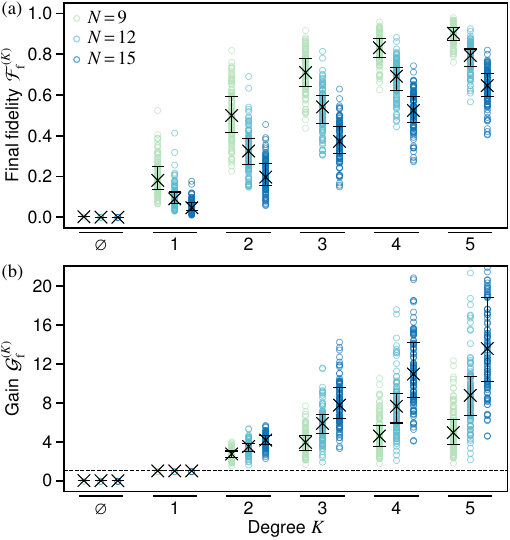}
\caption{Numerical test of the weighted variational method in ferromagnetic systems. We generate 100 instances for every system size $N = 9,12,15$ and perform the weighted variational method for $K = 1,\dots,5$ for each instance. We use the one-body driving and a short duration $\td = 0.01$. (a) Final fidelity $\mathcal{F}_{\mathrm{f}}^{(K)}$ for different values of degree $K$ and system size $N$, where $K = \varnothing$ is the absence of CD driving, $K = 1$ is the conventional variational method, and $K\protect\geq2$ are our proposal. Circle represents the final fidelity for each instance. Black $\times$ mark shows the median, and the error bar indicates the quartile range calculated for each $(N,K)$. (b) Gain $\mathcal{G}^{(K)}_{\mathrm{f}}$, i.e., the relative increase of the final fidelity compared to the conventional method. The final fidelity $\mathcal{F}_\mathrm{f}^{(K)}$ is divided by $\mathcal{F}_\mathrm{f}^{(1)}$ of the same instance. Black $\times$ mark and error bar are the median and quartile range of the gain. Some outliers are above the upper bound of the axis. 
\label{fig:ferromagnetic}}
\end{figure}

Figure \ref{fig:ferromagnetic} shows that the enhancement of the final fidelity is not limited to the specific instance (the set of realization of the random parameters $\{h_{i}\}$ and $\{J_{ij}\}$) but holds more generally. We generate 100 instances of ferromagnetic systems for each of the system sizes $N = 9,12,15$, calculate the driving protocols of the one-body driving for $K = 1,\dots,5$ for each instance, and perform a quantum simulation for every $K \in \{\varnothing,1,\dots,5\}$ for each instance. As shown in Fig.~\ref{fig:ferromagnetic}(a), the final fidelity $\mathcal{F}_{\mathrm{f}}^{(K)}$ varies from instance to instance, but it is within the same order of magnitude. The median of the final fidelity increases as the degree $K$ increases for every system size $N$. In Fig.~\ref{fig:ferromagnetic}(b), we show the gain $\mathcal{G}_{\mathrm{f}}^{(K)}$, i.e., the relative increase of the final fidelity. The gain is greater than unity in all instances and for all $K\geq2$, and the median gain increases as we increase $K$ for every system size $N$. For $K = 5$, the median of the gain $\mathcal{G}_{\mathrm{f}}^{(5)}$ is about $4.9$, $8.7$, and $14$ for $N = 9,12,15$, respectively, demonstrating a significant improvement in the final fidelity. 

Figure~\ref{fig:ferromagnetic} also shows the system-size dependence of the results, where the final fidelity $\mathcal{F}_{\mathrm{f}}^{(K)}$ and gain $\mathcal{G}_{\mathrm{f}}^{(K)}$ follow different trends. The median final fidelity is smaller for a larger system size $N$ for each $K \in \{\varnothing,1,\dots,5\}$ [Fig.~\ref{fig:ferromagnetic}(a)]. This behavior in the absence of CD driving ($K = \varnothing$) and with the conventional method ($K = 1$) is well known in the literature (see Ref.~\cite{Hartmann2022PolynomialScaling} for a detailed investigation), and it reflects the intrinsic---and presumably unavoidable---difficulty of realizing adiabatic processes with a limited number of driving fields. This fundamental difficulty persists in the weighted variational method, as suggested by our results with $K\geq2$. On the other hand, the median gain increases as the size $N$ increases for every fixed $K\geq2$ [Fig.~\ref{fig:ferromagnetic}(b)]. This result implies that the weighted variational method becomes more advantageous for larger systems in terms of gain. It suggests that our proposal is still effective for even larger systems than those tested here.

\begin{figure}
\includegraphics[width = 1\columnwidth]{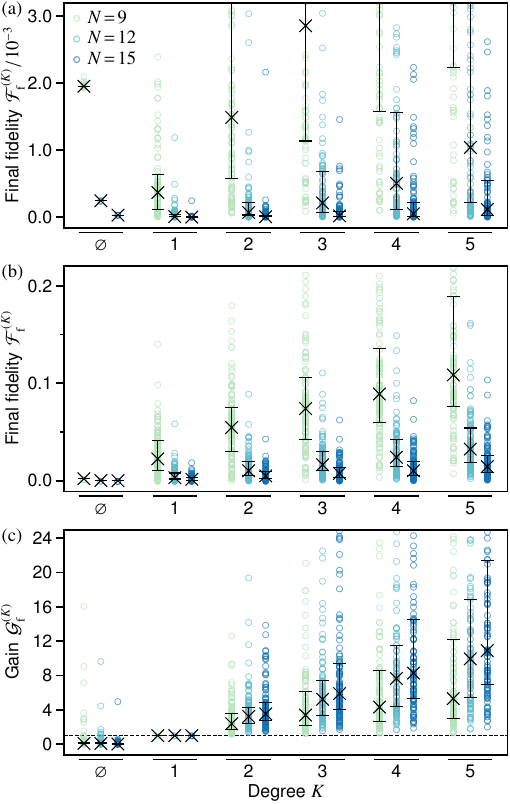}
\caption{Numerical test of the weighted variational method in antiferromagnetic systems. We generate $100$ instances for every system size $N = 9,12,15$ and perform simulations similarly to Fig.~\ref{fig:ferromagnetic}. (a) Final fidelity $\mathcal{F}_{\mathrm{f}}^{(K)}$ with the one-body driving, varying the system size $N$ and the degree $K$. (b) Final fidelity $\mathcal{F}_{\mathrm{f}}^{(K)}$ with the two-body driving. (c) Gain $\mathcal{G}_{\mathrm{f}}^{(K)}$ with the two-body driving. In all panels, some outliers are above the upper bound of the axes.
\label{fig:antiferromagnetic}}
\end{figure}

\begin{figure}
\includegraphics{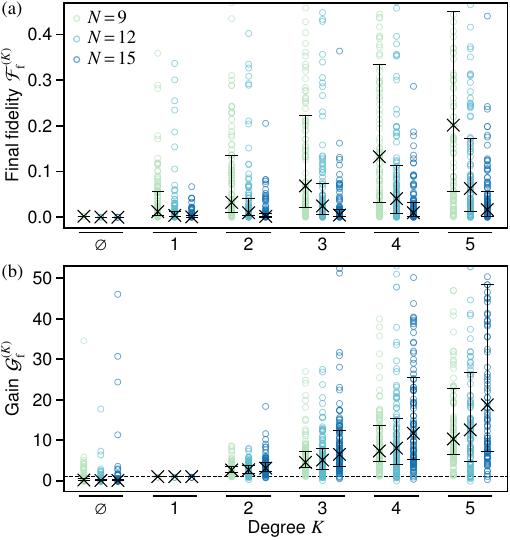}
\caption{Numerical test of the weighted variational method in spin-glass systems. We generate $100$ instances for every system size $N = 9,12,15$ and apply the weighted variational method with the one-body driving similarly to Fig.~\ref{fig:ferromagnetic}. (a) Final fidelity $\mathcal{F}_{\mathrm{f}}^{(K)}$ with varying the system size $N$ and the degree $K$. (b) Gain $\mathcal{G}_{\mathrm{f}}^{(K)}$. In both panels, some outliers are above the upper bound of the axes. 
\label{fig:spin-glass}}
\end{figure}

\subsection{Antiferromagnetic and spin-glass systems}

The ferromagnetic system presented in the previous section has no frustration or competition between the coupling and external fields. To test the general applicability of the weighted variational method to more disordered and chaotic systems, we perform a similar numerical test on antiferromagnetic and spin-glass systems. We randomly generate 300 instances of antiferromagnetic systems and 300 instances of spin-glass systems (100 each for $N = 9,12,15$), compute the driving coefficients, and perform quantum simulations similarly to the previous section. 
We plot representative driving coefficients in Fig.~S1 in Supplemental Material~\cite{Suppli}, which tends to show more complicated behavior than in Fig.~\ref{fig:ferromagnetic_weight_coeff}(b).

We first observe the time-dependent fidelity $\mathcal{F}^{(K)}(t)$ and the dependence of the final fidelity on protocol duration $\td$ for representative systems (Fig.~S2 in Supplemental Material~\cite{Suppli}), confirming that they are qualitatively similar to those in Fig.~\ref{fig:ferromagnetic_single}. Concretely, the fidelity $\mathcal{F}^{(K)}(t)$ drops from unity to a lower value over time evolution, and when we shorten the duration $\td$, the final fidelity $\mathcal{F}_\mathrm{f}^{(K)}$ converges to a constant value dependent on $K$. Thus, we can evaluate the overall performance by focusing on the final fidelity with the shortest duration $\td=0.01$, as we have done for ferromagnetic systems. 

In the antiferromagnetic systems with the one-body driving, conventional CD driving does harm to the adiabatic process, but the weighted variational method can assist the process [Fig.~\ref{fig:antiferromagnetic}(a)]. The final fidelity with the conventional method $\mathcal{F}_{\mathrm{f}}^{(1)}$ is smaller than the fidelity without CD driving $\mathcal{F}_{\mathrm{f}}^{(\varnothing)}$ in most instances (287 instances out of 300), meaning that the conventional method hinders the adiabatic process. Nevertheless, the median of $\mathcal{F}_{\mathrm{f}}^{(K)}$ increases as $K$ increases for all system sizes, and the final fidelity $\mathcal{F}_{\mathrm{f}}^{(5)}$ at $K = 5$ is greater than $\mathcal{F}_{\mathrm{f}}^{(\varnothing)}$ for about 75\% of the instances. Thus, the weighted variational method can recover the advantage of CD driving even when the conventional variational method does not lead to improvement.

We can circumvent this peculiar behavior of antiferromagnetic systems by using the two-body driving [Figs.~\ref{fig:antiferromagnetic}(b) and (c)]. The final fidelity with the conventional method $\mathcal{F}_{\mathrm{f}}^{(1)}$ is higher than the final fidelity without CD driving $\mathcal{F}_{\mathrm{f}}^{(\varnothing)}$ in most instances (283 instances out of 300), which implies that the conventional method is advantageous when used with the two-body driving. The final fidelity is further improved by the weighted variational method with $K\geq2$ [Fig.~\ref{fig:antiferromagnetic}(b)]. As shown in Fig.~\ref{fig:antiferromagnetic}(c), the gain $\mathcal{G}_{\mathrm{f}}^{(K)}$ exceeds unity in all instances for all $K\geq2$. The median gain reaches $5.3,9.9,11$ for $N = 9,12,15$, respectively, at $K = 5$.

For the spin-glass systems, the one-body driving is valid for a large fraction of instances, while the instance-to-instance variations of $\mathcal{F}_{\mathrm{f}}^{(K)}$ and $\mathcal{G}_{\mathrm{f}}^{(K)}$ are stronger than for the previous two classes of systems (Fig.~\ref{fig:spin-glass}). The conventional method $\mathcal{F}_{\mathrm{f}}^{(1)}$ is better than the bare time evolution $\mathcal{F}_{\mathrm{f}}^{(\varnothing)}$ for a large fraction of instances (259 instances out of 300), and the weighted variational method with a larger degree $K$ further improves the final fidelity [Fig.~\ref{fig:spin-glass}(a)]. The gain is greater than unity for all $K\geq2$ in most of the instances (295 instances out of 300), as shown in Fig.~\ref{fig:spin-glass}(c). The gain is more dramatic than that in the ferromagnetic and antiferromagnetic systems. The median of $\mathcal{G}_{\mathrm{f}}^{(5)}$ is about $10,13,19$ for $N = 9,12,15$, respectively.

In all the cases above, a larger system size results in a larger median gain $\mathcal{G}_{\mathrm{f}}^{(K)}$ for every fixed $K\geq2$ [Fig.~\ref{fig:antiferromagnetic}(c) and \ref{fig:spin-glass}(b)]. This fact suggests that our method remains powerful for chaotic and disordered systems of larger sizes, even beyond the sizes tested in this section.

\section{Properties of the driving coefficients from the weighted variational method
\label{sec:analysis}}

We analyze the properties of the driving coefficients obtained from the weighted variational method and discuss how these properties contribute to the observed improvement in the final fidelity, thereby bridging between the theoretical framework in Secs.~\ref{sec:general-framework} and \ref{sec:ground_state} and the numerical results in Sec.~\ref{sec:results}\@. As discussed in Sec.~\ref{subsec:weighted_variational_general}, the weighted action extends the conventional action in two aspects: it assigns more weights to important matrix elements and incorporates nonlocal terms. We examine how these features manifest in the resulting driving coefficients from both theoretical and numerical analyses. Throughout this section, we fix a single $\lambda$ and omit the $\lambda$-dependence of quantities such as $H(\lambda)$, $\epsilon_{n}(\lambda)$, $\Phi(\lambda)$, $\mathcal{S}_{\lambda}^{(K)}[V]$, $w_{mn}^{(K)}(\lambda)$, $w_{n}^{(K)}(\lambda)$, $\bm{\alpha}^{(K)}(\lambda)$, and $E_{\lambda}^{(K)}$, except in Eq.~\eqref{eq:speed_limit}.

\subsection{Effect of the weights on low-energy eigenstates }
\label{subsec:partial_action}

We analyze how the weights on low-energy states affect the action and improve the ground-state fidelity. 
We invoke the expansion of the weighted action into matrix elements [Eq.~\eqref{eq:actionK_explicit-general}]. We will show that the matrix elements associated with lower-energy states are closely related to the ground-state fidelity by using the quantum speed limit formula and an additional numerical calculation. This theoretical analysis supports the expectation that emphasizing these elements is favorable for improving the ground-state fidelity.

We decompose $\mathcal{S}^{(K)}[V]$ into contributions from each eigenstate $\mathcal{T}^{(K,n)}[V]$, which we call the \textit{partial action} (see Appendix~\ref{subsec:apdx_deriv-partial-action} for derivation), as
\begin{subequations}
\label{eq:partial_action_all}
\begin{align}
\mathcal{S}^{(K)}[V] & = \frac{2}{\hbar^{2}} \sum_{n = 1}^{D}w_{n}^{(K)}\mathcal{T}^{(K,n)}[V]+\mathrm{const.},
\label{eq:action_decompose}\\
\mathcal{T}^{(K,n)}[V]\, &  \!\coloneqq \!\!\sum_{m:\,w_{m}^{(K)} \leq  w_{n}^{(K)}}\!\!(\epsilon_{n}-\epsilon_{m})^{2}\chi_{mn}^{(K)}\frac{w_{mn}^{(K)}}{w_{n}^{(K)}}\bigl\vert[V-\Phi]_{mn}\bigr\vert^{2}.
\label{eq:partial_action}
\end{align}
\end{subequations}
Here, $w_{n}^{(K)} \equiv K^2 (\epsilon_{n}-E^{(K)})^{2K-2}$ is the state-dependent weight introduced in Sec.~\ref{subsec:ground_state_theory}, and we define $\chi_{mn}^{(K)}\coloneqq1/2$ if $w_{m}^{(K)} = w_{n}^{(K)}$ and $\chi_{mn}^{(K)}\coloneqq1$ otherwise. This decomposition shows that $\mathcal{S}^{(K)}[V]$ is understood as a weighted sum of the partial actions $\mathcal{T}^{(K,n)}[V]$ with the weight $w_n^{(K)}$. The partial action $\mathcal{T}^{(K,n)}[V]$ penalizes those deviations $[V-\Phi]_{mn}$ whose weight $w_m^{(K)}$ is smaller than $w_{n}^{(K)}$. 
The partial action depends on $K$, but the dependence is merely algebraic, with a scaling between $O(K^{0})$ and $O(K^{-2})$ (see Appendix~\ref{subsec:apdx_deriv-partial-action}). This algebraic dependence is much weaker than the exponential dependence of the weight $w_{n}^{(K)}$ on $K$. Therefore, the $K$-dependence of $\mathcal{S}^{(K)}[V]$ is predominantly captured by the weight $w_{n}^{(K)}$ in the decomposition in Eq.~\eqref{eq:action_decompose}.

Using this decomposition, we first discuss the behavior of $\mathcal{S}^{(K)}[V]$ in the $K \to \infty$ limit. For a sufficiently large $K$, the weight $w_{n}^{(K)}$ is concentrated on the ground state, and thus, the action $\mathcal{S}^{(K)}[V]$ is approximated by $2\hbar^{-2}w_{1}^{(K)}\mathcal{T}^{(K,1)}[V]$. Furthermore, $\mathcal{T}^{(K,1)}[V]$ is approximated by $K^{-2}(\epsilon_{1}-E^{(K)})^{2}\sum_{m:\,m\neq1}\vert[V-\Phi]_{1m}\vert^{2}$ (see Appendix~\ref{subsec:apdx_deriv-partial-action}).
Combining these two approximations, we observe that the action $\mathcal{S}^{(K)}[V]$ approaches 
\begin{equation}
\mathcal{S}^{(\infty)}[V] \coloneqq \frac{2}{\hbar^{2}}\sum_{n:\,n\neq1}\bigl\vert[V-\Phi]_{1n}\bigr\vert^{2}
\label{eq:actioninf_def}
\end{equation}
up to an overall normalization and a $V$-independent constant term in the $K \to \infty$ limit [we show another derivation of Eq.~\eqref{eq:actioninf_def} in Appendix~\ref{subsec:apdx_deriv-partial-action}]. This limiting action penalizes only $[V-\Phi]_{1n}$ for $n\geq 2$, i.e., the matrix elements of $(V-\Phi)$ between the ground and excited states. 

In fact, this penalty function in Eq.~\eqref{eq:actioninf_def} is suitable for suppressing nonadiabatic transitions from the ground state to the excited states. The escape of the population from the ground state is caused only by $[V-\Phi]_{1n}$ with $n \geq 2$, as seen from the Schr\"{o}dinger equation in Eq.~\eqref{eq:Schroedinger_coeff_driving}. Moreover, the final fidelity, denoted by $\vert c_{1}(\td)\vert^{2}$ in the notation of Eq.~\eqref{eq:Schroedinger_coeff_driving}, obeys the quantum speed limit formula [see Eq.~(31) of Ref.~\cite{Hatomura2021ControllingAndExploring}]:
\begin{equation}
\arccos\vert c_{1}(\td)\vert \leq \int_{0}^{\td}\frac{\vert\dot{\lambda}_{t}\vert}{\hbar}\sqrt{\sum_{n:\,n\neq1}\bigl\vert[V(\lambda_{t})-\Phi(\lambda_{t})]_{1n}\bigr\vert^{2}}dt.
\label{eq:speed_limit}
\end{equation}
The expression inside the square root in Eq.~\eqref{eq:speed_limit} is equal to $\hbar^{2}\mathcal{S}^{(\infty)}[V]/2$. Since $\arccos x$ is a decreasing function over $0 \leq x \leq1$, Eq.~\eqref{eq:speed_limit} provides a lower bound of the final fidelity $\vert c_1(\td)\vert$ in terms of $S^{(\infty)}[V]$. It shows that if the time integral of $\sqrt{\mathcal{S}^{(\infty)}[V]}$ is smaller, the final fidelity is closer to unity. In this sense, the action $\mathcal{S}^{(\infty)}[V]$ is a natural penalty function to determine driving coefficients that could maximize $\vert c_{1}(\td)\vert$. We call $\mathcal{S}^{(\infty)}[V]$ the \textit{ideal action} for convenience.

\begin{figure}
\includegraphics[width = 1\columnwidth]{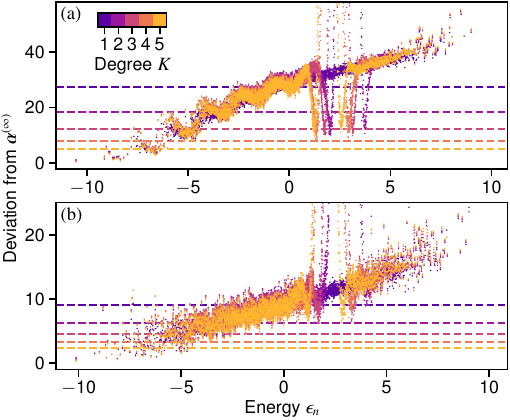}
\caption{Deviation of the driving coefficients from the ``ideal'' coefficient $\bm{\alpha}^{(\infty)}$, which minimizes the ideal action $\mathcal{S}^{(\infty)}[V(\bm{\alpha})]$. 
Colored dots represent the deviation $\Vert\bm{\alpha}^{(K,n)}-\bm{\alpha}^{(\infty)}\Vert^{2}_2$ plotted against the energy eigenvalue $\epsilon_n$, where $\bm{\alpha}^{(K,n)}$ is the minimizer of the partial action $\mathcal{T}^{(K,n)}[V(\bm{\alpha})]$.
Dashed horizontal lines denote the deviation $\Vert\bm{\alpha}^{(K)}-\bm{\alpha}^{(\infty)}\Vert^{2}_2$, where $\bm{\alpha}^{(K)}$ is the driving coefficient obtained from the weighted variational method of degree $K$. 
We fix $N = 12$ and $\lambda = 0.25$ as an example. (a) Ferromagnetic system with the one-body driving. (b) Spin-glass system with the one-body driving. 
\label{fig:partial_action}}
\end{figure}

Next, we focus on realistic values of $K$ such as $K\leq5$. For such a small $K$, the weight $w_{n}^{(K)}$ is not peaked enough to single out the $n=1$ partial action $\mathcal{T}^{(K,1)}[V]$ in Eq.~\eqref{eq:action_decompose}, and it is only mildly concentrated in the broad range of lower-energy states. Nevertheless, this mild concentration could still lead to an improvement in CD driving by the following mechanism. Let us hypothesize that $\mathcal{T}^{(K,n)}[V]$ depends smoothly on $n$, and thus $\mathcal{T}^{(K,n)}[V]$ with small $n$ is closer to $\mathcal{T}^{(K,1)}[V]$ than $\mathcal{T}^{(K,n)}[V]$ with large $n$. Then, even a mild concentration of $w_{n}^{(K)}$ in small $n$'s can shift $\mathcal{S}^{(K)}[V]$ toward $\mathcal{T}^{(K,1)}[V]$. The $n=1$ partial action $\mathcal{T}^{(K,1)}[V]$ is expected to be close to the ideal action because it contains the same set of terms as the ideal action ($[V-\Phi]_{1n}$ with $n\geq 2$). Therefore, under the above hypothesis, the weighted action is expected to approach the ideal action as $K$ increases, even for small $K$. This scenario provides a possible way to understand the improved final fidelity with small $K$.

This scenario is indeed numerically confirmed with our example systems (Fig.~\ref{fig:partial_action}). We define $\bm{\alpha}^{(\infty)} $ and $\bm{\alpha}^{(K,n)} $ as the minimizers of $\mathcal{S}^{(\infty)}[V(\bm{\alpha})]$ and $\mathcal{T}^{(K,n)}[V(\bm{\alpha})]$, respectively, and plot the deviation $\Vert\bm{\alpha}^{(K,n)}-\bm{\alpha}^{(\infty)}\Vert^{2}_2$ by varying $K$ and $n$, where $\Vert\cdot\Vert_2$ is the 2-norm. The figure shows that the deviation does not depend significantly on $K$, but it has a clear trend with $n$. The deviation is small for the ground state and tends to be larger for a larger $n$. This trend implies that $\mathcal{T}^{(K,n)}[V]$ with small $n$ is closer to the ideal action than $\mathcal{T}^{(K,n)}[V]$ with large $n$, consistent with the above scenario. Note that the deviation shows an irregular behavior around $\epsilon_{n} \simeq  E^{(K)}$ for $K\geq2$, but this irregularity does not significantly affect the overall action $\mathcal{S}^{(K)}[V]$ because the overall magnitude of $\mathcal{T}^{(K,n)}[V]$ is small for such $n$. This irregular behavior occurs because the weight $w_{n}^{(K)}$ is close to zero when $\epsilon_{n} \simeq E^{(K)}$, and thus $\mathcal{T}^{(K,n)}[V]$ contains only a small number of terms, making its minimization too sensitive to individual elements $[V-\Phi]_{mn}$. 

The scenario is also confirmed by looking at the deviation between the coefficient $\bm{\alpha}^{(K)}$ from the weighted variational method, which is the minimizer of $\mathcal{S}^{(K)}[V(\bm{\alpha})]$, and the ideal coefficient $\bm{\alpha}^{(\infty)}$. We plot the deviation $\Vert\bm{\alpha}^{(K)}-\bm{\alpha}^{(\infty)}\Vert^{2}_2$ in Fig.~\ref{fig:partial_action} (dashed horizontal lines). The plot shows that the driving coefficient $\bm{\alpha}^{(K)}$ approaches the ideal coefficient $\bm{\alpha}^{(\infty)}$ as $K$ increases, confirming the scenario above even for a small $K$.

\begin{figure}
\includegraphics[width = 1\columnwidth]{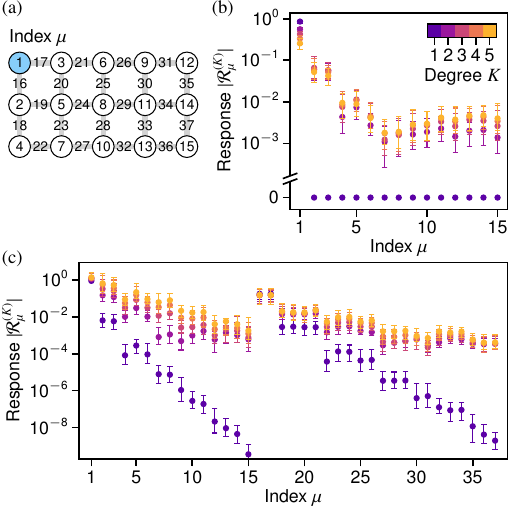}
\caption{Response of the driving coefficient $\alpha_{\mu}^{(K)}$ to a perturbation in the longitudinal magnetic field. We fix $N = 15$ ($N_{\mathrm{w}} = 5$ and $N_{\mathrm{h}} = 3$) and $\lambda = 0.25$. The index $\mu$ of the driving terms runs from 1 to 15 for the one-body driving [Eq.~\eqref{eq:one-body}]. The index runs from $1$ to $37$ for the two-body driving, in which $\mu\leq 15$ are the coefficients on the $Y_{i}$ terms, and $\mu\geq 16$ are on the $(Y_{i}Z_{j}+Z_{i}Y_{j})$ terms [Eq.~\eqref{eq:two-body}]. 
(a) Indices of the spins ($\mu = 1,\dots,15$) and the pairs of neighboring spins ($\mu = 16,\dots,37$) used in this figure. We give a perturbation $\delta\ln h_{1} = \ln 1.02$ at spin 1 (blue).
(b) Absolute value of the response function $\vert\mathcal{R}_{\mu}^{(K)}\vert$ for ferromagnetic systems with the one-body driving. Circle shows the median, and the error bar shows the quartile range of 100 instances. 
(c) Absolute value of the response function for antiferromagnetic systems with the two-body driving. The median and quartile range over 100 instances are shown. 
\label{fig:non_locality}}
\end{figure}

\subsection{Nonlocality of the driving coefficients}
\label{subsec:nonlocality}

Next, we demonstrate that the driving coefficients from the weighted variational method reflect the nonlocality of the weighted action. We capture the nonlocal dependence of the driving coefficients on the system parameters by giving a small perturbation to a system parameter and observing the resulting change in the driving coefficients. Concretely, we modify the longitudinal field at site 1 by $\ln h_{1} \to \ln h_{1}+\delta\ln h_{1}$ and compute the resulting change in the driving coefficients $\ln\vert\alpha_{\mu}^{(K)}\vert \to \ln\vert\alpha_{\mu}^{(K)}\vert+\delta\ln\vert\alpha_{\mu}^{(K)}\vert$ for every index $\mu$ of the driving terms. We define the response function as 
\begin{equation}
\mathcal{R}_{\mu}^{(K)} \coloneqq \frac{\delta\ln\vert\alpha_{\mu}^{(K)}\vert}{\delta\ln h_{1}},
\end{equation}
which varies from instance to instance. Here, we focus on the logarithmic change (fold change) so that the response function is independent of the magnitudes of $h_{1}$ and $\vert\alpha_{\mu}^{(K)}\vert$. Taking $N = 15$ as an example, we assign the indices $\mu$ as in Fig.~\ref{fig:non_locality}(a) so that a larger index corresponds to a driving field at a farther position from the perturbed spin. 

For the one-body driving, the response functions for the conventional method ($K = 1$) behave differently from the response functions for our proposal ($K\geq2$) [Fig.~\ref{fig:non_locality}(b)]. The response functions for $K = 1$ are exactly zero except for $\alpha_{1}^{(1)}$. In contrast, the response functions for $K\geq2$ are nonzero at all sites, which means that the driving coefficients with $K\geq2$ depend nonlocally on the parameters of distant spins. 

For the two-body driving, the response functions are nonzero for any $\mu$ and any $K\geq1$, but their magnitudes differ significantly between $K = 1$ and $K\geq2$ [Fig.~\ref{fig:non_locality}(c)]. The response functions for $K = 1$ drop exponentially with the spatial distance between the perturbed spin and the site on which the driving field acts. On the other hand, the response functions from the weighted variational method ($K\geq2$) remain at significantly higher values even when the distance increases. This difference implies that the weighted variational method can incorporate nonlocal information much more efficiently than the conventional method.

\section{Conclusion and outlook 
\label{sec:conclusion}}

In this paper, we have proposed the \textit{weighted variational method} to derive refined variational AGPs for general parameterized Hamiltonians, which significantly improves variational CD driving. The proposal combines a theoretical advance and a numerical algorithm (Sec.~\ref{sec:general-framework})\@. On the theoretical side, we find that the exact AGP has an infinite variety of algebraic characterizations involving fictitious Hamiltonians $\mathcal{P}_\lambda (H(\lambda))$, which results in an infinite number of weighted actions. Minimizing a weighted action under an operator ansatz gives an improved CD driving protocol. We have also developed a computer algebra algorithm to perform the minimization efficiently in polynomial time with respect to the system size.

We have applied our proposal to ground-state evolution (Sec.~\ref{sec:ground_state}) and demonstrated that it significantly improves the final fidelity in quantum Ising models (Sec.~\ref{sec:results})\@. The median improvement in final fidelity is more than tenfold for $N=15$ for all three classes of systems, and this improvement is robust across the vast majority of randomly generated instances, except for 5 out of 300 spin-glass instances.
We have analyzed its possible mechanisms from two aspects in Sec.~\ref{sec:analysis}\@. First, the weighted action emphasizes the partial actions associated with lower-energy states, whose minimizers are closer to the minimizer of the ideal action. The ideal action is a suitable penalty function for suppressing nonadiabatic transitions from the ground state, as understood by the quantum speed limit formula. Second, due to the nonlocality of the weighted action, the driving coefficients depend on the system parameters in a nonlocal way.

Our proposal is applicable to a wide range of systems, and we expect that it can improve almost all previous uses of conventional variational CD driving. 
In particular, our concrete C++ implementation of the ground-state evolution of spin-$1/2$ systems can be readily applied to various practical problems. It can be used to improve the preparation of nontrivial quantum states, such as the GHZ state~\cite{Cepaite2023CounterdiabaticOptimized} and the Kitaev ground state~\cite{Kumar2021CounterdiabaticRoute}, as well as to enhance computational tasks such as optimization~\cite{Hegade2021DigitizedAdiabatic,Hegade2022DigitizedCounterdiabatic, Hegade2022PortfolioOptimization, Hegade2023DigitizedCounterdiabatic, Guan2024SingleLayer, Romero2025BiasField} and adiabatic quantum computation. Testing these applications with a greater number of spins than those used in Sec.~\ref{sec:results} is an important future work. Our theoretical framework should also be straightforwardly applicable to the ground-state evolution of fermionic systems after implementing a computer algebra for fermions. A more nontrivial future application is to quantum processes other than ground-state evolution. Such processes include the adiabatic preparation of the highest energy state and a mid-spectrum eigenstate~\cite{Meier2020CounterdiabaticControl}, as well as the adiabatic evolution of a Gibbs state to assist heat engines~\cite{Hartmann2020ManyBody,Hartmann2020MultiSpin}. These processes would be improved by designing weighted actions that emphasize the relevant parts of the energy spectrum. 
For example, a weighted action for preparing a mid-spectrum eigenstate could be designed by regarding the problem as the preparation of the low-energy state of $[H(\lambda)-e_\lambda]^2$, where a constant $e_\lambda$ is chosen close to the energy of the target state, and using our framework of ground-state evolution in Sec.~\ref{sec:ground_state}. 

The weighted variational method can be combined with various existing frameworks to further improve CD driving. For example, some literature considers optimizations of the intermediate driving path between fixed initial and final states by equipping the Hamiltonian with extra degrees of freedom to be optimized~\cite{Prielinger2021TwoParameterCounter, Mbeng2022RotatedAnsatz, Cepaite2023CounterdiabaticOptimized}. This approach can be easily combined with the weighted variational method. Although the extra optimization requires us to compute driving coefficients multiple times, this computational cost can be reduced by separating parameter-independent traces of operators with parameter-dependent coefficients, as has been discussed in  Sec.~\ref{subsec:algorithm_general}\@. It is also an interesting direction to incorporate weighted actions into existing frameworks with tensor networks~\cite{Kim2024PRXQuantum, McKeever2024PRXQuantum}, digital quantum simulation~\cite{Hegade2021DigitizedAdiabatic,Hegade2022DigitizedCounterdiabatic}, and Floquet-based methods~\cite{Petiziol2024QuantumControl,Petiziol2018FastAdiabatic,Petiziol2019AcceleratingAdiabatic,Claeys2019FloquetEngineering, Schindler2024CounterdiabaticDriving}, which enhance variational CD driving in ways that complement our proposal.
In particular, the Floquet engineering can produce effective coupling terms that are hard to implement directly. The coefficients on these effective terms can be improved using our method.

From a computational point of view, we leave two problems for future work. First, as mentioned above, a computer algebra framework for fermionic systems should be explicitly constructed and implemented. This can be achieved by combining an existing treatment of fermionic operators based on normal-order products~\cite{Xie2022VariationalCounterdiabatic} with our algorithm for spin-$1/2$ systems. Second, while our algorithm works for systems without any regular structure, it may be possible to find a more efficient algorithm by taking advantage of regular lattice structures. In fact, the traces of the powers of the Hamiltonian $\tr(H^k)$ for $k=1,2,\dots$ for such systems have been efficiently evaluated using graph-theoretic considerations in high-temperature expansion of Gibbs states~\cite{Domb1972ModernQuantumMechanics,Schmidt2011EighthOrder}. Similar considerations may be useful for evaluating $Q^{(\mathcal{P})}_{\mu\nu}(\lambda)$ and $r^{(\mathcal{P})}_\mu(\lambda)$ more efficiently.

While we have focused entirely on the application of variational AGPs to CD driving, our results may also have implications for the roles of variational AGPs as probes of fundamental properties of quantum many-body systems. The exact AGP, along with closely related concepts such as the quantum geometric tensor~\cite{del2012AssistedFinite,Funo2017UniversalWork,Shingu2025GeometricalScheduling,Dengis2025MultimodeNoon} and the regularized AGP~\cite{Pandey2020AdiabaticEigenstate}, is known to be a sensitive probe of quantum phase transitions~\cite{Zanardi2007InformationTheoretic} and quantum chaos~\cite{Pandey2020AdiabaticEigenstate}. Inheriting this role, variational AGPs have been used to probe phase transitions~\cite{Hatomura2021ControllingAndExploring,Takahashi2024ShortcutsToAdiabaticity}, quantum chaos~\cite{Bhattacharjee2023ALanczosApproach}, and macroscopic singularities~\cite{Sugiura2021AdiabaticLandscape}. These roles of variational AGPs could be enhanced by the weighted variational method. For example, the variational AGP obtained from the weighted variational method for ground-state evolution (Sec.~\ref{sec:ground_state}) should reflect the properties of the ground state more sensitively than the conventional method, and thus it may be more sensitive to quantum phase transitions, which are most prominent in the ground state. Exploring the impact of the weighted variational method in these fundamental contexts is another interesting direction for future research.

\begin{acknowledgments}
N.O. thanks Lewis Ruks, Sosuke Ito, Ken Hiura, and Kohei Yoshimura for discussions. This work was supported by JSPS KAKENHI Grant Number 23KJ0732 and JST Moonshot R\&D Grant Number JPMJMS2061.
\end{acknowledgments}

\section*{Data Availability}

The data that support the findings of this article are openly available~\cite{GitHub}.

\appendix

\section{Appendix to the general framework 
\label{sec:apdx_general}}

\subsection{Locality of the conventional action 
\label{subsec:apdx_nonlocal}}

We show that the conventional action is given by a sum of local terms, assuming that $H(\lambda)$, $\partial_{\lambda}H(\lambda)$, and $A_{\mu}$'s are $k$-local operators with $k$ small and independent of the system size. More precisely, for a fixed $\lambda$, we assume that we can write $H(\lambda) = h_{1}+h_{2}+\cdots$, $\partial_{\lambda}H(\lambda) = l_{1}+l_{2}+\cdots$, and $A_{\mu} = a_{\mu1}+a_{\mu2}+\cdots$, where $h_{i}$, $l_{i}$, and $a_{\mu i}$ are operators that act nontrivially on at most $k$ sites. Based on this assumption, we can expand the action $\mathcal{S}_{\lambda}^{(1)}[V(\bm{\alpha})]$ in Eq.~\eqref{eq:S_min_2-1} as
\begin{align}
 \mathcal{S}_{\lambda}^{(1)}[V(\bm{\alpha})] &  = -\hbar^{-2}\sum_{\mu ij}\sum_{\nu i'j'}\tr\left\{ [h_{i},a_{\mu j}][h_{i'},a_{\nu j'}]\right\} \alpha_{\mu}\alpha_{\nu}\nonumber \\
 & \quad-2i\hbar^{-1} \sum_{\mu ij}\sum_{i'}\tr\left\{ [h_{i},a_{\mu j}]l_{i'}\right\} \alpha_{\mu}+\mathrm{const.}
\end{align}
In this expansion, $[h_{i},a_{\mu j}]$ is zero if the domain of $h_{i}$ does not overlap the domain of $a_{\mu j}$. Otherwise, $[h_{i},a_{\mu j}]$ is at most a $(2k-1)$-local operator. Furthermore, if the domains of $[h_{i}, a_{\mu j}]$ and $l_{i'}$ do not overlap, the trace $\tr\{ [h_{i}, a_{\mu j}]l_{i'}\}$ is proportional to the product $\tr\{[h_{i}, a_{\mu j}]\} \tr (l_{i'}) $, which is zero due to $\tr\{[h_{i}, a_{\mu j}]\}=0$. Otherwise, the operator $[h_{i},a_{\mu j}]l_{i'}$ is at most $(3k-2)$-local, and thus the trace $\tr\{ [h_{i},a_{\mu j}]l_{i'}\} $ is determined by at most $(3k-2)$-local information. Similarly, the trace $\tr\{ [h_{i}, a_{\mu j}][h_{i'}, a_{\nu j'}]\} $ is either zero or dependent on at most $(4k-3)$-local information. Thus, the action $\mathcal{S}_{\lambda}^{(1)}[V(\bm{\alpha})]$ consists of local terms except for the ``const.''~term, which does not affect the minimization.

This locality of $\mathcal{S}_{\lambda}^{(1)}[V(\bm{\alpha})]$ does not necessarily imply that the resulting coefficient $\alpha_{\mu}$ depends only on local information. Nevertheless, this is true if $\sum_{iji'j'}\tr\{ [h_{i}, a_{\mu j}][h_{i'}, a_{\nu j'}]\}  = 0$ for any $\mu \neq \nu$, in which case the determination of $\alpha_\mu$ is decoupled from the determination of $\alpha_\nu$. This situation arises, for example, for locally interacting spin systems with the one-body driving~\cite{Sels2017Minimizing, Hartmann2019RapidCounter}.

\subsection{Derivation of \texorpdfstring{Eqs.~\eqref{eq:algebraic_char-general}, \eqref{eq:variational_char-general}, and \eqref{eq:actionK_explicit-general}}
{Eqs.~(11), (12), and (14)}
\label{subsec:apdx_derivations}}

We derive Eqs.~\eqref{eq:algebraic_char-general}, \eqref{eq:variational_char-general}, and \eqref{eq:actionK_explicit-general}, which are central to the general framework in Sec.~\ref{subsec:weighted_variational_general}\@.
We fix a single value of $\lambda$ and omit the $\lambda$ dependencies of quantities such as 
$H(\lambda)$, $\epsilon_{n}(\lambda)$, $\ket{\phi_{n}(\lambda)}$, $\Phi(\lambda)$, $\mathcal{S}^{(\mathcal{P})}_\lambda[V]$, and $\mathcal{P}_{\lambda}(x)$
for notational simplicity.

To derive the algebraic characterization of the exact AGP in Eq.~\eqref{eq:algebraic_char-general}, we first calculate $\partial_\lambda \bigl(\ketbra{\phi_{n}}{\phi_{n}}\bigr)$ and $\partial'_{\lambda}\mathcal{P}(H)$. The derivative of the projector $\ketbra{\phi_{n}}{\phi_{n}}$ is 
\begin{align}
 & \partial_{\lambda}\bigl(\ketbra{\phi_{n}}{\phi_{n}}\bigr)
 \nonumber \\
 &  = \ketbra{\phi_{n}}{\partial_{\lambda}\phi_{n}} + \ketbra{\partial_{\lambda}\phi_{n}}{\phi_{n}}
 \nonumber \\
 &  = \sum_{m}\ket{\phi_{n}}\braket{\partial_{\lambda}\phi_{n}}{\phi_{m}}\bra{\phi_{m}}
 + \sum_{m}\ket{\phi_{m}}\braket{\phi_{m}}{\partial_{\lambda}\phi_{n}}\bra{\phi_{n}}
 \nonumber \\
 &  = i\hbar^{-1} \sum_{m}\bigl(\Phi_{nm}\ketbra{\phi_{n}}{\phi_{m}} - \Phi_{mn}\ketbra{\phi_{m}}{\phi_{n}}\bigr),
\label{eq:deriv_projection}
\end{align}
where the last equality uses the definition of $\Phi$ in Eq.~\eqref{eq:Hcd} and $\braket{\partial_{\lambda}\phi_{m}}{\phi_{n}}+\braket{\phi_{m}}{\partial_{\lambda}\phi_{n}} = \partial_{\lambda}[\braket{\phi_{m}}{\phi_{n}}] = 0$. Using Eq.~\eqref{eq:deriv_projection}, we can calculate the $\lambda$-derivative of the fictitious Hamiltonian $\mathcal{P}(H) = \sum_{n}\mathcal{P}(\epsilon_{n})\ketbra{\phi_{n}}{\phi_{n}}$ as
\begin{align}
\partial'_{\lambda}\mathcal{P}(H) &  = \sum_{n}\bigl[\partial'_{\lambda}\mathcal{P}(\epsilon_{n})\bigr]\ketbra{\phi_{n}}{\phi_{n}}\nonumber \\
 & \quad+i\hbar^{-1}\sum_{nm}\mathcal{P}(\epsilon_{n}) \bigl(\Phi_{nm}\ketbra{\phi_{n}}{\phi_{m}} -\Phi_{mn}\ketbra{\phi_{m}}{\phi_{n}} \bigr)\nonumber \\
 &  = \sum_{n}\bigl[\partial'_{\lambda}\mathcal{P}(\epsilon_{n})\bigr]\ketbra{\phi_{n}}{\phi_{n}}\nonumber \\
 & \quad+i\hbar^{-1}\sum_{nm}\bigl[\mathcal{P}(\epsilon_{n})-\mathcal{P}(\epsilon_{m})\bigr]\Phi_{nm}\ketbra{\phi_{n}}{\phi_{m}}\nonumber \\
 &  = \sum_{n}\bigl[\partial'_{\lambda}\mathcal{P}(\epsilon_{n})\bigr]\ketbra{\phi_{n}}{\phi_{n}}+i\hbar^{-1}[\mathcal{P}(H),\Phi].
\label{eq:deriv_PH}
\end{align}
This relation shows that $\partial'_{\lambda}\mathcal{P}(H)-i\hbar^{-1}[\mathcal{P}(H),\Phi]$ is diagonal in the Hamiltonian basis, leading to the algebraic characterization in Eq.~\eqref{eq:algebraic_char-general}.

Next, we show that the algebraic characterization in Eq.~\eqref{eq:algebraic_char-general} is equivalent to the variational expression in Eq.~\eqref{eq:variational_char-general}, which extends the derivation of the conventional variational method in Refs.~\cite{Sels2017Minimizing,Kolodrubetz2017GeometryAndNonAdiabatic}. The algebraic characterization says that $\partial'_{\lambda}\mathcal{P}(H)-i\hbar^{-1}[\mathcal{P}(H),\Phi]$ is diagonal with respect to the Hamiltonian basis. The diagonal elements come solely from $\partial'_{\lambda}\mathcal{P}(H)$ since $[\mathcal{P}(H),\Phi]$ has no diagonal elements. Let $\Psi \coloneqq \sum_n [\partial'_{\lambda}\mathcal{P}(H)]_{nn} \ketbra{\phi_n}{\phi_n} $ denote the diagonal part of $\partial'_\lambda\mathcal{P}(H)$. Then, the algebraic characterization is equivalent to 
\begin{equation}
\partial'_{\lambda}\mathcal{P}(H)-i\hbar^{-1}[\mathcal{P}(H),\Phi] - \Psi = 0.
\label{eq:deriv_algebraic_char_1}
\end{equation}
This characterization of $\Phi$ is further equivalent to 
\begin{equation}
\Phi \in \argmin_{V}\Vert\partial'_{\lambda}\mathcal{P}(H)-i\hbar^{-1}[\mathcal{P}(H),V]-\Psi\Vert^{2}.
\label{eq:deriv_algebraic_char_2}
\end{equation}
We expand the Hilbert--Schmidt norm as 
\begin{align}
 & \Vert\partial'_{\lambda}\mathcal{P}(H)-i\hbar^{-1}[\mathcal{P}(H),V]-\Psi\Vert^{2}\nonumber \\
 &  = \Vert\partial'_{\lambda}\mathcal{P}(H)-i\hbar^{-1}[\mathcal{P}(H),V]\Vert^{2}
 -2\tr\left\{ \partial'_{\lambda}\mathcal{P}(H)\Psi\right\}
 \nonumber \\
 & \quad +2i\hbar^{-1}\tr\left\{ [\mathcal{P}(H),V]\Psi\right\}  +\tr(\Psi^{2})\nonumber \\
 &  = \Vert\partial'_{\lambda}\mathcal{P}(H)-i\hbar^{-1}[\mathcal{P}(H),V]\Vert^{2}+\mathrm{const}.,
 \label{eq:deriv_algebraic_char_3}
\end{align}
where ``const.''~refers to terms independent of $V$, and we use $\tr\bigl\{[\mathcal{P}(H),V]\Psi\bigr\} = \tr\bigl\{[\Psi,\mathcal{P}(H)]V\bigr\} = 0$ in the second equality because $\Psi$ and $\mathcal{P}(H)$ are both diagonal in the basis of $H$. Combining Eqs.~\eqref{eq:deriv_algebraic_char_2} and \eqref{eq:deriv_algebraic_char_3} proves that the algebraic characterization in Eq.~\eqref{eq:algebraic_char-general} is equivalent to the variational principle in Eq.~\eqref{eq:variational_char-general}.

Finally, we derive the expansion of $\mathcal{S}^{(\mathcal{P})}[V]$ in Eq.~\eqref{eq:actionK_explicit-general}. Using the expression of $\partial'_\lambda\mathcal{P}(H)$ in Eq.~\eqref{eq:deriv_PH}, we can rewrite the operator inside the Hilbert--Schmidt norm as
\begin{align}
 & \partial'_{\lambda}\mathcal{P}(H)-i\hbar^{-1}[\mathcal{P}(H),V]\nonumber \\
 &  = \sum_{n}\bigl[\partial'_{\lambda}\mathcal{P}(\epsilon_{n})\bigr]\ketbra{\phi_{n}}{\phi_{n}}\nonumber \\
 & \quad+i\hbar^{-1}\sum_{nm}\bigl[\mathcal{P}(\epsilon_{m})-\mathcal{P}(\epsilon_{n})\bigr][\Phi_{mn}-V_{mn}]\ketbra{\phi_{m}}{\phi_{n}}.
\end{align}
Since the Hilbert--Schmidt norm equals the sum of the squared absolute values of the matrix elements, the action is given by
\begin{align}
 \mathcal{S}^{(\mathcal{P})}[V] &= \sum_{n}\bigl[\partial'_{\lambda}\mathcal{P}(\epsilon_{n})\bigr]^{2} 
 \notag \\
 &\quad + \hbar^{-2}\sum_{\smash{nm}}\bigl[\mathcal{P}(\epsilon_{m})-\mathcal{P}(\epsilon_{n})\bigr]^{2}\vert\Phi_{mn}-V_{mn}\vert^{2},
\end{align}
which proves Eq.~\eqref{eq:actionK_explicit-general}.

\subsection{Inclusion of a conserved charge 
\label{subsec:apdx_conserved_charge}}

When the Hamiltonian has a conserved charge, our formulation can incorporate it to derive a wider variety of weighted actions. Let $J(\lambda)$ be an operator that satisfies $[H(\lambda),J(\lambda)] = 0$, and we expand it in the Hamiltonian basis as 
\begin{equation}
J(\lambda) = \sum_{n}j_{n}(\lambda)\ketbra{\phi_{n}(\lambda)}{\phi_{n}(\lambda)}.
\end{equation}
Introducing an arbitrary bivariate degree-$K$ polynomial $\mathcal{P}_{\lambda}(x,y) = \sum_{k = 0}^{K}\sum_{l = 0}^{k}p_{l,k-l}(\lambda)x^{l}y^{k-l}$, we can use $\mathcal{P}_\lambda(H(\lambda),J(\lambda))$ as a fictitious Hamiltonian in our framework. We can then determine driving coefficients by minimizing a weighted action of the form
\begin{align}
\mathcal{S}_{\lambda}^{(\mathcal{P})}[V] &  = \Vert\partial'_{\lambda}\mathcal{P}_{\lambda}(H(\lambda),J(\lambda))-i\hbar^{-1}[\mathcal{P}_{\lambda}(H(\lambda),J(\lambda)),V]\Vert^{2}\nonumber \\[0.3em]
 &  = \hbar^{-2}\smash{\sum_{nm}}\Bigl\{[\mathcal{P}_{\lambda}(\epsilon_{m}(\lambda),j_{m}(\lambda))-\mathcal{P}_{\lambda}(\epsilon_{n}(\lambda),j_{n}(\lambda))]^{2}\nonumber \\
 & \qquad\qquad\qquad\times\bigl\vert[V-\Phi(\lambda)]_{mn}\bigr\vert^{2}\Bigr\}+\mathrm{const.}
 \label{eq:conserved-charge-expand}
\end{align}
This action can be useful for assigning weights more selectively to relevant eigenstate pairs.

For example, let us consider that $J(\lambda) \equiv J$ is a parity operator independent of $\lambda$ with the eigenvalues $j_n(\lambda) \equiv j_n =\pm 1$. The eigenstates are then split into even-parity ($j_n = +1$) and odd-parity ($j_n=-1$) states. Let us also assume that the driving operator $A_\mu$ has even parity for all $\mu$, meaning that $JA_\mu J=A_\mu$. Then, the matrix element $[V(\bm \alpha)-\Phi(\lambda)]_{mn}$ is nonzero only when the states $m$ and $n$ have the same parity. In this case, we can use $\mathcal{P}_\lambda(H(\lambda),J)=[H(\lambda)-E_\lambda^{(K)}]^K(J + 1)$ to prioritize the suppression of nonadiabatic transitions between even-parity states. With this choice, $\mathcal{P}_\lambda(\epsilon_n(\lambda),j_n)$ is zero for any odd-parity $n$, and thus the weighted action only penalizes the nonadiabatic transitions between even-parity states.

Note that the idea of focusing on a conserved subspace of $H(\lambda)$ has already been considered for the conventional variational method on a specific system.
Reference~\cite{Passarelli2020CounterdiabaticDriving} studies the $p$-spin Hamiltonian on spin-1/2 systems, which conserves the total spin. The study compares the conventional variational method applied to the subspace of the maximum total spin with that applied to the whole Hilbert space, finding that the former outperforms the latter for a severely underparameterized ansatz. Since including $J(\lambda)$ in our proposal is considered a much generalized version of this specific result, we expect that the inclusion of $J(\lambda)$ would indeed improve the driving fidelity.

\subsection{Efficient algorithm for the entire range of \texorpdfstring{$\lambda$}{lambda}
\label{subsec:apdx_time-dependent-general}}

We discuss an efficient algorithm to compute the driving protocol over the whole range of $\lambda$, assuming that the Hamiltonian has the form of Eq.~\eqref{eq:Hamiltonian_fF}. We first introduce some notation. Let $\tilde{f}_{1}(\lambda),\tilde{f}_{2}(\lambda),\dots$ be the list of all the possible $k$-fold products of the coefficients $\{f_{1}(\lambda),\dots,f_{\Gamma}(\lambda)\}$ with $k = 0,\dots,K$. For example, when $K = 2$ and $\Gamma = 2$, the list is $(\tilde{f}_{1}(\lambda),\dots,\tilde{f}_{6}(\lambda)) = (f_{1}(\lambda)^{2},f_{1}(\lambda)f_{2}(\lambda),f_{2}(\lambda)^{2},f_{1}(\lambda),f_{2}(\lambda),1)$. The order of the list does not matter. We define $d_{g}$ as the total number of $f_{\gamma}(\lambda)$'s contained in $\tilde{f}_{g}(\lambda)$. In the above example, the numbers are $(d_{1},\dots,d_{6}) = (2,2,2,1,1,0)$. We then define $\lambda$-independent operators $\tilde{F}_{1},\tilde{F}_{2},\dots$ by the relation,
\begin{equation}
H(\lambda)^{k} = \left(\sum_{\gamma = 1}^{\Gamma}f_{\gamma}(\lambda)F_{\gamma}\right)^{k} \eqqcolon \sum_{g:\,d_{g} = k}\tilde{f}_{g}(\lambda)\tilde{F}_{g}
\label{eq_gen:F_g}
\end{equation}
for $k = 0,\dots,K$. In other words, $\tilde{F}_{g}$ is defined as the terms in $H(\lambda)^{d_{g}}$ whose $\lambda$-dependence is given by $\tilde{f}_{g}(\lambda)$. In the example above, the operators are $(\tilde{F}_{1},\dots,\tilde{F}_{6}) = ((F_{1})^{2},F_{1}F_{2}+F_{2}F_{1},(F_{2})^{2},F_{1},F_{2},I)$.

Using these definitions, we can conveniently separate $Q_{\mu\nu}^{(\mathcal{P})}(\lambda)$ and $r_{\mu}^{(\mathcal{P})}(\lambda)$ into $\lambda$-dependent coefficients and traces of $\lambda$-independent operators. By combining Eqs.~\eqref{eq:polynomial} and \eqref{eq_gen:F_g}, the fictitious Hamiltonian is rewritten as $\mathcal{P}_\lambda(H(\lambda)) = \sum_g p_{d_{g}}(\lambda)\tilde{f}_g(\lambda) \tilde F_g $. Inserting this expression into the definitions of $Q_{\mu\nu}^{(\mathcal{P})}(\lambda)$ and $r_{\mu}^{(\mathcal{P})}(\lambda)$ in Eq.~\eqref{eq:def_Qr-general}, we get%
\begin{subequations}
\label{eq:time-dependent-Qr_expansion}
\begin{align}
\!\! Q_{\mu\nu}^{(\mathcal{P})}(\lambda) &  = -\sum_{gg'}p_{d_{g}}(\lambda)p_{d_{g'}}(\lambda)\tilde{f}_{g}(\lambda)\tilde{f}_{g'}(\lambda)\tilde{Q}_{\mu\nu,gg'},\\
\!\! r_{\mu}^{(\mathcal{P})}(\lambda) &  = i\hbar\sum_{gg'}p_{d_{g}}(\lambda)p_{d_{g'}}(\lambda)[\partial_{\lambda}\tilde{f}_{g}(\lambda)]\tilde{f}_{g'}(\lambda)\tilde{r}_{\mu,gg'},
\end{align}
\end{subequations}
where we define%
\begin{subequations}
\label{eq:time-dependent-Qr_def}
\begin{align}
\tilde{Q}_{\mu\nu,gg'} &  \coloneqq \tr\left\{ [\tilde{F}_{g},A_{\mu}][\tilde{F}_{g'},A_{\nu}]\right\} ,\\
\tilde{r}_{\mu,gg'} &  \coloneqq \tr\left\{ \tilde{F}_{g}[\tilde{F}_{g'},A_{\mu}]\right\}.
\end{align}
\end{subequations}
These expressions allow efficient computation of $Q_{\mu\nu}^{(\mathcal{P})}(\lambda)$ and $r_{\mu}^{(\mathcal{P})}(\lambda)$. To calculate them over the entire range of $\lambda$, we only need to calculate the traces $\tilde{Q}_{\mu\nu,gg'}$ and $\tilde{r}_{\mu,gg'}$ once. 

\section{Appendix to the ground-state evolution 
\label{sec:apdx_GS}}

\subsection{Derivation of the properties of the weight
\label{subsec:apdx_deriv-weight}}

We derive the properties of the weight $w_{mn}^{(K)}$ for ground-state evolution in Sec.~\ref{subsec:ground_state_theory}\@. We omit the $\lambda$ dependencies of $\epsilon_{n}(\lambda)$, $w_{mn}^{(K)}(\lambda)$, $w_{n}^{(K)}(\lambda)$, and $E_\lambda^{(K)}$ for conciseness.

First, we derive the explicit expression of $w_{mn}^{(K)}$ in Eq.~\eqref{eq:weight-GS} as 
\begin{align}
 w_{mn}^{(K)} &  = \left[\frac{(\epsilon_{n}-E^{(K)})^{K}-(\epsilon_{m}-E^{(K)})^{K}}{(\epsilon_{n}-E^{(K)})-(\epsilon_{m}-E^{(K)})}\right]^{2}
 \nonumber \\
 &  = \left[\sum_{s = 0}^{K-1}(\epsilon_{n}-E^{(K)})^{K-1-s}(\epsilon_{m}-E^{(K)})^{s}\right]^{2}
 \nonumber \\
 &  = \sum_{s = 0}^{K-1}\sum_{u = 0}^{K-1}(\epsilon_{n}-E^{(K)})^{(K-1-s)+u}(\epsilon_{m}-E^{(K)})^{(K-1-u)+s}
 \nonumber \\
 & = \!\sum_{\smash{s = -(K-1)}}^{K-1}\!(K-\vert s\vert)(\epsilon_{n}-E^{(K)})^{K-1-s}(\epsilon_{m}-E^{(K)})^{K-1+s},
\label{eq:weight-GS_deriv}
\end{align}
where we combined the terms with the same value of $s-u$ in the last equality.

Next, we prove the inequality in Eq.~\eqref{eq:weight_upper}. Each term in the last line of Eq.~\eqref{eq:weight-GS_deriv} is upper bounded as
\begin{align}
    & (\epsilon_{n}-E^{(K)})^{K-1-s}(\epsilon_{m}-E^{(K)})^{K-1+s}
    \nonumber \\
    &\quad\leq \vert\epsilon_{n}-E^{(K)}\vert^{K-1-s} \vert\epsilon_{m}-E^{(K)}\vert^{K-1+s}
    \nonumber \\
    &\quad \leq \max \bigl\{ \vert\epsilon_{n}-E^{(K)}\vert^{2K-2}, \vert\epsilon_{m}-E^{(K)}\vert^{2K-2}\bigr\}.
\end{align}
Inserting this upper bound into Eq.~\eqref{eq:weight-GS_deriv} and using $\sum_{s=-(K-1)}^{K-1}(K-\vert s\vert)=K^2$ gives
\begin{equation}
    w_{mn}^{(K)} \leq \max\bigl\{w_n^{(K)},w_m^{(K)}\bigr\},
    \label{eq:weight-upper_deriv}
\end{equation}
which reproduces Eq.~\eqref{eq:weight_upper}. Furthermore, when $(\epsilon_{n}-E^{(K)})$ and $(\epsilon_{m}-E^{(K)})$ have the same sign, each term in the last line of Eq.~\eqref{eq:weight-GS_deriv} is nonnegative, and thus we have a lower bound,
\begin{align}
    & (\epsilon_{n}-E^{(K)})^{K-1-s}(\epsilon_{m}-E^{(K)})^{K-1+s}
    \nonumber \\
    &\quad = \vert\epsilon_{n}-E^{(K)}\vert^{K-1-s} \vert\epsilon_{m}-E^{(K)}\vert^{K-1+s}
    \nonumber \\
    &\quad \geq \min \bigl\{ \vert\epsilon_{n}-E^{(K)}\vert^{2K-2}, \vert\epsilon_{m}-E^{(K)}\vert^{2K-2}\bigr\}.
\end{align}
Inserting this lower bound into Eq.~\eqref{eq:weight-GS_deriv} and combining with Eq.~\eqref{eq:weight-upper_deriv}, we obtain 
\begin{equation}
\min\{w_{n}^{(K)},w_{m}^{(K)}\} \leq  w_{mn}^{(K)} \leq  \max\{w_{n}^{(K)},w_{m}^{(K)}\},
\label{eq:weight-sandwitch_deriv}
\end{equation}
which shows that $w_{mn}^{(K)}$ lies between $w_{n}^{(K)}$ and $w_{m}^{(K)}$.

\begin{figure}
\includegraphics[width=\columnwidth]{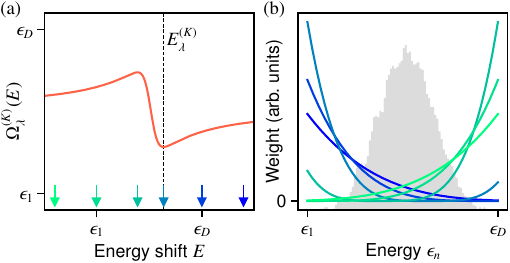}
\caption{Typical behavior of the function $\Omega_{\lambda}^{(K)}(E)$ and the weight $[\epsilon_n(\lambda)-E]^{2K-2}$, exemplified with a ferromagnetic system of size $N = 12$ with $K = 3$ at $\lambda = 0.25$. (a) Typical shape of the function $\Omega_{\lambda}^{(K)}(E)$ (red), which has one maximum and one minimum. The optimal energy shift $E_{\lambda}^{(K)}$ is determined as the value of $E$ that minimizes $\Omega_{\lambda}^{(K)}(E)$ (black dashed line). The energies $\epsilon_{1}$ and $\epsilon_{D}$ denote the lowest and highest energy eigenvalues. (b) To understand this shape of $\Omega_{\lambda}^{(K)}(E)$, we plot the weight $[\epsilon_n(\lambda)-E]^{2K-2}$ for six different values of $E$ (green to blue). Green corresponds to a smaller $E$, and blue corresponds to a larger $E$, as indicated by the arrows in panel (a). Each weight is normalized so that the area under the curve is constant for better visualization, noting that the overall normalization is unimportant. Gray histogram shows the density of states.
\label{fig:Omega_function}}
\end{figure}

\subsection{Determination of the energy shift 
\label{subsec:apdx_Omega}}

We discuss the typical behavior of the function $\Omega_{\lambda}^{(K)}(E)$ and the optimal energy shift $E_{\lambda}^{(K)}$ in Eq.~\eqref{eq:E_min}. We also discuss a numerical procedure for minimizing $\Omega_{\lambda}^{(K)}(E)$ to determine $E_{\lambda}^{(K)}$. We regard $\lambda$ as fixed throughout this subsection.

In Fig.~\ref{fig:Omega_function}(a), we show a typical shape of the function $\Omega_{\lambda}^{(K)}(E)$, which has one maximum and one minimum. To understand this shape, we recall the definition of $\Omega_{\lambda}^{(K)}(E)$ in Eq.~\eqref{eq:E_min_2}, which says that $\Omega_{\lambda}^{(K)}(E)$ is the weighted average of the energy eigenvalues $\epsilon_n(\lambda)$ over the weight $[\epsilon_n(\lambda )-E]^{2K-2}$. We plot the weight for six values of $E$ in Fig.~\ref{fig:Omega_function}(b). When $E$ takes a large negative value $[\epsilon_1(\lambda) - E] \gg [\epsilon_D(\lambda)-\epsilon_1(\lambda)]$, the weight is approximated as $[\epsilon_n(\lambda)-E]^{2K-2} \simeq [\epsilon_1(\lambda) - E]^{2K-2}$, and thus, it is almost uniform over the entire range of the energy spectrum. As $E$ is gradually increased, the weight becomes an increasing function of the energy eigenvalue over the energy spectrum, putting weight to the high-energy states [green curves in Fig.~\ref{fig:Omega_function}(b)]. As $E$ is further increased, the minimum point of the weight passes through the middle of the energy spectrum. Then, the weight turns into a decreasing function of the energy eigenvalue over the energy spectrum, and the weight concentrates on the low-energy states [blue curves in Fig.~\ref{fig:Omega_function}(b)]. Finally, as $E$ becomes larger $[E-\epsilon_D(\lambda)] \gg [\epsilon_D(\lambda) - \epsilon_1(\lambda)]$, the weight becomes almost uniform again. This course of change forms the maximum and minimum of $\Omega_{\lambda}^{(K)}(E)$.

The optimal energy shift $E =E_{\lambda}^{(K)}$ is determined so that the weight $[\epsilon_n(\lambda)-E]^{2K-2}$ is the most concentrated on the low-energy states. Due to the shape of $\Omega_{\lambda}^{(K)}(E)$, the resulting $E_{\lambda}^{(K)}$ is typically higher than the center of the energy spectrum. This behavior is confirmed by Fig.~\ref{fig:ferromagnetic_weight_coeff}(a).

The numerical minimization of $\Omega_{\lambda}^{(K)}(E)$ is a one-dimensional optimization problem, and it can be stably solved by the following procedure. We start with Eq.~\eqref{eq:E_cost_function}, which shows that $\Omega_{\lambda}^{(K)}(E)$ is a rational function of $E$. We analytically calculate $d\Omega_{\lambda}^{(K)}/dE$, which will again be a rational function of $E$, and we equate the numerator to zero. The numerator is a polynomial of $E$, and the resulting polynomial equation is expected to have two real roots because $\Omega_{\lambda}^{(K)}(E)$ has two extremum points. We need to solve the larger root of the polynomial equation. For $K = 2$, the polynomial equation is quadratic and analytically solvable, and the larger solution gives $E^{(2)}_\lambda$. For $K\geq3$, we can use standard numerical methods, such as the Newton method, to get $E_{\lambda}^{(K)}$. As the initial value of the Newton method, we can use the analytical value of $E^{(2)}_\lambda$, which is expected to be close to $E_{\lambda}^{(K)}$ with $K\geq3$.

\subsection{Efficient algorithm for the entire range of \texorpdfstring{$\lambda$}{lambda}
\label{subsec:apdx_time-dependent-GS}}

When the Hamiltonian has the form of Eq.~\eqref{eq:Hamiltonian_fF}, we can efficiently evaluate $\omega_{k}(\lambda) = \tr[H(\lambda){}^{k}]$ for $k = 1,\dots,2K-1$ by separating the $\lambda$-dependent coefficients and $\lambda$-independent operators, as has been done for $Q_{\mu\nu}^{(\mathcal{P})}(\lambda)$ and $r_{\mu}^{(\mathcal{P})}(\lambda)$ in the general framework in Appendix~\ref{subsec:apdx_time-dependent-general}\@. We use Eq.~\eqref{eq_gen:F_g} to expand $\omega_{k}(\lambda) = \tr[H(\lambda)^{k}]$. For $k = 1,\dots,K$, we obtain
\begin{equation}
\omega_{k}(\lambda) = {\displaystyle \sum_{g:\,d_{g} = k}}\tilde{f}_{g}(\lambda)\tilde{\omega}_{g},\qquad\tilde{\omega}_{g} \coloneqq \tr(\tilde{F}_{g}),
\label{eq_min:omega_former}
\end{equation}
where $\tilde{f}_g(\lambda)$ and $\tilde{F}_g$ are the scalar coefficient and operator, respectively, defined by the expansion of $H(\lambda)^k$ in Eq.~\eqref{eq_gen:F_g}. Similarly, for $k = K+1,\dots,2K-1$, we obtain%
\begin{subequations}
\label{eq_min:omega_latter}
\begin{align}
\omega_{k}(\lambda) &  = {\displaystyle \sum_{g:\,d_{g} = K}\,\sum_{g':\,d_{g'} = k-K}}\!\!\tilde{f}_{g}(\lambda)\tilde{f}_{g'}(\lambda)\tilde{\omega}_{gg'},\\
\tilde{\omega}_{gg'}\, & \! \coloneqq \tr(\tilde{F}_{g}\tilde{F}_{g'}),
\end{align}
\end{subequations}
where we used $\omega_k(\lambda) = \tr[H(\lambda)^K H(\lambda)^{k-K}]$. Once we compute the traces $\tilde{\omega}_{g}$ and $\tilde{\omega}_{gg'}$, we can easily calculate $\omega_{k}(\lambda)$ over the entire range of $\lambda$.

\subsection{Derivation of the partial action decomposition and the ideal action
\label{subsec:apdx_deriv-partial-action}}

We derive the partial action decomposition and its properties used in Sec.~\ref{subsec:partial_action}\@. We fix $\lambda$ and omit the $\lambda$ dependencies of quantities.

To decompose the action into partial actions in Eq.~\eqref{eq:partial_action_all}, we combine the expansion of the action in Eq.~\eqref{eq:actionK_explicit-general} and the definition of the weight in Eq.~\eqref{eq:weight_def-general} to obtain
\begin{equation}
    \mathcal{S}^{(K)}[V] 
    = \frac{1}{\hbar^{2}} \sum_{nm} (\epsilon_m - \epsilon_n)^2 w_{mn}^{(K)} \bigl\vert[V-\Phi]_{mn}\bigr\vert^{2}+\mathrm{const.} 
\end{equation}
The summand is symmetric under the exchange of $m$ and $n$, and it is zero for $m=n$. Therefore, we can combine the terms with $(n,m)$ and $(m,n)$ to rewrite
\begin{align}
    &\mathcal{S}^{(K)}[V] 
    = \frac{2}{\hbar^{2}} \sum_{\substack{nm\\w_m^{(K)} \leq w_n^{(K)}}} (\epsilon_m - \epsilon_n)^2 \chi_{mn}^{(K)} w_{mn}^{(K)} \bigl\vert[V-\Phi]_{mn}\bigr\vert^{2} 
    \nonumber \\[-1em]
    &\hspace{15em}+\mathrm{const.},
    \label{eq:partial_action_deriv}
\end{align}
where we inserted $\chi_{mn}^{(K)}$ because if $m$ and $n$ satisfy $w_m^{(K)} = w_n^{(K)}$ and $m\neq n$, both $(m,n)$ and $(n,m)$ appear in the sum $\sum_{nm:w_m^{(K)}\leq w_n^{(K)}}$. The decomposition in Eq.~\eqref{eq:partial_action_all} easily follows from Eq.~\eqref{eq:partial_action_deriv}.

Next, we show that the $K$-dependence of the partial action $\mathcal{T}^{(K,n)}[V]$ is between $O(K^0)$ and $O(K^{-2})$. For simplicity, we neglect the $K$-dependence of $E^{(K)}$ because the dependence is subtle [see, e.g., Fig.~\ref{fig:ferromagnetic_weight_coeff}(a)]. Then, the $K$-dependence of $\mathcal{T}^{(K,n)}[V]$ is due to the ratio $w_{mn}^{(K)}/w_n^{(K)}$. This ratio is written as
\begin{align}
    \frac{w_{mn}^{(K)}}{w_{n}^{(K)}} 
    = \frac{1}{K^2} \left[ \sum_{s = 0}^{K-1} \left( \frac{\epsilon_{m}-E^{(K)}}{\epsilon_{n}-E^{(K)}} \right)^{s}\right]^{2},
    \label{eq:weight-ratio}
\end{align}
where we used the second line of Eq.~\eqref{eq:weight-GS_deriv}. The condition $w_m^{(K)} \leq  w_n^{(K)}$ implies $\vert \epsilon_m - E^{(K)}\vert \leq \vert \epsilon_n -E^{(K)}\vert$, and thus, the sum in Eq.~\eqref{eq:weight-ratio} is a finite geometric sum with a common ratio $(\epsilon_m-E^{(K)})/(\epsilon_n-E^{(K)}) \in [-1,1]$. From Eq.~\eqref{eq:weight-ratio}, we deduce the following scaling of the ratio $w_{mn}^{(K)}/w_n^{(K)}$ in three extreme cases: (i) if $\vert \epsilon_m - E^{(K)}\vert \ll \vert \epsilon_n -E^{(K)}\vert$, the geometric sum converges quickly for a small value of $K$, and thus the ratio scales as $O(K^{-2})$; (ii) if $ (\epsilon_m - E^{(K)}) \simeq  (\epsilon_n -E^{(K)})$ and $K$ is not too large, the geometric sum is roughly proportional to $K$, and the ratio scales as $O(K^{0})$; However, as $K$ becomes larger, the geometric sum starts to converge, and the scaling changes to $O(K^{-2})$; (iii) if $ (\epsilon_m - E^{(K)}) \simeq  -(\epsilon_n -E^{(K)})$, the sign alternates in the geometric sum, and the value of $w_{mn}^{(K)}/w_{n}^{(K)}$ oscillates for a small $K$; Nevertheless, the scaling eventually becomes $O(K^{-2})$. A more general case lies between these extreme cases. The scaling of $\mathcal{T}^{(K,n)}[V]$ is determined by the sum of finitely many terms with these scalings, and thus, the overall scaling also lies between $O(K^{0})$ and $O(K^{-2})$.

The asymptotic formula of $\mathcal{T}^{(K,n)}[V]$ is obtained from a similar consideration. Under the condition $w^{(K)}_n \geq w^{(K)}_m$, we can assume $(\epsilon_n-E^{(K)})^K \gg (\epsilon_m-E^{(K)})^K$ for a sufficiently large $K$. Using this relation in the second line of Eq.~\eqref{eq:weight-GS}, we can approximate the ratio $w_{mn}^{(K)}/w_n^{(K)}$ as
\begin{align}
 \frac{w_{mn}^{(K)}}{w_{n}^{(K)}} 
  &\simeq \frac{1}{K^2(\epsilon_n-E^{(K)})^{2K-2}}\frac{(\epsilon_n - E^{(K)})^{2K}}{(\epsilon_m - \epsilon_n)^2}
  \nonumber \\
  &=  \frac{1}{K^{2}}\frac{(\epsilon_{n}-E^{(K)})^{2}}{(\epsilon_{m}-\epsilon_{n})^{2}}.
 \label{eq:weight-asymptotic}
\end{align}
Inserting this approximation, the partial action $\mathcal{T}^{(K,1)}[V]$ is approximated by $K^{-2}(\epsilon_{1}-E^{(K)})^{2}\sum_{m:\,m\neq1}\vert[V-\Phi]_{1m}\vert^{2}$, where we used the fact that $w_1^{(K)}$ becomes larger than any other $w_m^{(K)}$ for a sufficiently large $K$. This asymptotic formula of $\mathcal{T}^{(K,1)}[V]$ has been used in Sec.~\ref{subsec:partial_action} to show that $\mathcal{S}^{(K)}[V]$ approaches the ideal action up to normalization.

Note that the ideal action can be derived more directly without relying on the partial action decomposition. In the limit $K \to \infty$, the fictitious Hamiltonian for the ground-state driving,
\begin{equation}
\mathcal{P}^{\mathrm{GS},K}(H) = (H-E^{(K)})^{K} = \sum_{n = 1}^{D}(\epsilon_{n}-E^{(K)})^{K}\ketbra{\phi_{n}}{\phi_{n}},
\end{equation}
approaches $(\epsilon_1 - E^{(K)})^K \ketbra{\phi_n}{\phi_n} $, which is proportional to the projection operator onto the ground state. Thus, $\mathcal{P}(\epsilon_n)$ approaches $(\epsilon_1 - E^{(K)})^K \delta_{n1}$. Inserting this into Eq.~\eqref{eq:actionK_explicit-general} reproduces the ideal action in Eq.~\eqref{eq:actioninf_def} up to normalization and the constant term.

\section{Details of the computer algebra for general spin-1/2 systems 
\label{sec:apdx_algebraic_computation}}

We discuss the details of the computer algebra algorithm and analyze its computational time. We focus entirely on spin-$1/2$ systems for concreteness, while a similar analysis should be possible for spin systems with any spin quantum number and fermionic systems.

\subsection{Data structure and elementary operations 
\label{subsec:apdx_elementary_operation}}

We propose a computer algebra framework of spin-$1/2$ systems. 
The idea of using algebraic computations in the quantum mechanics of spin-$1/2$ systems is not new; it has appeared repeatedly, but rather sporadically, in the literature~\cite{Baylis1996PauliAlgebra,Filip2010SdCasSpinDynamics,YiZhuangMathematicaPackages,Steeb2010QuantumMechanics,Barone2024CounterdiabaticOptimized,Loizeau2025QuantumMany}.
Our framework inherits the basic concepts from these previous frameworks, but is more efficient and better suited for our purpose. 

We introduce some notation for convenience. We use $A,B,\ldots$ to refer to a general operator acting on the Hilbert space of a system of $N$ Pauli spins. An operator is expanded as a sum of terms, each consisting of a scalar coefficient and a tensor product of Pauli operators. For example,
\begin{equation}
A = 1.5X_{1}Z_{3}-0.3X_{2}Y_{3}Y_{4}+2.4Z_{4}Z_{5}Z_{6},
\label{eq_alg:exampleA}
\end{equation}
is an operator composed of three terms. We use $\sigma,\tau,\dots$ to denote tensor products of Pauli operators, including $I$ as a special case. We define $z_{A,\sigma}$ as the scalar coefficient before $\sigma$ in $A$. For example, $z_{A,\sigma} = 1.5$ for $\sigma = X_{1}Z_{3}$ in Eq.~\eqref{eq_alg:exampleA}. We define $\vert A\vert$ as the number of terms in $A$, e.g., $\vert A\vert = 3$ in the example in Eq.~\eqref{eq_alg:exampleA}. Below, we assume that all terms are $k$-local, with $k$ small and independent of the system size. Based on this assumption, we will disregard the number of Pauli operators per term and only count the number of terms to evaluate the computational time.

Our computer algebra framework stores an operator using a data structure called a hash table~\cite{Cormen2009IntroductionToAlgorithms}. A hash table is a set of correspondences from a ``key'' to a ``value,'' which, in our case, is from a tensor product $\sigma$ to the scalar coefficient $z_{A,\sigma}$. A tensor product of Pauli operators is represented by a list of Pauli operator kinds and indices, sorted in ascending order of the spin index. For example, $X_{1}Z_{3}$ is represented by a list $[(X,1),(Z,3)]$, and the entire operator $A$ in Eq.~\eqref{eq_alg:exampleA} is stored as
\begin{align}
\bigl\{ & [(X,1),(Z,3)]\mapsto1.5,\quad[(X,2),(Y,3),(Y,4)] \mapsto -0.3,\nonumber \\
 & [(Z,4),(Z,5),(Z,6)]\mapsto2.4\bigr\}.
\label{eq_alg:hash_table}
\end{align}
A hash table allows a quick lookup of its elements~\cite{Cormen2009IntroductionToAlgorithms}. Given an arbitrary $\sigma$ and a hash-table representation of an arbitrary operator $A$, it only takes $O(1)$ time to determine whether $\sigma$ is contained in $A$ and, if so, to obtain the scalar coefficient $z_{A,\sigma}$. This search time is independent of $\vert A\vert$ for ideal hash tables.

We can efficiently perform elementary algebraic operations with this data structure. The addition of two operators, $A+B$, is completed in $O(\vert A\vert+\vert B\vert)$ time by the following procedure. We first copy the entire hash table representing $A$ in $O(\vert A\vert)$ time, which we write as $A'$. Next, we iterate through the terms of $B$ and add each term to $A'$. More precisely, for each term $\sigma \mapsto  z_{B,\sigma}$ in $B$, we determine whether $\sigma$ appears in $A'$ or not in $O(1)$ time. If $\sigma$ is found in $A'$, we update the scalar coefficient $z_{A',\sigma}$ to $z_{A',\sigma}+z_{B,\sigma}$ in $O(1)$ time. If $\sigma$ is not found in $A'$, we insert $\sigma \mapsto z_{B,\sigma}$ into $A'$ in $O(1)$ time. At the end of this iteration through $B$, the hash table $A'$ stores the result of $A+B$. This iteration operates in $O(\vert B\vert)$ time. Thus, the entire computation for $A+B$ is performed in $O(\vert A\vert+\vert B\vert)$ time.

The multiplication of two operators, $A$ and $B$, is completed in $O(\vert A\vert\vert B\vert)$ time by the following procedure. We prepare an empty hash table for storing the result. We pick up one term $\sigma \mapsto  z_{A,\sigma}$ from $A$ and one term $\tau \mapsto  z_{B,\tau}$ from $B$, multiply the two tensor products $\sigma$ and $\tau$ according to the rules such as $X_{i}Y_{i} = iZ_{i}$, $Y_{i}Z_{i} = iX_{i},$ $Z_{i}X_{i} = iY_{i}$ and $X_{i}X_{i} = Y_{i}Y_{i} = Z_{i}Z_{i} = I_{i}$, multiply their scalar coefficients, and insert $\sigma\tau \mapsto  z_{A,\sigma}z_{B,\tau}$ into the hash table for storing the result. This procedure is done in $O(1)$ time. We repeat this procedure for all pairs of terms. Since there are $\vert A\vert\vert B\vert$ pairs, the total computation time is $O(\vert A\vert\vert B\vert)$. This computation time is also expressed as $O(\vert AB\vert)$ because the resulting product $AB$ usually contains $\vert AB\vert = O(\vert A\vert\vert B\vert)$ terms.

We can similarly analyze the computation time of other operations, such as subtraction and scalar multiplication. In general, the computation time of any operation is roughly proportional to the number of Pauli operators appearing during the calculation.

\subsection{Traces and commutators 
\label{subsec:apdx_trace_commutator}}

We can efficiently compute traces and commutators by special algorithms that take less time than naive approaches. These algorithms will be crucial for efficiently calculating weighted actions.

The trace $\tr(AB)$ for two operators $A$ and $B$ can be computed in $O(\min\{\vert A\vert,\vert B\vert\})$ time, which is faster than computing the product $AB$. To achieve this performance, we use the fact that $\tr(\sigma) = 0$ for any tensor product of Pauli operators $\sigma \neq  I$, except that $\tr(I) = 2^{N}$. Moreover, the product $\sigma\tau$ becomes the identity operator if and only if $\sigma = \tau$. Therefore, the trace $\tr(AB)$ is given by $\tr(AB) = 2^{N}\sum_{\sigma}z_{A,\sigma}z_{B,\sigma}$, where the sum is over all $\sigma$'s that appear in both $A$ and $B$. If $\vert A\vert \leq \vert B\vert$, we can calculate $2^{-N}\tr(AB)$ by picking up each term $\sigma \mapsto  z_{A,\sigma}$ in $A$, looking up whether $\sigma$ appears in $B$, and if so, adding $z_{A,\sigma}z_{B,\sigma}$ to the result. The single lookup process is done in $O(1)$ time regardless of $\vert B\vert$ using a hash table, and thus, the total computational time is $O(\vert A\vert)$. If $\vert A\vert \geq \vert B\vert$, we can swap the roles of $A$ and $B$ to complete the calculation in $O(\vert B\vert)$ time. Therefore, we can compute $\tr(AB)$ in $O(\min\{\vert A\vert,\vert B\vert\})$ time. 

Below, we use the star symbol $*$ between the two operators to clarify how we calculate the trace of a product of operators. For example, $\tr(A*BC)$ means that we evaluate $\tr(ABC)$ by regarding $ABC$ as the product of $A$ and $BC$. In this case, we must compute the operator $A$ and the product $BC$ before evaluating the trace, but we need not compute the product $ABC$. The computation of the trace thus takes $O(\min\{\vert A\vert,\vert BC\vert\})$ time. Similarly, $\tr(AB*C)$ denotes that we evaluate $\tr(ABC)$ by preparing $AB$ and $C$, and the computation time of the trace is $O(\min\{\vert AB\vert,\vert C\vert\})$.

The commutator $[B, A]$ for two operators $B$ and $A$ also has a fast algorithm. The commutator of two terms is zero if they have no spins in common, e.g., $[X_{1}Y_{2}Z_{3}, X_{4}Y_{5}] = 0$. Therefore, we do not need to iterate through all pairs of terms of $B$ and $A$, but we can skip most of them. This shortcut is achieved by the following procedure. First, we assign integers $1,2,\dots,\vert A\vert$ to the terms of $A$ in an arbitrary order, and we prepare the set of indices $\Theta_{i}(A) = \{l\mid\text{$l$th term of $A$ acts on spin $i$}\}$ for $i = 1,\dots, N$. This is done by iterating through the terms of $A$ once, taking $O(\vert A\vert)$ time. Next, we prepare an empty hash table to store the result of $[B, A]$. We then iterate through each term $\sigma \mapsto z_{B,\sigma}$ of $B$. If, for example, $\sigma$ acts on spins $i = 3,4$, we compute the commutator between $\sigma$ and the terms of $A$ whose indices are in $\Theta_{3}(A)\cup\Theta_{4}(A)$, and we add them to the hash table for storing the result.

This procedure for computing $[B, A]$ runs in $O(N^{-1}\vert A\vert\vert B\vert)$ time under the following assumptions in addition to the $k$-local assumption for all operators. We assume that $A$ consists of $O(N)$ or more terms, and the terms are spread evenly over an $O(N)$ number of spins so that each spin is acted by at most $O(N^{-1}\vert A\vert)$ terms. Thus, the size of the index set $\vert\Theta_{i}(A)\vert$ is at most $O(N^{-1}\vert A\vert)$. For every term in $B$, we compute the commutator with the union of at most $k$ sets out of $\Theta_{1}(A),\dots,\Theta_{N}(A)$ because $B$ is a $k$-local operator. The union contains at most $O(N^{-1}\vert A\vert)$ terms, where we neglect $k$ because $k$ is assumed to be small. Thus, the iteration through $B$ takes $O(N^{-1}\vert A\vert\vert B\vert)$ time in total. This analysis also shows that the number of terms in the resulting commutator $\vert[B,A]\vert$ is at most $O(N^{-1}\vert A\vert\vert B\vert)$.

By extending this algorithm, we can also efficiently compute the commutators $[B,A_{\mu}]$ between an operator $B$ and a series of operators $A_{1},A_{2},\dots$ simultaneously. To do so, we prepare the set $\Theta_{i}(\{A_\mu\}) = \{(\mu,l)\mid\text{$l$th term of $A_{\mu}$ acts on spin $i$}\}$, and we compute the commutators $[B,A_{1}],[B,A_{2}],\ldots$ simultaneously by iterating through the terms of $B$ once, similarly to the above procedure. The computation time is estimated by replacing $\vert A\vert$ with $\sum_{\mu}\vert A_{\mu}\vert$ in the previous paragraph, assuming that $A_{1}, A_{2},\dots$ have $O(N)$ or more terms in total, and the terms spread over $O(N)$ number of spins without concentrating on a few spins. Thus, the total computation takes $O(N^{-1}\vert B\vert\sum_{\mu}\vert A_{\mu}\vert)$ time, and the resulting number of terms $\sum_{\mu}\vert[B,A_{\mu}]\vert$ is at most $O(N^{-1}\vert B\vert\sum_{\mu}\vert A_{\mu}\vert)$.

\subsection{Algorithm and its time complexity for the general framework 
\label{subsec:apdx_algorithm-general}}

We detail the entire algorithm for computing the optimal driving coefficients $\bm{\alpha}^{(\mathcal{P})}(\lambda)$ in the general framework in Sec.~\ref{sec:general-framework} and evaluate its computation time. In computational time analysis, we only count the power of $N$ and neglect prefactors that depend mildly on $K$, since the values of $K$ considered in this paper is small.
We use notation from Appendices~\ref{subsec:apdx_elementary_operation} and \ref{subsec:apdx_trace_commutator}: the symbol $\vert A\vert$ denotes the number of terms in $A$, and we use $\tr(A*B)$ to emphasize that we evaluate the trace by preparing $A$ and $B$ without computing the entire product $AB$. 

We present two algorithms, Algorithm 1 for calculating $\bm{\alpha}^{(\mathcal{P})}(\lambda)$ for a single value of $\lambda$, and Algorithm 2 for calculating $\bm{\alpha}^{(\mathcal{P})}(\lambda)$ over the entire range of $\lambda$. Algorithm 2 assumes that the Hamiltonian has the form of Eq.~\eqref{eq:Hamiltonian_fF} while Algorithm 1 does not require this assumption. 

\textit{Algorithm 1.}---For a fixed $\lambda$, this algorithm takes the algebraic representations of $H(\lambda)$ and $\partial_{\lambda}H(\lambda)$, the algebraic representations of $A_{\mu}$ for $\mu = 1,\dots,M$, and the scalar coefficients $p_{k}(\lambda)$ for $k = 0,\dots,K$ as input. The algorithm proceeds as follows:

(1) Compute the powers of the Hamiltonian, $H(\lambda),H(\lambda)^{2},\ldots,H(\lambda)^{K}$, by iteratively multiplying $H(\lambda)$. 
The number of terms scales as $\vert H(\lambda)^{k}\vert = O(N^{k})$, and therefore, this step takes $O(N^{K})$ time in total.

(2) Compute the time derivative of the powers of the Hamiltonian, $\partial_{\lambda}[H(\lambda)^{2}],\ldots,\partial_{\lambda}[H(\lambda)^{K}]$, by the Leibnitz rule,
\begin{equation}
\partial{}_{\lambda}[H(\lambda)^{k}] = \sum_{l = 0}^{k-1}H(\lambda)^{l}[\partial_{\lambda}H(\lambda)]H(\lambda)^{k-1-l}.
\label{eq_alg:lambda_derivative-general}
\end{equation}
Due to the scaling $\vert H(\lambda)^{l}[\partial_{\lambda}H(\lambda)]H(\lambda)^{k-1-l}\vert = O(N^{k})$, this step takes $O(N^{K})$ time in total.

(3) Compute the fictitious Hamiltonian $\mathcal{P}_{\lambda}(H(\lambda))$ and its derivative $\partial'_{\lambda}\mathcal{P}_{\lambda}(H(\lambda))$,%
\begin{subequations}
\begin{align}
\mathcal{P}_{\lambda}(H(\lambda)) &  = \sum_{k = 0}^{K}p_{k}(\lambda)H(\lambda)^{k},\\
\partial'_{\lambda}\mathcal{P}_{\lambda}(H(\lambda)) &  = \sum_{k = 1}^{K}p_{k}(\lambda)\partial{}_{\lambda}[H(\lambda)^{k}],
\end{align}
\end{subequations}
by summing up the terms. These are done in $O(N^{K})$ time.

(4) Compute the commutators $[\mathcal{P}_\lambda(H(\lambda)),A_{\mu}]$ simultaneously for all $\mu$. Since $\vert\mathcal{P}_\lambda(H(\lambda))\vert = O(N^{K})$ and $\sum_{\mu}\vert A_{\mu}\vert = O(N)$, the computation time is 
\begin{equation}
    O\biggl( N^{-1} \vert\mathcal{P}_\lambda (H(\lambda))\vert \sum_\mu \vert A_\mu \vert \biggr) =O(N^{-1}N^{K}N) = O(N^{K}), 
\end{equation}
and the resulting commutators contain $O(N^{K})$ terms in total.

(5) For each $\mu$, compute the commutators $[[\mathcal{P}_\lambda (H(\lambda)),A_{\mu}],A_{\nu}]$ for $\nu=1,\dots,\mu$ simultaneously. Using $\sum_{\mu}\vert[\mathcal{P}_\lambda (H(\lambda)),A_{\mu}]\vert = O(N^{K})$ and $\sum_{\nu = 1}^\mu \vert A_{\nu}\vert = O(N)$ for large enough $\mu$, the total computation time is 
\begin{align}
    \sum_{\mu}O\biggl(N^{-1}\vert[\mathcal{P}_\lambda (H(\lambda)),A_{\mu}]\vert\sum_{\nu:\,\nu\leq \mu}\vert A_{\nu}\vert\biggr) \qquad 
    \notag \\
    = O(N^{-1}N^{K}N) = O(N^{K}),    
\end{align}
and the total number of resulting terms is $O(N^{K})$.

(6) Compute $r_{\mu}^{(\mathcal{P})}(\lambda)$ for all $\mu$ by
\begin{equation}
r_{\mu}^{(\mathcal{P})}(\lambda) = i\hbar\,\tr\left\{ \partial'_{\lambda}\mathcal{P}_\lambda(H(\lambda))*[\mathcal{P}_\lambda(H(\lambda)),A_{\mu}]\right\} .
\label{eq_alg:r_single-general}
\end{equation}
Since $\vert [\mathcal{P}_\lambda(H(\lambda)),A_{\mu}] \vert$ is less than $\vert \partial'_{\lambda}\mathcal{P}(H(\lambda)) \vert$, the calculation of $r_\mu^{(\mathcal{P})}(\lambda)$ for a single $\mu$ is done in $O\bigl(\vert[\mathcal{P}_\lambda(H(\lambda)),A_{\mu}]\vert\bigr)$ time. Thus, the calculation for all $\mu$ takes time $\sum_{\mu} O\bigl( \vert[\mathcal{P}_\lambda(H(\lambda)),A_{\mu}]\vert\bigr) = O(N^{K})$.

(7) Compute $Q_{\mu\nu}^{(\mathcal{P})}(\lambda)$ for all $(\mu,\nu)$ with $\mu \geq \nu$ by
\begin{equation}
Q_{\mu\nu}^{(\mathcal{P})}(\lambda) = \tr\left\{ \mathcal{P}_\lambda (H(\lambda))*[[\mathcal{P}_\lambda(H(\lambda)),A_{\mu}],A_{\nu}]\right\} ,
\label{eq_alg:Q_single-general}
\end{equation}
where we use $-\tr([A,B]C) = \tr(A[C,B])$ to deduce Eq.~\eqref{eq_alg:Q_single-general} from Eq.~\eqref{eq:def_Qr_Q-general}. 
Since $\vert [[\mathcal{P}_\lambda(H(\lambda)),A_{\mu}],A_{\nu}] \vert$ is less than $\vert  \mathcal{P}_\lambda (H(\lambda)) \vert $, the calculation for a single $(\mu,\nu)$ takes $O(\vert [[\mathcal{P}_\lambda(H(\lambda)),A_{\mu}],A_{\nu}] \vert)$ time. Thus, the calculation for all $(\mu,\nu)$ with $\mu \geq \nu$ takes the time $\sum_{\mu,\nu:\,\mu\geq\nu}O\bigl( \vert[[\mathcal{P}_\lambda(H(\lambda)),A_{\mu}],A_{\nu}]\vert\bigr) = O(N^{K})$. Note that $Q_{\mu\nu}^{(\mathcal{P})}(\lambda)$ for $\mu < \nu$ is known from the symmetry $Q_{\mu\nu}^{(\mathcal{P})}(\lambda) = Q_{\nu\mu}^{(\mathcal{P})}(\lambda)$. 

(8) Solve the $M$-variate linear equation in Eq.~\eqref{eq:linear_equation-general} to determine $\bm{\alpha}^{(\mathcal{P})}(\lambda)$. Output $\bm{\alpha}^{(\mathcal{P})}(\lambda)$. 

In total, the time for the algebraic computation (steps (1)--(7)) scales as $O(N^{K})$. Step (8) takes $O(M^3)$ or $O(RM^2)$ time depending on the numerical algorithm, as discussed in Sec.~\ref{subsec:algorithm_general}.

\textit{Algorithm 2.}---This algorithm assumes that the Hamiltonian has the form of Eq.~\eqref{eq:Hamiltonian_fF}. We use symbols and definitions in Appendix~\ref{subsec:apdx_time-dependent-general}\@. 

The algorithm takes the following objects as input: the algebraic representations of the operators $F_{\gamma}$ for $\gamma = 1,\dots,\Gamma$ and $A_{\mu}$ for $\mu = 1,\dots,M$; the $\lambda$-dependent coefficients $f_{\gamma}(\lambda)$ for $\gamma = 1,\dots,\Gamma$, their $\lambda$-derivatives $\partial_{\lambda}f_{\gamma}(\lambda)$ for $\gamma = 1,\dots,\Gamma$, and the $\lambda$-dependent coefficients of the polynomial $p_{k}(\lambda)$ for $k = 0,\dots,K$, all as functions of $\lambda$ so that the program can evaluate the function values at an arbitrary $\lambda$; a list of $\lambda$'s for which we want to compute the driving coefficients.

The algorithm consists of two stages. In the first stage, it computes the traces of $\lambda$-independent operators by the following procedure:

(1) Compute the operators $\tilde{F}_{g}$ for all $g$ by multiplication and addition of $F_{\gamma}$'s.

(2) For each $g$, compute the commutators $[\tilde{F}_{g},A_{\mu}]$ simultaneously for all $\mu$.

(3) For each $g$ and $\mu$, compute the commutators $[[\tilde{F}_{g},A_{\mu}],A_{\nu}]$ for $\nu=1,\dots,\mu$ simultaneously.

(4) For all $(g,g')$, compute 
\begin{equation}
\tilde{r}_{\mu,gg'}  = \tr\left\{ \tilde{F}_{g}*[\tilde{F}_{g'},A_{\mu}]\right\}
\label{eq_alg:r_entire-general}
\end{equation}
for all $\mu$, and compute
\begin{equation}
\tilde{Q}_{\mu\nu,gg'} = -\tr\left\{ \tilde{F}_{g'}*[[\tilde{F}_{g},A_{\mu}],A_{\nu}]\right\} 
\label{eq_alg:Q_entire-general}
\end{equation}
for all $(\mu,\nu)$ such that $\mu \geq \nu$. 

In the second stage, the algorithm repeats the following calculation for each $\lambda$ in the list of $\lambda$'s:

(5) Compute the scalar coefficient $\tilde{f}_{g}(\lambda)$ for all $g$ by multiplying $f_{\gamma}(\lambda)$'s, and compute $\partial_{\lambda}\tilde{f}_{g}(\lambda)$ for all $g$ using the Leibniz rule.

(6) Compute $r_{\mu}^{(\mathcal{P})}(\lambda)$ for all $\mu$ and $Q_{\mu\nu}^{(\mathcal{P})}(\lambda)$ for all $(\mu,\nu)$ with $\mu \geq \nu$ by Eq.~\eqref{eq:time-dependent-Qr_expansion}. Note that $Q_{\mu\nu}^{(\mathcal{P})}(\lambda)$ for $\mu < \nu$ is known from the symmetry $Q_{\mu\nu}^{(\mathcal{P})}(\lambda) = Q_{\nu\mu}^{(\mathcal{P})}(\lambda)$.

(7) Solve the linear equation in Eq.~\eqref{eq:linear_equation-general} to determine $\bm{\alpha}^{(\mathcal{P})}(\lambda)$. Output $\bm{\alpha}^{(\mathcal{P})}(\lambda)$.

The computation complexity for the first stage, which deals with algebraic computation, scales as $O(N^K)$. This is understood by analyzing the computation time for each step similarly to the corresponding step in Algorithm 1. Concretely, step (1) in Algorithm 2 corresponds to step (1) in Algorithm 1, (2) corresponds to (4), (3) corresponds to (5), and (4) corresponds to (6) and (7).

The computation cost of the second stage, which involves only scalar computation, is dominated by step (7), requiring $O(M^3)$ or $O(RM^2)$ time depending on the numerical algorithm. Step (5) takes $O(M^0)$ time, and step (6) takes $O(M^2)$ time, which are faster than step (7).

\subsection{Algorithm for ground-state evolution 
\label{subsec:apdx_algorithm-GS}}

This subsection presents entire algorithms specialized for ground-state evolution in Sec.~\ref{sec:ground_state}\@. These algorithms combine the general algorithms in Appendix~\ref{subsec:apdx_algorithm-general} with the minimization of $\Omega_{\lambda}^{(K)}(E)$ discussed in Appendices~\ref{subsec:apdx_Omega} and \ref{subsec:apdx_time-dependent-GS}. The general algorithms are slightly simplified by taking advantage of the simple form of the polynomial, $\mathcal{P}^{\mathrm{GS}, K}(H(\lambda)) = [H(\lambda)-E_{\lambda}^{(K)}]^{K}$. As in Appendix~\ref{subsec:apdx_algorithm-general}, we present two algorithms, Algorithm 3 for a single value of $\lambda$ and Algorithm 4 for an entire range of $\lambda$. Our C++ code~\cite{GitHub} implements Algorithm 3 and Algorithm 4.

\textit{Algorithm 3.}---For a fixed $\lambda$, this algorithm takes the algebraic representations of $H(\lambda)$, $\partial_{\lambda}H(\lambda)$, and $A_{\mu}$ for $\mu = 1,\dots,M$ as input. The algorithm proceeds as follows:

(1) Compute the powers $H(\lambda),H(\lambda)^{2},\dots,H(\lambda)^{K}$ by iteratively multiplying $H(\lambda)$. 

(2) Compute $\omega_{k}(\lambda) = \tr[H(\lambda)^{k}*I]$ for $k = 1,\dots,K$ and $\omega_{k}(\lambda) = \tr[H(\lambda)^{K}*H(\lambda)^{k-K}]$ for $k = K+1,\dots,2K-1$. 

(3) Minimize the function $\Omega_{\lambda}^{(K)}(E)$ to determine $E_{\lambda}^{(K)}$. The details of the minimization are discussed in Appendix~\ref{subsec:apdx_Omega}\@.

(4) Compute $[H(\lambda) - E_\lambda^{(K)}]^{k}$ for $k=2,\dots,K$ by
\begin{equation}
[H(\lambda) - E_\lambda^{(K)}]^{k} = \sum_{l = 0}^{k}{\textstyle\binom{k}{l}}H(\lambda)^l(-E_{\lambda}^{(K)})^{k-l}.
\end{equation}

(5) Compute $\partial'_{\lambda}\bigl\{[H(\lambda) - E_\lambda^{(K)}]^{K}\bigr\}$ by 
\begin{align}
&\partial'_{\lambda}\bigl\{[H(\lambda) - E_\lambda^{(K)}]^{K}\bigr\} 
\nonumber \\
&= \sum_{l = 0}^{K-1} [H(\lambda) - E_\lambda^{(K)}]^{l}[\partial_{\lambda}H(\lambda)] [H(\lambda) - E_\lambda^{(K)}]^{K-1-l}.
\label{eq_alg:lambda_derivative}
\end{align}

(6) Compute the commutators $[[H(\lambda) - E_\lambda^{(K)}]^{K},A_{\mu}]$ simultaneously for all $\mu$.

(7) For each $\mu$, compute the commutators $[[[H(\lambda) - E_\lambda^{(K)}]^{K},A_{\mu}],A_{\nu}]$ simultaneously for all $\nu=1,\dots,\mu$.

(8) Compute $r_{\mu}^{(K)}(\lambda)$ for all $\mu$ by Eq.~\eqref{eq_alg:r_single-general}. 

(9) Compute $Q_{\mu\nu}^{(K)}(\lambda)$ for all $(\mu,\nu)$ with $\mu \geq \nu$ by Eq.~\eqref{eq_alg:Q_single-general}. 

(10) Solve the linear equation in Eq.~\eqref{eq:linear_equation-general} to determine $\bm{\alpha}^{(K)}(\lambda)$. Output $\bm{\alpha}^{(K)}(\lambda)$.

The total computation time for Algorithm 3 scales as $O(N^{K})$ for the algebraic computation (steps (1)--(9)) and $O(M^3)$ or $O(RM^2)$ for the scalar computation (step (10)), similarly to Algorithm 1. Concretely, the computation time for steps (1) and (4)--(10) are the same as for the corresponding steps in Algorithm 1. The computation for step (2) takes $O(N^{K-1})$ time, and step (3) takes $O(N^{0})$ time, making them negligible compared to the rest of the algorithm.

\textit{Algorithm 4.}---This algorithm assumes that the Hamiltonian has the form of Eq.~\eqref{eq:Hamiltonian_fF}. We use symbols and definitions in Appendices~\ref{subsec:apdx_time-dependent-general}
and \ref{subsec:apdx_time-dependent-GS}\@. 

The algorithm takes the following things as the input: algebraic representations of the operators $F_{\gamma}$ for $\gamma = 1,\dots,\Gamma$ and $A_{\mu}$ for $\mu = 1,\dots,M$; the $\lambda$-dependent coefficients $f_{\gamma}(\lambda)$ for $\gamma = 1,\dots,\Gamma$ and their $\lambda$-derivatives $\partial_{\lambda}f_{\gamma}(\lambda)$ for $\gamma = 1,\dots,\Gamma$, both as functions so that the program can evaluate the function values at an arbitrary $\lambda$; a list of $\lambda$'s for which we want to compute the driving coefficients.

The algorithm consists of two stages. In the first stage, it computes the traces of $\lambda$-independent operators:

(1) Compute the operators $\tilde{F}_{g}$ for all $g$ by multiplication and addition of $F_{\gamma}$'s.

(2) Compute the traces $\tilde{\omega}_{g} = \tr(\tilde{F}_{g}*I)$ for all $g$ and $\tilde{\omega}_{gg'} = \tr(\tilde{F}_{g}*\tilde{F}_{g'})$ for all $(g,g')$.

(3) For every $g$, compute the commutators $[\tilde{F}_{g},A_{\mu}]$ simultaneously for all $\mu$.

(4) For every $g$ and every $\mu$, compute the commutators $[[\tilde{F}_{g},A_{\mu}],A_{\nu}]$ simultaneously for all $\nu = 1,\dots,\mu$.

(5) For all $(g,g')$, compute $\tilde{r}_{\mu,gg'}$ by Eq.~\eqref{eq_alg:r_entire-general} for all $\mu$, and compute $\tilde{Q}_{\mu\nu,gg'}$ by Eq.~\eqref{eq_alg:Q_entire-general} for all $(\mu,\nu)$ such that $\mu \geq \nu$.

In the second stage, the algorithm repeats the following calculation for each $\lambda$ in the list of $\lambda$'s:

(6) Compute the scalar coefficient $\tilde{f}_{g}(\lambda)$ for all $g$ by multiplying $f_{\gamma}(\lambda)$'s, and compute $\partial_{\lambda}\tilde{f}_{g}(\lambda)$ for all $g$ using the Leibniz rule.

(7) Compute $\omega_{k}(\lambda)$ for $k = 1,\dots,2K-1$ by Eqs.~\eqref{eq_min:omega_former} and \eqref{eq_min:omega_latter}. 

(8) Minimize the function $\Omega_{\lambda}^{(K)}(E)$ to determine $E_{\lambda}^{(K)}$. The details of the minimization are discussed in Appendix~\ref{subsec:apdx_Omega}\@.

(9) Compute the scalar coefficients $p_{k}(\lambda) = \binom{K}{k}(-E_{\lambda}^{(K)})^{K-k}$ for $k = 0,\dots,K$.

(10) Compute $r_{\mu}^{(K)}(\lambda)$ for all $\mu$ and $Q_{\mu\nu}^{(K)}(\lambda)$ for all $\mu \geq \nu$ by Eq.~\eqref{eq:time-dependent-Qr_expansion}. 

(11) Solve the linear equation in Eq.~\eqref{eq:linear_equation-general} to determine $\bm{\alpha}^{(K)}(\lambda)$. Output $\bm{\alpha}^{(K)}(\lambda)$.

The computation time of Algorithm 4 scales as $O(N^{K})$ for the first stage and $O(M^3)$ or $O(RM^2)$ for the second stage, similarly to the above three algorithms.


\begin{thebibliography}{91}%
\makeatletter
\providecommand \@ifxundefined [1]{%
 \@ifx{#1\undefined}
}%
\providecommand \@ifnum [1]{%
 \ifnum #1\expandafter \@firstoftwo
 \else \expandafter \@secondoftwo
 \fi
}%
\providecommand \@ifx [1]{%
 \ifx #1\expandafter \@firstoftwo
 \else \expandafter \@secondoftwo
 \fi
}%
\providecommand \natexlab [1]{#1}%
\providecommand \enquote  [1]{``#1''}%
\providecommand \bibnamefont  [1]{#1}%
\providecommand \bibfnamefont [1]{#1}%
\providecommand \citenamefont [1]{#1}%
\providecommand \href@noop [0]{\@secondoftwo}%
\providecommand \href [0]{\begingroup \@sanitize@url \@href}%
\providecommand \@href[1]{\@@startlink{#1}\@@href}%
\providecommand \@@href[1]{\endgroup#1\@@endlink}%
\providecommand \@sanitize@url [0]{\catcode `\\12\catcode `\$12\catcode `\&12\catcode `\#12\catcode `\^12\catcode `\_12\catcode `\%12\relax}%
\providecommand \@@startlink[1]{}%
\providecommand \@@endlink[0]{}%
\providecommand \url  [0]{\begingroup\@sanitize@url \@url }%
\providecommand \@url [1]{\endgroup\@href {#1}{\urlprefix }}%
\providecommand \urlprefix  [0]{URL }%
\providecommand \Eprint [0]{\href }%
\providecommand \doibase [0]{https://doi.org/}%
\providecommand \selectlanguage [0]{\@gobble}%
\providecommand \bibinfo  [0]{\@secondoftwo}%
\providecommand \bibfield  [0]{\@secondoftwo}%
\providecommand \translation [1]{[#1]}%
\providecommand \BibitemOpen [0]{}%
\providecommand \bibitemStop [0]{}%
\providecommand \bibitemNoStop [0]{.\EOS\space}%
\providecommand \EOS [0]{\spacefactor3000\relax}%
\providecommand \BibitemShut  [1]{\csname bibitem#1\endcsname}%
\let\auto@bib@innerbib\@empty
\bibitem [{\citenamefont {Acín}\ \emph {et~al.}(2018)\citenamefont {Acín}, \citenamefont {Bloch}, \citenamefont {Buhrman}, \citenamefont {Calarco}, \citenamefont {Eichler}, \citenamefont {Eisert}, \citenamefont {Esteve}, \citenamefont {Gisin}, \citenamefont {Glaser}, \citenamefont {Jelezko}, \citenamefont {Kuhr}, \citenamefont {Lewenstein}, \citenamefont {Riedel}, \citenamefont {Schmidt}, \citenamefont {Thew}, \citenamefont {Wallraff}, \citenamefont {Walmsley},\ and\ \citenamefont {Wilhelm}}]{Acin2018TheQuantumTechnologies}%
  \BibitemOpen
  \bibfield  {author} {\bibinfo {author} {\bibfnamefont {A.}~\bibnamefont {Acín}}, \bibinfo {author} {\bibfnamefont {I.}~\bibnamefont {Bloch}}, \bibinfo {author} {\bibfnamefont {H.}~\bibnamefont {Buhrman}}, \bibinfo {author} {\bibfnamefont {T.}~\bibnamefont {Calarco}}, \bibinfo {author} {\bibfnamefont {C.}~\bibnamefont {Eichler}}, \bibinfo {author} {\bibfnamefont {J.}~\bibnamefont {Eisert}}, \bibinfo {author} {\bibfnamefont {D.}~\bibnamefont {Esteve}}, \bibinfo {author} {\bibfnamefont {N.}~\bibnamefont {Gisin}}, \bibinfo {author} {\bibfnamefont {S.~J.}\ \bibnamefont {Glaser}}, \bibinfo {author} {\bibfnamefont {F.}~\bibnamefont {Jelezko}}, \bibinfo {author} {\bibfnamefont {S.}~\bibnamefont {Kuhr}}, \bibinfo {author} {\bibfnamefont {M.}~\bibnamefont {Lewenstein}}, \bibinfo {author} {\bibfnamefont {M.~F.}\ \bibnamefont {Riedel}}, \bibinfo {author} {\bibfnamefont {P.~O.}\ \bibnamefont {Schmidt}}, \bibinfo {author} {\bibfnamefont {R.}~\bibnamefont {Thew}}, \bibinfo {author} {\bibfnamefont {A.}~\bibnamefont
  {Wallraff}}, \bibinfo {author} {\bibfnamefont {I.}~\bibnamefont {Walmsley}},\ and\ \bibinfo {author} {\bibfnamefont {F.~K.}\ \bibnamefont {Wilhelm}},\ }\bibfield  {title} {\bibinfo {title} {The quantum technologies roadmap: a {E}uropean community view},\ }\href {https://doi.org/10.1088/1367-2630/aad1ea} {\bibfield  {journal} {\bibinfo  {journal} {New J. Phys.}\ }\textbf {\bibinfo {volume} {20}},\ \bibinfo {pages} {080201} (\bibinfo {year} {2018})}\BibitemShut {NoStop}%
\bibitem [{\citenamefont {Glaser}\ \emph {et~al.}(2015)\citenamefont {Glaser}, \citenamefont {Boscain}, \citenamefont {Calarco}, \citenamefont {Koch}, \citenamefont {Köckenberger}, \citenamefont {Kosloff}, \citenamefont {Kuprov}, \citenamefont {Luy}, \citenamefont {Schirmer}, \citenamefont {Schulte-Herbrüggen}, \citenamefont {Sugny},\ and\ \citenamefont {Wilhelm}}]{Glaser2015TrainingSchrodingers}%
  \BibitemOpen
  \bibfield  {author} {\bibinfo {author} {\bibfnamefont {S.~J.}\ \bibnamefont {Glaser}}, \bibinfo {author} {\bibfnamefont {U.}~\bibnamefont {Boscain}}, \bibinfo {author} {\bibfnamefont {T.}~\bibnamefont {Calarco}}, \bibinfo {author} {\bibfnamefont {C.~P.}\ \bibnamefont {Koch}}, \bibinfo {author} {\bibfnamefont {W.}~\bibnamefont {Köckenberger}}, \bibinfo {author} {\bibfnamefont {R.}~\bibnamefont {Kosloff}}, \bibinfo {author} {\bibfnamefont {I.}~\bibnamefont {Kuprov}}, \bibinfo {author} {\bibfnamefont {B.}~\bibnamefont {Luy}}, \bibinfo {author} {\bibfnamefont {S.}~\bibnamefont {Schirmer}}, \bibinfo {author} {\bibfnamefont {T.}~\bibnamefont {Schulte-Herbrüggen}}, \bibinfo {author} {\bibfnamefont {D.}~\bibnamefont {Sugny}},\ and\ \bibinfo {author} {\bibfnamefont {F.~K.}\ \bibnamefont {Wilhelm}},\ }\bibfield  {title} {\bibinfo {title} {Training {S}chrödinger's cat: quantum optimal control},\ }\href {https://doi.org/10.1140/epjd/e2015-60464-1} {\bibfield  {journal} {\bibinfo  {journal} {Eur. Phys. J. D}\
  }\textbf {\bibinfo {volume} {69}},\ \bibinfo {pages} {279} (\bibinfo {year} {2015})}\BibitemShut {NoStop}%
\bibitem [{\citenamefont {Arute}\ \emph {et~al.}(2019)\citenamefont {Arute}, \citenamefont {Arya}, \citenamefont {Babbush}, \citenamefont {Bacon}, \citenamefont {Bardin}, \citenamefont {Barends}, \citenamefont {Biswas}, \citenamefont {Boixo}, \citenamefont {Brandao}, \citenamefont {Buell}, \citenamefont {Burkett}, \citenamefont {Chen}, \citenamefont {Chen}, \citenamefont {Chiaro}, \citenamefont {Collins}, \citenamefont {Courtney}, \citenamefont {Dunsworth}, \citenamefont {Farhi}, \citenamefont {Foxen}, \citenamefont {Fowler}, \citenamefont {Gidney}, \citenamefont {Giustina}, \citenamefont {Graff}, \citenamefont {Guerin}, \citenamefont {Habegger}, \citenamefont {Harrigan}, \citenamefont {Hartmann}, \citenamefont {Ho}, \citenamefont {Hoffmann}, \citenamefont {Huang}, \citenamefont {Humble}, \citenamefont {Isakov}, \citenamefont {Jeffrey}, \citenamefont {Jiang}, \citenamefont {Kafri}, \citenamefont {Kechedzhi}, \citenamefont {Kelly}, \citenamefont {Klimov}, \citenamefont {Knysh}, \citenamefont {Korotkov},
  \citenamefont {Kostritsa}, \citenamefont {Landhuis}, \citenamefont {Lindmark}, \citenamefont {Lucero}, \citenamefont {Lyakh}, \citenamefont {Mandrà}, \citenamefont {McClean}, \citenamefont {McEwen}, \citenamefont {Megrant}, \citenamefont {Mi}, \citenamefont {Michielsen}, \citenamefont {Mohseni}, \citenamefont {Mutus}, \citenamefont {Naaman}, \citenamefont {Neeley}, \citenamefont {Neill}, \citenamefont {Niu}, \citenamefont {Ostby}, \citenamefont {Petukhov}, \citenamefont {Platt}, \citenamefont {Quintana}, \citenamefont {Rieffel}, \citenamefont {Roushan}, \citenamefont {Rubin}, \citenamefont {Sank}, \citenamefont {Satzinger}, \citenamefont {Smelyanskiy}, \citenamefont {Sung}, \citenamefont {Trevithick}, \citenamefont {Vainsencher}, \citenamefont {Villalonga}, \citenamefont {White}, \citenamefont {Yao}, \citenamefont {Yeh}, \citenamefont {Zalcman}, \citenamefont {Neven},\ and\ \citenamefont {Martinis}}]{Arute2019QuantumSupremacy}%
  \BibitemOpen
  \bibfield  {author} {\bibinfo {author} {\bibfnamefont {F.}~\bibnamefont {Arute}}, \bibinfo {author} {\bibfnamefont {K.}~\bibnamefont {Arya}}, \bibinfo {author} {\bibfnamefont {R.}~\bibnamefont {Babbush}}, \bibinfo {author} {\bibfnamefont {D.}~\bibnamefont {Bacon}}, \bibinfo {author} {\bibfnamefont {J.~C.}\ \bibnamefont {Bardin}}, \bibinfo {author} {\bibfnamefont {R.}~\bibnamefont {Barends}}, \bibinfo {author} {\bibfnamefont {R.}~\bibnamefont {Biswas}}, \bibinfo {author} {\bibfnamefont {S.}~\bibnamefont {Boixo}}, \bibinfo {author} {\bibfnamefont {F.~G. S.~L.}\ \bibnamefont {Brandao}}, \bibinfo {author} {\bibfnamefont {D.~A.}\ \bibnamefont {Buell}}, \bibinfo {author} {\bibfnamefont {B.}~\bibnamefont {Burkett}}, \bibinfo {author} {\bibfnamefont {Y.}~\bibnamefont {Chen}}, \bibinfo {author} {\bibfnamefont {Z.}~\bibnamefont {Chen}}, \bibinfo {author} {\bibfnamefont {B.}~\bibnamefont {Chiaro}}, \bibinfo {author} {\bibfnamefont {R.}~\bibnamefont {Collins}}, \bibinfo {author} {\bibfnamefont {W.}~\bibnamefont
  {Courtney}}, \bibinfo {author} {\bibfnamefont {A.}~\bibnamefont {Dunsworth}}, \bibinfo {author} {\bibfnamefont {E.}~\bibnamefont {Farhi}}, \bibinfo {author} {\bibfnamefont {B.}~\bibnamefont {Foxen}}, \bibinfo {author} {\bibfnamefont {A.}~\bibnamefont {Fowler}}, \bibinfo {author} {\bibfnamefont {C.}~\bibnamefont {Gidney}}, \bibinfo {author} {\bibfnamefont {M.}~\bibnamefont {Giustina}}, \bibinfo {author} {\bibfnamefont {R.}~\bibnamefont {Graff}}, \bibinfo {author} {\bibfnamefont {K.}~\bibnamefont {Guerin}}, \bibinfo {author} {\bibfnamefont {S.}~\bibnamefont {Habegger}}, \bibinfo {author} {\bibfnamefont {M.~P.}\ \bibnamefont {Harrigan}}, \bibinfo {author} {\bibfnamefont {M.~J.}\ \bibnamefont {Hartmann}}, \bibinfo {author} {\bibfnamefont {A.}~\bibnamefont {Ho}}, \bibinfo {author} {\bibfnamefont {M.}~\bibnamefont {Hoffmann}}, \bibinfo {author} {\bibfnamefont {T.}~\bibnamefont {Huang}}, \bibinfo {author} {\bibfnamefont {T.~S.}\ \bibnamefont {Humble}}, \bibinfo {author} {\bibfnamefont {S.~V.}\ \bibnamefont
  {Isakov}}, \bibinfo {author} {\bibfnamefont {E.}~\bibnamefont {Jeffrey}}, \bibinfo {author} {\bibfnamefont {Z.}~\bibnamefont {Jiang}}, \bibinfo {author} {\bibfnamefont {D.}~\bibnamefont {Kafri}}, \bibinfo {author} {\bibfnamefont {K.}~\bibnamefont {Kechedzhi}}, \bibinfo {author} {\bibfnamefont {J.}~\bibnamefont {Kelly}}, \bibinfo {author} {\bibfnamefont {P.~V.}\ \bibnamefont {Klimov}}, \bibinfo {author} {\bibfnamefont {S.}~\bibnamefont {Knysh}}, \bibinfo {author} {\bibfnamefont {A.}~\bibnamefont {Korotkov}}, \bibinfo {author} {\bibfnamefont {F.}~\bibnamefont {Kostritsa}}, \bibinfo {author} {\bibfnamefont {D.}~\bibnamefont {Landhuis}}, \bibinfo {author} {\bibfnamefont {M.}~\bibnamefont {Lindmark}}, \bibinfo {author} {\bibfnamefont {E.}~\bibnamefont {Lucero}}, \bibinfo {author} {\bibfnamefont {D.}~\bibnamefont {Lyakh}}, \bibinfo {author} {\bibfnamefont {S.}~\bibnamefont {Mandrà}}, \bibinfo {author} {\bibfnamefont {J.~R.}\ \bibnamefont {McClean}}, \bibinfo {author} {\bibfnamefont {M.}~\bibnamefont {McEwen}},
  \bibinfo {author} {\bibfnamefont {A.}~\bibnamefont {Megrant}}, \bibinfo {author} {\bibfnamefont {X.}~\bibnamefont {Mi}}, \bibinfo {author} {\bibfnamefont {K.}~\bibnamefont {Michielsen}}, \bibinfo {author} {\bibfnamefont {M.}~\bibnamefont {Mohseni}}, \bibinfo {author} {\bibfnamefont {J.}~\bibnamefont {Mutus}}, \bibinfo {author} {\bibfnamefont {O.}~\bibnamefont {Naaman}}, \bibinfo {author} {\bibfnamefont {M.}~\bibnamefont {Neeley}}, \bibinfo {author} {\bibfnamefont {C.}~\bibnamefont {Neill}}, \bibinfo {author} {\bibfnamefont {M.~Y.}\ \bibnamefont {Niu}}, \bibinfo {author} {\bibfnamefont {E.}~\bibnamefont {Ostby}}, \bibinfo {author} {\bibfnamefont {A.}~\bibnamefont {Petukhov}}, \bibinfo {author} {\bibfnamefont {J.~C.}\ \bibnamefont {Platt}}, \bibinfo {author} {\bibfnamefont {C.}~\bibnamefont {Quintana}}, \bibinfo {author} {\bibfnamefont {E.~G.}\ \bibnamefont {Rieffel}}, \bibinfo {author} {\bibfnamefont {P.}~\bibnamefont {Roushan}}, \bibinfo {author} {\bibfnamefont {N.~C.}\ \bibnamefont {Rubin}}, \bibinfo
  {author} {\bibfnamefont {D.}~\bibnamefont {Sank}}, \bibinfo {author} {\bibfnamefont {K.~J.}\ \bibnamefont {Satzinger}}, \bibinfo {author} {\bibfnamefont {V.}~\bibnamefont {Smelyanskiy}}, \bibinfo {author} {\bibfnamefont {K.~J.}\ \bibnamefont {Sung}}, \bibinfo {author} {\bibfnamefont {M.~D.}\ \bibnamefont {Trevithick}}, \bibinfo {author} {\bibfnamefont {A.}~\bibnamefont {Vainsencher}}, \bibinfo {author} {\bibfnamefont {B.}~\bibnamefont {Villalonga}}, \bibinfo {author} {\bibfnamefont {T.}~\bibnamefont {White}}, \bibinfo {author} {\bibfnamefont {Z.~J.}\ \bibnamefont {Yao}}, \bibinfo {author} {\bibfnamefont {P.}~\bibnamefont {Yeh}}, \bibinfo {author} {\bibfnamefont {A.}~\bibnamefont {Zalcman}}, \bibinfo {author} {\bibfnamefont {H.}~\bibnamefont {Neven}},\ and\ \bibinfo {author} {\bibfnamefont {J.~M.}\ \bibnamefont {Martinis}},\ }\bibfield  {title} {\bibinfo {title} {Quantum supremacy using a programmable superconducting processor},\ }\href {https://doi.org/10.1038/s41586-019-1666-5} {\bibfield  {journal}
  {\bibinfo  {journal} {Nature}\ }\textbf {\bibinfo {volume} {574}},\ \bibinfo {pages} {505} (\bibinfo {year} {2019})}\BibitemShut {NoStop}%
\bibitem [{\citenamefont {Wu}\ \emph {et~al.}(2021)\citenamefont {Wu}, \citenamefont {Bao}, \citenamefont {Cao}, \citenamefont {Chen}, \citenamefont {Chen}, \citenamefont {Chen}, \citenamefont {Chung}, \citenamefont {Deng}, \citenamefont {Du}, \citenamefont {Fan}, \citenamefont {Gong}, \citenamefont {Guo}, \citenamefont {Guo}, \citenamefont {Guo}, \citenamefont {Han}, \citenamefont {Hong}, \citenamefont {Huang}, \citenamefont {Huo}, \citenamefont {Li}, \citenamefont {Li}, \citenamefont {Li}, \citenamefont {Li}, \citenamefont {Liang}, \citenamefont {Lin}, \citenamefont {Lin}, \citenamefont {Qian}, \citenamefont {Qiao}, \citenamefont {Rong}, \citenamefont {Su}, \citenamefont {Sun}, \citenamefont {Wang}, \citenamefont {Wang}, \citenamefont {Wu}, \citenamefont {Xu}, \citenamefont {Yan}, \citenamefont {Yang}, \citenamefont {Yang}, \citenamefont {Ye}, \citenamefont {Yin}, \citenamefont {Ying}, \citenamefont {Yu}, \citenamefont {Zha}, \citenamefont {Zhang}, \citenamefont {Zhang}, \citenamefont {Zhang}, \citenamefont
  {Zhang}, \citenamefont {Zhao}, \citenamefont {Zhao}, \citenamefont {Zhou}, \citenamefont {Zhu}, \citenamefont {Lu}, \citenamefont {Peng}, \citenamefont {Zhu},\ and\ \citenamefont {Pan}}]{Wu2021StrongQuantum}%
  \BibitemOpen
  \bibfield  {author} {\bibinfo {author} {\bibfnamefont {Y.}~\bibnamefont {Wu}}, \bibinfo {author} {\bibfnamefont {W.-S.}\ \bibnamefont {Bao}}, \bibinfo {author} {\bibfnamefont {S.}~\bibnamefont {Cao}}, \bibinfo {author} {\bibfnamefont {F.}~\bibnamefont {Chen}}, \bibinfo {author} {\bibfnamefont {M.-C.}\ \bibnamefont {Chen}}, \bibinfo {author} {\bibfnamefont {X.}~\bibnamefont {Chen}}, \bibinfo {author} {\bibfnamefont {T.-H.}\ \bibnamefont {Chung}}, \bibinfo {author} {\bibfnamefont {H.}~\bibnamefont {Deng}}, \bibinfo {author} {\bibfnamefont {Y.}~\bibnamefont {Du}}, \bibinfo {author} {\bibfnamefont {D.}~\bibnamefont {Fan}}, \bibinfo {author} {\bibfnamefont {M.}~\bibnamefont {Gong}}, \bibinfo {author} {\bibfnamefont {C.}~\bibnamefont {Guo}}, \bibinfo {author} {\bibfnamefont {C.}~\bibnamefont {Guo}}, \bibinfo {author} {\bibfnamefont {S.}~\bibnamefont {Guo}}, \bibinfo {author} {\bibfnamefont {L.}~\bibnamefont {Han}}, \bibinfo {author} {\bibfnamefont {L.}~\bibnamefont {Hong}}, \bibinfo {author} {\bibfnamefont
  {H.-L.}\ \bibnamefont {Huang}}, \bibinfo {author} {\bibfnamefont {Y.-H.}\ \bibnamefont {Huo}}, \bibinfo {author} {\bibfnamefont {L.}~\bibnamefont {Li}}, \bibinfo {author} {\bibfnamefont {N.}~\bibnamefont {Li}}, \bibinfo {author} {\bibfnamefont {S.}~\bibnamefont {Li}}, \bibinfo {author} {\bibfnamefont {Y.}~\bibnamefont {Li}}, \bibinfo {author} {\bibfnamefont {F.}~\bibnamefont {Liang}}, \bibinfo {author} {\bibfnamefont {C.}~\bibnamefont {Lin}}, \bibinfo {author} {\bibfnamefont {J.}~\bibnamefont {Lin}}, \bibinfo {author} {\bibfnamefont {H.}~\bibnamefont {Qian}}, \bibinfo {author} {\bibfnamefont {D.}~\bibnamefont {Qiao}}, \bibinfo {author} {\bibfnamefont {H.}~\bibnamefont {Rong}}, \bibinfo {author} {\bibfnamefont {H.}~\bibnamefont {Su}}, \bibinfo {author} {\bibfnamefont {L.}~\bibnamefont {Sun}}, \bibinfo {author} {\bibfnamefont {L.}~\bibnamefont {Wang}}, \bibinfo {author} {\bibfnamefont {S.}~\bibnamefont {Wang}}, \bibinfo {author} {\bibfnamefont {D.}~\bibnamefont {Wu}}, \bibinfo {author} {\bibfnamefont
  {Y.}~\bibnamefont {Xu}}, \bibinfo {author} {\bibfnamefont {K.}~\bibnamefont {Yan}}, \bibinfo {author} {\bibfnamefont {W.}~\bibnamefont {Yang}}, \bibinfo {author} {\bibfnamefont {Y.}~\bibnamefont {Yang}}, \bibinfo {author} {\bibfnamefont {Y.}~\bibnamefont {Ye}}, \bibinfo {author} {\bibfnamefont {J.}~\bibnamefont {Yin}}, \bibinfo {author} {\bibfnamefont {C.}~\bibnamefont {Ying}}, \bibinfo {author} {\bibfnamefont {J.}~\bibnamefont {Yu}}, \bibinfo {author} {\bibfnamefont {C.}~\bibnamefont {Zha}}, \bibinfo {author} {\bibfnamefont {C.}~\bibnamefont {Zhang}}, \bibinfo {author} {\bibfnamefont {H.}~\bibnamefont {Zhang}}, \bibinfo {author} {\bibfnamefont {K.}~\bibnamefont {Zhang}}, \bibinfo {author} {\bibfnamefont {Y.}~\bibnamefont {Zhang}}, \bibinfo {author} {\bibfnamefont {H.}~\bibnamefont {Zhao}}, \bibinfo {author} {\bibfnamefont {Y.}~\bibnamefont {Zhao}}, \bibinfo {author} {\bibfnamefont {L.}~\bibnamefont {Zhou}}, \bibinfo {author} {\bibfnamefont {Q.}~\bibnamefont {Zhu}}, \bibinfo {author} {\bibfnamefont {C.-Y.}\
  \bibnamefont {Lu}}, \bibinfo {author} {\bibfnamefont {C.-Z.}\ \bibnamefont {Peng}}, \bibinfo {author} {\bibfnamefont {X.}~\bibnamefont {Zhu}},\ and\ \bibinfo {author} {\bibfnamefont {J.-W.}\ \bibnamefont {Pan}},\ }\bibfield  {title} {\bibinfo {title} {Strong quantum computational advantage using a superconducting quantum processor},\ }\href {https://doi.org/10.1103/physrevlett.127.180501} {\bibfield  {journal} {\bibinfo  {journal} {Phys. Rev. Lett.}\ }\textbf {\bibinfo {volume} {127}},\ \bibinfo {pages} {180501} (\bibinfo {year} {2021})}\BibitemShut {NoStop}%
\bibitem [{\citenamefont {Bluvstein}\ \emph {et~al.}(2023)\citenamefont {Bluvstein}, \citenamefont {Evered}, \citenamefont {Geim}, \citenamefont {Li}, \citenamefont {Zhou}, \citenamefont {Manovitz}, \citenamefont {Ebadi}, \citenamefont {Cain}, \citenamefont {Kalinowski}, \citenamefont {Hangleiter}, \citenamefont {Bonilla~Ataides}, \citenamefont {Maskara}, \citenamefont {Cong}, \citenamefont {Gao}, \citenamefont {Sales~Rodriguez}, \citenamefont {Karolyshyn}, \citenamefont {Semeghini}, \citenamefont {Gullans}, \citenamefont {Greiner}, \citenamefont {Vuletić},\ and\ \citenamefont {Lukin}}]{Bluvstein2023LogicalQuantum}%
  \BibitemOpen
  \bibfield  {author} {\bibinfo {author} {\bibfnamefont {D.}~\bibnamefont {Bluvstein}}, \bibinfo {author} {\bibfnamefont {S.~J.}\ \bibnamefont {Evered}}, \bibinfo {author} {\bibfnamefont {A.~A.}\ \bibnamefont {Geim}}, \bibinfo {author} {\bibfnamefont {S.~H.}\ \bibnamefont {Li}}, \bibinfo {author} {\bibfnamefont {H.}~\bibnamefont {Zhou}}, \bibinfo {author} {\bibfnamefont {T.}~\bibnamefont {Manovitz}}, \bibinfo {author} {\bibfnamefont {S.}~\bibnamefont {Ebadi}}, \bibinfo {author} {\bibfnamefont {M.}~\bibnamefont {Cain}}, \bibinfo {author} {\bibfnamefont {M.}~\bibnamefont {Kalinowski}}, \bibinfo {author} {\bibfnamefont {D.}~\bibnamefont {Hangleiter}}, \bibinfo {author} {\bibfnamefont {J.~P.}\ \bibnamefont {Bonilla~Ataides}}, \bibinfo {author} {\bibfnamefont {N.}~\bibnamefont {Maskara}}, \bibinfo {author} {\bibfnamefont {I.}~\bibnamefont {Cong}}, \bibinfo {author} {\bibfnamefont {X.}~\bibnamefont {Gao}}, \bibinfo {author} {\bibfnamefont {P.}~\bibnamefont {Sales~Rodriguez}}, \bibinfo {author} {\bibfnamefont
  {T.}~\bibnamefont {Karolyshyn}}, \bibinfo {author} {\bibfnamefont {G.}~\bibnamefont {Semeghini}}, \bibinfo {author} {\bibfnamefont {M.~J.}\ \bibnamefont {Gullans}}, \bibinfo {author} {\bibfnamefont {M.}~\bibnamefont {Greiner}}, \bibinfo {author} {\bibfnamefont {V.}~\bibnamefont {Vuletić}},\ and\ \bibinfo {author} {\bibfnamefont {M.~D.}\ \bibnamefont {Lukin}},\ }\bibfield  {title} {\bibinfo {title} {Logical quantum processor based on reconfigurable atom arrays},\ }\href {https://doi.org/10.1038/s41586-023-06927-3} {\bibfield  {journal} {\bibinfo  {journal} {Nature}\ }\textbf {\bibinfo {volume} {626}},\ \bibinfo {pages} {58} (\bibinfo {year} {2023})}\BibitemShut {NoStop}%
\bibitem [{\citenamefont {Bukov}\ \emph {et~al.}(2018)\citenamefont {Bukov}, \citenamefont {Day}, \citenamefont {Sels}, \citenamefont {Weinberg}, \citenamefont {Polkovnikov},\ and\ \citenamefont {Mehta}}]{Bukov2018ReinforcementLearning}%
  \BibitemOpen
  \bibfield  {author} {\bibinfo {author} {\bibfnamefont {M.}~\bibnamefont {Bukov}}, \bibinfo {author} {\bibfnamefont {A.~G.~R.}\ \bibnamefont {Day}}, \bibinfo {author} {\bibfnamefont {D.}~\bibnamefont {Sels}}, \bibinfo {author} {\bibfnamefont {P.}~\bibnamefont {Weinberg}}, \bibinfo {author} {\bibfnamefont {A.}~\bibnamefont {Polkovnikov}},\ and\ \bibinfo {author} {\bibfnamefont {P.}~\bibnamefont {Mehta}},\ }\bibfield  {title} {\bibinfo {title} {Reinforcement learning in different phases of quantum control},\ }\href {https://doi.org/10.1103/physrevx.8.031086} {\bibfield  {journal} {\bibinfo  {journal} {Phys. Rev. X}\ }\textbf {\bibinfo {volume} {8}},\ \bibinfo {pages} {031086} (\bibinfo {year} {2018})}\BibitemShut {NoStop}%
\bibitem [{\citenamefont {Orús}(2019)}]{Orus2019TensorNetworks}%
  \BibitemOpen
  \bibfield  {author} {\bibinfo {author} {\bibfnamefont {R.}~\bibnamefont {Orús}},\ }\bibfield  {title} {\bibinfo {title} {Tensor networks for complex quantum systems},\ }\href {https://doi.org/10.1038/s42254-019-0086-7} {\bibfield  {journal} {\bibinfo  {journal} {Nat. Rev. Phys.}\ }\textbf {\bibinfo {volume} {1}},\ \bibinfo {pages} {538} (\bibinfo {year} {2019})}\BibitemShut {NoStop}%
\bibitem [{\citenamefont {Kolodrubetz}\ \emph {et~al.}(2017)\citenamefont {Kolodrubetz}, \citenamefont {Sels}, \citenamefont {Mehta},\ and\ \citenamefont {Polkovnikov}}]{Kolodrubetz2017GeometryAndNonAdiabatic}%
  \BibitemOpen
  \bibfield  {author} {\bibinfo {author} {\bibfnamefont {M.}~\bibnamefont {Kolodrubetz}}, \bibinfo {author} {\bibfnamefont {D.}~\bibnamefont {Sels}}, \bibinfo {author} {\bibfnamefont {P.}~\bibnamefont {Mehta}},\ and\ \bibinfo {author} {\bibfnamefont {A.}~\bibnamefont {Polkovnikov}},\ }\bibfield  {title} {\bibinfo {title} {Geometry and non-adiabatic response in quantum and classical systems},\ }\href {https://doi.org/10.1016/j.physrep.2017.07.001} {\bibfield  {journal} {\bibinfo  {journal} {Phys. Rep.}\ }\textbf {\bibinfo {volume} {697}},\ \bibinfo {pages} {1} (\bibinfo {year} {2017})}\BibitemShut {NoStop}%
\bibitem [{\citenamefont {Demirplak}\ and\ \citenamefont {Rice}(2003)}]{Demirplak2003AdiabaticPopulation}%
  \BibitemOpen
  \bibfield  {author} {\bibinfo {author} {\bibfnamefont {M.}~\bibnamefont {Demirplak}}\ and\ \bibinfo {author} {\bibfnamefont {S.~A.}\ \bibnamefont {Rice}},\ }\bibfield  {title} {\bibinfo {title} {Adiabatic population transfer with control fields},\ }\href {https://doi.org/10.1021/jp030708a} {\bibfield  {journal} {\bibinfo  {journal} {J. Phys. Chem. A}\ }\textbf {\bibinfo {volume} {107}},\ \bibinfo {pages} {9937} (\bibinfo {year} {2003})}\BibitemShut {NoStop}%
\bibitem [{\citenamefont {Demirplak}\ and\ \citenamefont {Rice}(2005)}]{Demirplak2005AssistedAdiabatic}%
  \BibitemOpen
  \bibfield  {author} {\bibinfo {author} {\bibfnamefont {M.}~\bibnamefont {Demirplak}}\ and\ \bibinfo {author} {\bibfnamefont {S.~A.}\ \bibnamefont {Rice}},\ }\bibfield  {title} {\bibinfo {title} {Assisted adiabatic passage revisited},\ }\href {https://doi.org/10.1021/jp040647w} {\bibfield  {journal} {\bibinfo  {journal} {J. Phys. Chem. B}\ }\textbf {\bibinfo {volume} {109}},\ \bibinfo {pages} {6838} (\bibinfo {year} {2005})}\BibitemShut {NoStop}%
\bibitem [{\citenamefont {Demirplak}\ and\ \citenamefont {Rice}(2008)}]{Demirplak2008OnTheConsistencyExtremal}%
  \BibitemOpen
  \bibfield  {author} {\bibinfo {author} {\bibfnamefont {M.}~\bibnamefont {Demirplak}}\ and\ \bibinfo {author} {\bibfnamefont {S.~A.}\ \bibnamefont {Rice}},\ }\bibfield  {title} {\bibinfo {title} {On the consistency, extremal, and global properties of counterdiabatic fields},\ }\href {https://doi.org/10.1063/1.2992152} {\bibfield  {journal} {\bibinfo  {journal} {J. Chem. Phys.}\ }\textbf {\bibinfo {volume} {129}},\ \bibinfo {pages} {154111} (\bibinfo {year} {2008})}\BibitemShut {NoStop}%
\bibitem [{\citenamefont {Berry}(2009)}]{Berry2009Transitionless}%
  \BibitemOpen
  \bibfield  {author} {\bibinfo {author} {\bibfnamefont {M.~V.}\ \bibnamefont {Berry}},\ }\bibfield  {title} {\bibinfo {title} {Transitionless quantum driving},\ }\href {https://doi.org/10.1088/1751-8113/42/36/365303} {\bibfield  {journal} {\bibinfo  {journal} {J. Phys. A: Math. Theor.}\ }\textbf {\bibinfo {volume} {42}},\ \bibinfo {pages} {365303} (\bibinfo {year} {2009})}\BibitemShut {NoStop}%
\bibitem [{\citenamefont {Campbell}\ and\ \citenamefont {Deffner}(2017)}]{Campbell2017TradeOffBetween}%
  \BibitemOpen
  \bibfield  {author} {\bibinfo {author} {\bibfnamefont {S.}~\bibnamefont {Campbell}}\ and\ \bibinfo {author} {\bibfnamefont {S.}~\bibnamefont {Deffner}},\ }\bibfield  {title} {\bibinfo {title} {Trade-off between speed and cost in shortcuts to adiabaticity},\ }\href {https://doi.org/10.1103/physrevlett.118.100601} {\bibfield  {journal} {\bibinfo  {journal} {Phys. Rev. Lett.}\ }\textbf {\bibinfo {volume} {118}},\ \bibinfo {pages} {100601} (\bibinfo {year} {2017})}\BibitemShut {NoStop}%
\bibitem [{\citenamefont {Santos}\ and\ \citenamefont {Sarandy}(2015)}]{Santos2015SuperadiabaticControlled}%
  \BibitemOpen
  \bibfield  {author} {\bibinfo {author} {\bibfnamefont {A.~C.}\ \bibnamefont {Santos}}\ and\ \bibinfo {author} {\bibfnamefont {M.~S.}\ \bibnamefont {Sarandy}},\ }\bibfield  {title} {\bibinfo {title} {Superadiabatic controlled evolutions and universal quantum computation},\ }\href {https://doi.org/10.1038/srep15775} {\bibfield  {journal} {\bibinfo  {journal} {Sci. Rep.}\ }\textbf {\bibinfo {volume} {5}},\ \bibinfo {pages} {15775} (\bibinfo {year} {2015})}\BibitemShut {NoStop}%
\bibitem [{\citenamefont {Funo}\ \emph {et~al.}(2017)\citenamefont {Funo}, \citenamefont {Zhang}, \citenamefont {Chatou}, \citenamefont {Kim}, \citenamefont {Ueda},\ and\ \citenamefont {del Campo}}]{Funo2017UniversalWork}%
  \BibitemOpen
  \bibfield  {author} {\bibinfo {author} {\bibfnamefont {K.}~\bibnamefont {Funo}}, \bibinfo {author} {\bibfnamefont {J.-N.}\ \bibnamefont {Zhang}}, \bibinfo {author} {\bibfnamefont {C.}~\bibnamefont {Chatou}}, \bibinfo {author} {\bibfnamefont {K.}~\bibnamefont {Kim}}, \bibinfo {author} {\bibfnamefont {M.}~\bibnamefont {Ueda}},\ and\ \bibinfo {author} {\bibfnamefont {A.}~\bibnamefont {del Campo}},\ }\bibfield  {title} {\bibinfo {title} {Universal work fluctuations during shortcuts to adiabaticity by counterdiabatic driving},\ }\href {https://doi.org/10.1103/physrevlett.118.100602} {\bibfield  {journal} {\bibinfo  {journal} {Phys. Rev. Lett.}\ }\textbf {\bibinfo {volume} {118}},\ \bibinfo {pages} {100602} (\bibinfo {year} {2017})}\BibitemShut {NoStop}%
\bibitem [{\citenamefont {Abah}\ \emph {et~al.}(2019)\citenamefont {Abah}, \citenamefont {Puebla}, \citenamefont {Kiely}, \citenamefont {De~Chiara}, \citenamefont {Paternostro},\ and\ \citenamefont {Campbell}}]{Abah2019EnergeticCost}%
  \BibitemOpen
  \bibfield  {author} {\bibinfo {author} {\bibfnamefont {O.}~\bibnamefont {Abah}}, \bibinfo {author} {\bibfnamefont {R.}~\bibnamefont {Puebla}}, \bibinfo {author} {\bibfnamefont {A.}~\bibnamefont {Kiely}}, \bibinfo {author} {\bibfnamefont {G.}~\bibnamefont {De~Chiara}}, \bibinfo {author} {\bibfnamefont {M.}~\bibnamefont {Paternostro}},\ and\ \bibinfo {author} {\bibfnamefont {S.}~\bibnamefont {Campbell}},\ }\bibfield  {title} {\bibinfo {title} {Energetic cost of quantum control protocols},\ }\href {https://doi.org/10.1088/1367-2630/ab4c8c} {\bibfield  {journal} {\bibinfo  {journal} {New J. Phys.}\ }\textbf {\bibinfo {volume} {21}},\ \bibinfo {pages} {103048} (\bibinfo {year} {2019})}\BibitemShut {NoStop}%
\bibitem [{\citenamefont {Ruschhaupt}\ \emph {et~al.}(2012)\citenamefont {Ruschhaupt}, \citenamefont {Chen}, \citenamefont {Alonso},\ and\ \citenamefont {Muga}}]{Ruschhaupt2012OptimallyRobust}%
  \BibitemOpen
  \bibfield  {author} {\bibinfo {author} {\bibfnamefont {A.}~\bibnamefont {Ruschhaupt}}, \bibinfo {author} {\bibfnamefont {X.}~\bibnamefont {Chen}}, \bibinfo {author} {\bibfnamefont {D.}~\bibnamefont {Alonso}},\ and\ \bibinfo {author} {\bibfnamefont {J.~G.}\ \bibnamefont {Muga}},\ }\bibfield  {title} {\bibinfo {title} {Optimally robust shortcuts to population inversion in two-level quantum systems},\ }\href {https://doi.org/10.1088/1367-2630/14/9/093040} {\bibfield  {journal} {\bibinfo  {journal} {New J. Phys.}\ }\textbf {\bibinfo {volume} {14}},\ \bibinfo {pages} {093040} (\bibinfo {year} {2012})}\BibitemShut {NoStop}%
\bibitem [{\citenamefont {Takahashi}(2013)}]{Takahashi2013HowFastAndRobust}%
  \BibitemOpen
  \bibfield  {author} {\bibinfo {author} {\bibfnamefont {K.}~\bibnamefont {Takahashi}},\ }\bibfield  {title} {\bibinfo {title} {How fast and robust is the quantum adiabatic passage?},\ }\href {https://doi.org/10.1088/1751-8113/46/31/315304} {\bibfield  {journal} {\bibinfo  {journal} {J. Phys. A: Math. Theor.}\ }\textbf {\bibinfo {volume} {46}},\ \bibinfo {pages} {315304} (\bibinfo {year} {2013})}\BibitemShut {NoStop}%
\bibitem [{\citenamefont {Chen}\ \emph {et~al.}(2010)\citenamefont {Chen}, \citenamefont {Lizuain}, \citenamefont {Ruschhaupt}, \citenamefont {Guéry-Odelin},\ and\ \citenamefont {Muga}}]{Chen2010ShortcutToAdiabatic}%
  \BibitemOpen
  \bibfield  {author} {\bibinfo {author} {\bibfnamefont {X.}~\bibnamefont {Chen}}, \bibinfo {author} {\bibfnamefont {I.}~\bibnamefont {Lizuain}}, \bibinfo {author} {\bibfnamefont {A.}~\bibnamefont {Ruschhaupt}}, \bibinfo {author} {\bibfnamefont {D.}~\bibnamefont {Guéry-Odelin}},\ and\ \bibinfo {author} {\bibfnamefont {J.~G.}\ \bibnamefont {Muga}},\ }\bibfield  {title} {\bibinfo {title} {Shortcut to adiabatic passage in two- and three-level atoms},\ }\href {https://doi.org/10.1103/physrevlett.105.123003} {\bibfield  {journal} {\bibinfo  {journal} {Phys. Rev. Lett.}\ }\textbf {\bibinfo {volume} {105}},\ \bibinfo {pages} {123003} (\bibinfo {year} {2010})}\BibitemShut {NoStop}%
\bibitem [{\citenamefont {Muga}\ \emph {et~al.}(2010)\citenamefont {Muga}, \citenamefont {Chen}, \citenamefont {Ibáñez}, \citenamefont {Lizuain},\ and\ \citenamefont {Ruschhaupt}}]{Muga2010TransitionlessQuantum}%
  \BibitemOpen
  \bibfield  {author} {\bibinfo {author} {\bibfnamefont {J.~G.}\ \bibnamefont {Muga}}, \bibinfo {author} {\bibfnamefont {X.}~\bibnamefont {Chen}}, \bibinfo {author} {\bibfnamefont {S.}~\bibnamefont {Ibáñez}}, \bibinfo {author} {\bibfnamefont {I.}~\bibnamefont {Lizuain}},\ and\ \bibinfo {author} {\bibfnamefont {A.}~\bibnamefont {Ruschhaupt}},\ }\bibfield  {title} {\bibinfo {title} {Transitionless quantum drivings for the harmonic oscillator},\ }\href {https://doi.org/10.1088/0953-4075/43/8/085509} {\bibfield  {journal} {\bibinfo  {journal} {J. Phys. B}\ }\textbf {\bibinfo {volume} {43}},\ \bibinfo {pages} {085509} (\bibinfo {year} {2010})}\BibitemShut {NoStop}%
\bibitem [{\citenamefont {del Campo}\ \emph {et~al.}(2012)\citenamefont {del Campo}, \citenamefont {Rams},\ and\ \citenamefont {Zurek}}]{del2012AssistedFinite}%
  \BibitemOpen
  \bibfield  {author} {\bibinfo {author} {\bibfnamefont {A.}~\bibnamefont {del Campo}}, \bibinfo {author} {\bibfnamefont {M.~M.}\ \bibnamefont {Rams}},\ and\ \bibinfo {author} {\bibfnamefont {W.~H.}\ \bibnamefont {Zurek}},\ }\bibfield  {title} {\bibinfo {title} {Assisted finite-rate adiabatic passage across a quantum critical point: Exact solution for the quantum {I}sing model},\ }\href {https://doi.org/10.1103/physrevlett.109.115703} {\bibfield  {journal} {\bibinfo  {journal} {Phys. Rev. Lett.}\ }\textbf {\bibinfo {volume} {109}},\ \bibinfo {pages} {115703} (\bibinfo {year} {2012})}\BibitemShut {NoStop}%
\bibitem [{\citenamefont {Jarzynski}(2013)}]{Jarzynski2013GeneratingShortcuts}%
  \BibitemOpen
  \bibfield  {author} {\bibinfo {author} {\bibfnamefont {C.}~\bibnamefont {Jarzynski}},\ }\bibfield  {title} {\bibinfo {title} {Generating shortcuts to adiabaticity in quantum and classical dynamics},\ }\href {https://doi.org/10.1103/physreva.88.040101} {\bibfield  {journal} {\bibinfo  {journal} {Phys. Rev. A}\ }\textbf {\bibinfo {volume} {88}},\ \bibinfo {pages} {040101(R)} (\bibinfo {year} {2013})}\BibitemShut {NoStop}%
\bibitem [{\citenamefont {del Campo}(2013)}]{del2013ShortcutsToAdiabaticity}%
  \BibitemOpen
  \bibfield  {author} {\bibinfo {author} {\bibfnamefont {A.}~\bibnamefont {del Campo}},\ }\bibfield  {title} {\bibinfo {title} {Shortcuts to adiabaticity by counterdiabatic driving},\ }\href {https://doi.org/10.1103/physrevlett.111.100502} {\bibfield  {journal} {\bibinfo  {journal} {Phys. Rev. Lett.}\ }\textbf {\bibinfo {volume} {111}},\ \bibinfo {pages} {100502} (\bibinfo {year} {2013})}\BibitemShut {NoStop}%
\bibitem [{\citenamefont {Deffner}\ \emph {et~al.}(2014)\citenamefont {Deffner}, \citenamefont {Jarzynski},\ and\ \citenamefont {del Campo}}]{Deffner2014ClassicalAndQuantum}%
  \BibitemOpen
  \bibfield  {author} {\bibinfo {author} {\bibfnamefont {S.}~\bibnamefont {Deffner}}, \bibinfo {author} {\bibfnamefont {C.}~\bibnamefont {Jarzynski}},\ and\ \bibinfo {author} {\bibfnamefont {A.}~\bibnamefont {del Campo}},\ }\bibfield  {title} {\bibinfo {title} {Classical and quantum shortcuts to adiabaticity for scale-invariant driving},\ }\href {https://doi.org/10.1103/physrevx.4.021013} {\bibfield  {journal} {\bibinfo  {journal} {Phys. Rev. X}\ }\textbf {\bibinfo {volume} {4}},\ \bibinfo {pages} {021013} (\bibinfo {year} {2014})}\BibitemShut {NoStop}%
\bibitem [{\citenamefont {Okuyama}\ and\ \citenamefont {Takahashi}(2016)}]{Okuyama2016FromClassical}%
  \BibitemOpen
  \bibfield  {author} {\bibinfo {author} {\bibfnamefont {M.}~\bibnamefont {Okuyama}}\ and\ \bibinfo {author} {\bibfnamefont {K.}~\bibnamefont {Takahashi}},\ }\bibfield  {title} {\bibinfo {title} {From classical nonlinear integrable systems to quantum shortcuts to adiabaticity},\ }\href {https://doi.org/10.1103/physrevlett.117.070401} {\bibfield  {journal} {\bibinfo  {journal} {Phys. Rev. Lett.}\ }\textbf {\bibinfo {volume} {117}},\ \bibinfo {pages} {070401} (\bibinfo {year} {2016})}\BibitemShut {NoStop}%
\bibitem [{\citenamefont {Hatomura}\ and\ \citenamefont {Mori}(2018)}]{Hatomura2018ShortcutsToAdiabatic}%
  \BibitemOpen
  \bibfield  {author} {\bibinfo {author} {\bibfnamefont {T.}~\bibnamefont {Hatomura}}\ and\ \bibinfo {author} {\bibfnamefont {T.}~\bibnamefont {Mori}},\ }\bibfield  {title} {\bibinfo {title} {Shortcuts to adiabatic classical spin dynamics mimicking quantum annealing},\ }\href {https://doi.org/10.1103/physreve.98.032136} {\bibfield  {journal} {\bibinfo  {journal} {Phys. Rev. E}\ }\textbf {\bibinfo {volume} {98}},\ \bibinfo {pages} {032136} (\bibinfo {year} {2018})}\BibitemShut {NoStop}%
\bibitem [{\citenamefont {Sels}\ and\ \citenamefont {Polkovnikov}(2017)}]{Sels2017Minimizing}%
  \BibitemOpen
  \bibfield  {author} {\bibinfo {author} {\bibfnamefont {D.}~\bibnamefont {Sels}}\ and\ \bibinfo {author} {\bibfnamefont {A.}~\bibnamefont {Polkovnikov}},\ }\bibfield  {title} {\bibinfo {title} {Minimizing irreversible losses in quantum systems by local counterdiabatic driving},\ }\href {https://doi.org/10.1073/pnas.1619826114} {\bibfield  {journal} {\bibinfo  {journal} {Proc. Natl. Acad. Sci. USA}\ }\textbf {\bibinfo {volume} {114}},\ \bibinfo {pages} {E3909} (\bibinfo {year} {2017})}\BibitemShut {NoStop}%
\bibitem [{\citenamefont {Hartmann}\ and\ \citenamefont {Lechner}(2019)}]{Hartmann2019RapidCounter}%
  \BibitemOpen
  \bibfield  {author} {\bibinfo {author} {\bibfnamefont {A.}~\bibnamefont {Hartmann}}\ and\ \bibinfo {author} {\bibfnamefont {W.}~\bibnamefont {Lechner}},\ }\bibfield  {title} {\bibinfo {title} {Rapid counter-diabatic sweeps in lattice gauge adiabatic quantum computing},\ }\href {https://doi.org/10.1088/1367-2630/ab14a0} {\bibfield  {journal} {\bibinfo  {journal} {New J. Phys.}\ }\textbf {\bibinfo {volume} {21}},\ \bibinfo {pages} {043025} (\bibinfo {year} {2019})}\BibitemShut {NoStop}%
\bibitem [{\citenamefont {Hartmann}\ \emph {et~al.}(2022)\citenamefont {Hartmann}, \citenamefont {Mbeng},\ and\ \citenamefont {Lechner}}]{Hartmann2022PolynomialScaling}%
  \BibitemOpen
  \bibfield  {author} {\bibinfo {author} {\bibfnamefont {A.}~\bibnamefont {Hartmann}}, \bibinfo {author} {\bibfnamefont {G.~B.}\ \bibnamefont {Mbeng}},\ and\ \bibinfo {author} {\bibfnamefont {W.}~\bibnamefont {Lechner}},\ }\bibfield  {title} {\bibinfo {title} {Polynomial scaling enhancement in the ground-state preparation of {I}sing spin models via counterdiabatic driving},\ }\href {https://doi.org/10.1103/physreva.105.022614} {\bibfield  {journal} {\bibinfo  {journal} {Phys. Rev. A}\ }\textbf {\bibinfo {volume} {105}},\ \bibinfo {pages} {022614} (\bibinfo {year} {2022})}\BibitemShut {NoStop}%
\bibitem [{\citenamefont {Passarelli}\ \emph {et~al.}(2020)\citenamefont {Passarelli}, \citenamefont {Cataudella}, \citenamefont {Fazio},\ and\ \citenamefont {Lucignano}}]{Passarelli2020CounterdiabaticDriving}%
  \BibitemOpen
  \bibfield  {author} {\bibinfo {author} {\bibfnamefont {G.}~\bibnamefont {Passarelli}}, \bibinfo {author} {\bibfnamefont {V.}~\bibnamefont {Cataudella}}, \bibinfo {author} {\bibfnamefont {R.}~\bibnamefont {Fazio}},\ and\ \bibinfo {author} {\bibfnamefont {P.}~\bibnamefont {Lucignano}},\ }\bibfield  {title} {\bibinfo {title} {Counterdiabatic driving in the quantum annealing of the $p$-spin model: A variational approach},\ }\href {https://doi.org/10.1103/physrevresearch.2.013283} {\bibfield  {journal} {\bibinfo  {journal} {Phys. Rev. Res.}\ }\textbf {\bibinfo {volume} {2}},\ \bibinfo {pages} {013283} (\bibinfo {year} {2020})}\BibitemShut {NoStop}%
\bibitem [{\citenamefont {Prielinger}\ \emph {et~al.}(2021)\citenamefont {Prielinger}, \citenamefont {Hartmann}, \citenamefont {Yamashiro}, \citenamefont {Nishimura}, \citenamefont {Lechner},\ and\ \citenamefont {Nishimori}}]{Prielinger2021TwoParameterCounter}%
  \BibitemOpen
  \bibfield  {author} {\bibinfo {author} {\bibfnamefont {L.}~\bibnamefont {Prielinger}}, \bibinfo {author} {\bibfnamefont {A.}~\bibnamefont {Hartmann}}, \bibinfo {author} {\bibfnamefont {Y.}~\bibnamefont {Yamashiro}}, \bibinfo {author} {\bibfnamefont {K.}~\bibnamefont {Nishimura}}, \bibinfo {author} {\bibfnamefont {W.}~\bibnamefont {Lechner}},\ and\ \bibinfo {author} {\bibfnamefont {H.}~\bibnamefont {Nishimori}},\ }\bibfield  {title} {\bibinfo {title} {Two-parameter counter-diabatic driving in quantum annealing},\ }\href {https://doi.org/10.1103/physrevresearch.3.013227} {\bibfield  {journal} {\bibinfo  {journal} {Phys. Rev. Res.}\ }\textbf {\bibinfo {volume} {3}},\ \bibinfo {pages} {013227} (\bibinfo {year} {2021})}\BibitemShut {NoStop}%
\bibitem [{\citenamefont {Kumar}\ \emph {et~al.}(2021)\citenamefont {Kumar}, \citenamefont {Sharma},\ and\ \citenamefont {Tripathi}}]{Kumar2021CounterdiabaticRoute}%
  \BibitemOpen
  \bibfield  {author} {\bibinfo {author} {\bibfnamefont {S.}~\bibnamefont {Kumar}}, \bibinfo {author} {\bibfnamefont {S.}~\bibnamefont {Sharma}},\ and\ \bibinfo {author} {\bibfnamefont {V.}~\bibnamefont {Tripathi}},\ }\bibfield  {title} {\bibinfo {title} {Counterdiabatic route for preparation of state with long-range topological order},\ }\href {https://doi.org/10.1103/physrevb.104.245113} {\bibfield  {journal} {\bibinfo  {journal} {Phys. Rev. B}\ }\textbf {\bibinfo {volume} {104}},\ \bibinfo {pages} {245113} (\bibinfo {year} {2021})}\BibitemShut {NoStop}%
\bibitem [{\citenamefont {Barone}\ \emph {et~al.}(2024)\citenamefont {Barone}, \citenamefont {Kiss}, \citenamefont {Grossi}, \citenamefont {Vallecorsa},\ and\ \citenamefont {Mandarino}}]{Barone2024CounterdiabaticOptimized}%
  \BibitemOpen
  \bibfield  {author} {\bibinfo {author} {\bibfnamefont {F.~P.}\ \bibnamefont {Barone}}, \bibinfo {author} {\bibfnamefont {O.}~\bibnamefont {Kiss}}, \bibinfo {author} {\bibfnamefont {M.}~\bibnamefont {Grossi}}, \bibinfo {author} {\bibfnamefont {S.}~\bibnamefont {Vallecorsa}},\ and\ \bibinfo {author} {\bibfnamefont {A.}~\bibnamefont {Mandarino}},\ }\bibfield  {title} {\bibinfo {title} {Counterdiabatic optimized driving in quantum phase sensitive models},\ }\href {https://doi.org/10.1088/1367-2630/ad313e} {\bibfield  {journal} {\bibinfo  {journal} {New J. Phys.}\ }\textbf {\bibinfo {volume} {26}},\ \bibinfo {pages} {033031} (\bibinfo {year} {2024})}\BibitemShut {NoStop}%
\bibitem [{\citenamefont {Grabarits}\ \emph {et~al.}(2026)\citenamefont {Grabarits}, \citenamefont {Balducci},\ and\ \citenamefont {del Campo}}]{Andras2024FightingExponentially}%
  \BibitemOpen
  \bibfield  {author} {\bibinfo {author} {\bibfnamefont {A.}~\bibnamefont {Grabarits}}, \bibinfo {author} {\bibfnamefont {F.}~\bibnamefont {Balducci}},\ and\ \bibinfo {author} {\bibfnamefont {A.}~\bibnamefont {del Campo}},\ }\bibfield  {title} {\bibinfo {title} {Fighting exponentially small gaps by counterdiabatic driving},\ }\href {https://doi.org/10.1103/tgzt-dy3h} {\bibfield  {journal} {\bibinfo  {journal} {PRX Quantum}\ }\textbf {\bibinfo {volume} {7}},\ \bibinfo {pages} {010322} (\bibinfo {year} {2026})}\BibitemShut {NoStop}%
\bibitem [{\citenamefont {Xie}\ \emph {et~al.}(2022)\citenamefont {Xie}, \citenamefont {Seki},\ and\ \citenamefont {Yunoki}}]{Xie2022VariationalCounterdiabatic}%
  \BibitemOpen
  \bibfield  {author} {\bibinfo {author} {\bibfnamefont {Q.}~\bibnamefont {Xie}}, \bibinfo {author} {\bibfnamefont {K.}~\bibnamefont {Seki}},\ and\ \bibinfo {author} {\bibfnamefont {S.}~\bibnamefont {Yunoki}},\ }\bibfield  {title} {\bibinfo {title} {Variational counterdiabatic driving of the {H}ubbard model for ground-state preparation},\ }\href {https://doi.org/10.1103/physrevb.106.155153} {\bibfield  {journal} {\bibinfo  {journal} {Phys. Rev. B}\ }\textbf {\bibinfo {volume} {106}},\ \bibinfo {pages} {155153} (\bibinfo {year} {2022})}\BibitemShut {NoStop}%
\bibitem [{\citenamefont {Zhou}\ \emph {et~al.}(2020)\citenamefont {Zhou}, \citenamefont {Ji}, \citenamefont {Nie}, \citenamefont {Yang}, \citenamefont {Chen}, \citenamefont {Bian},\ and\ \citenamefont {Peng}}]{Zhou2020ExperimentalRealization}%
  \BibitemOpen
  \bibfield  {author} {\bibinfo {author} {\bibfnamefont {H.}~\bibnamefont {Zhou}}, \bibinfo {author} {\bibfnamefont {Y.}~\bibnamefont {Ji}}, \bibinfo {author} {\bibfnamefont {X.}~\bibnamefont {Nie}}, \bibinfo {author} {\bibfnamefont {X.}~\bibnamefont {Yang}}, \bibinfo {author} {\bibfnamefont {X.}~\bibnamefont {Chen}}, \bibinfo {author} {\bibfnamefont {J.}~\bibnamefont {Bian}},\ and\ \bibinfo {author} {\bibfnamefont {X.}~\bibnamefont {Peng}},\ }\bibfield  {title} {\bibinfo {title} {Experimental realization of shortcuts to adiabaticity in a nonintegrable spin chain by local counterdiabatic driving},\ }\href {https://doi.org/10.1103/physrevapplied.13.044059} {\bibfield  {journal} {\bibinfo  {journal} {Phys. Rev. Appl.}\ }\textbf {\bibinfo {volume} {13}},\ \bibinfo {pages} {044059} (\bibinfo {year} {2020})}\BibitemShut {NoStop}%
\bibitem [{\citenamefont {Meier}\ \emph {et~al.}(2020)\citenamefont {Meier}, \citenamefont {Ngan}, \citenamefont {Sels},\ and\ \citenamefont {Gadway}}]{Meier2020CounterdiabaticControl}%
  \BibitemOpen
  \bibfield  {author} {\bibinfo {author} {\bibfnamefont {E.~J.}\ \bibnamefont {Meier}}, \bibinfo {author} {\bibfnamefont {K.}~\bibnamefont {Ngan}}, \bibinfo {author} {\bibfnamefont {D.}~\bibnamefont {Sels}},\ and\ \bibinfo {author} {\bibfnamefont {B.}~\bibnamefont {Gadway}},\ }\bibfield  {title} {\bibinfo {title} {Counterdiabatic control of transport in a synthetic tight-binding lattice},\ }\href {https://doi.org/10.1103/physrevresearch.2.043201} {\bibfield  {journal} {\bibinfo  {journal} {Phys. Rev. Res.}\ }\textbf {\bibinfo {volume} {2}},\ \bibinfo {pages} {043201} (\bibinfo {year} {2020})}\BibitemShut {NoStop}%
\bibitem [{\citenamefont {Hegade}\ \emph {et~al.}(2021{\natexlab{a}})\citenamefont {Hegade}, \citenamefont {Paul}, \citenamefont {Ding}, \citenamefont {Sanz}, \citenamefont {Albarrán-Arriagada}, \citenamefont {Solano},\ and\ \citenamefont {Chen}}]{Hegade2021ShortcutsToAdiabaticity}%
  \BibitemOpen
  \bibfield  {author} {\bibinfo {author} {\bibfnamefont {N.~N.}\ \bibnamefont {Hegade}}, \bibinfo {author} {\bibfnamefont {K.}~\bibnamefont {Paul}}, \bibinfo {author} {\bibfnamefont {Y.}~\bibnamefont {Ding}}, \bibinfo {author} {\bibfnamefont {M.}~\bibnamefont {Sanz}}, \bibinfo {author} {\bibfnamefont {F.}~\bibnamefont {Albarrán-Arriagada}}, \bibinfo {author} {\bibfnamefont {E.}~\bibnamefont {Solano}},\ and\ \bibinfo {author} {\bibfnamefont {X.}~\bibnamefont {Chen}},\ }\bibfield  {title} {\bibinfo {title} {Shortcuts to adiabaticity in digitized adiabatic quantum computing},\ }\href {https://doi.org/10.1103/physrevapplied.15.024038} {\bibfield  {journal} {\bibinfo  {journal} {Phys. Rev. Appl.}\ }\textbf {\bibinfo {volume} {15}},\ \bibinfo {pages} {024038} (\bibinfo {year} {2021}{\natexlab{a}})}\BibitemShut {NoStop}%
\bibitem [{\citenamefont {Hartmann}\ \emph {et~al.}(2020{\natexlab{a}})\citenamefont {Hartmann}, \citenamefont {Mukherjee}, \citenamefont {Niedenzu},\ and\ \citenamefont {Lechner}}]{Hartmann2020ManyBody}%
  \BibitemOpen
  \bibfield  {author} {\bibinfo {author} {\bibfnamefont {A.}~\bibnamefont {Hartmann}}, \bibinfo {author} {\bibfnamefont {V.}~\bibnamefont {Mukherjee}}, \bibinfo {author} {\bibfnamefont {W.}~\bibnamefont {Niedenzu}},\ and\ \bibinfo {author} {\bibfnamefont {W.}~\bibnamefont {Lechner}},\ }\bibfield  {title} {\bibinfo {title} {Many-body quantum heat engines with shortcuts to adiabaticity},\ }\href {https://doi.org/10.1103/physrevresearch.2.023145} {\bibfield  {journal} {\bibinfo  {journal} {Phys. Rev. Res.}\ }\textbf {\bibinfo {volume} {2}},\ \bibinfo {pages} {023145} (\bibinfo {year} {2020}{\natexlab{a}})}\BibitemShut {NoStop}%
\bibitem [{\citenamefont {Hartmann}\ \emph {et~al.}(2020{\natexlab{b}})\citenamefont {Hartmann}, \citenamefont {Mukherjee}, \citenamefont {Mbeng}, \citenamefont {Niedenzu},\ and\ \citenamefont {Lechner}}]{Hartmann2020MultiSpin}%
  \BibitemOpen
  \bibfield  {author} {\bibinfo {author} {\bibfnamefont {A.}~\bibnamefont {Hartmann}}, \bibinfo {author} {\bibfnamefont {V.}~\bibnamefont {Mukherjee}}, \bibinfo {author} {\bibfnamefont {G.~B.}\ \bibnamefont {Mbeng}}, \bibinfo {author} {\bibfnamefont {W.}~\bibnamefont {Niedenzu}},\ and\ \bibinfo {author} {\bibfnamefont {W.}~\bibnamefont {Lechner}},\ }\bibfield  {title} {\bibinfo {title} {Multi-spin counter-diabatic driving in many-body quantum otto refrigerators},\ }\href {https://doi.org/10.22331/q-2020-12-24-377} {\bibfield  {journal} {\bibinfo  {journal} {Quantum}\ }\textbf {\bibinfo {volume} {4}},\ \bibinfo {pages} {377} (\bibinfo {year} {2020}{\natexlab{b}})}\BibitemShut {NoStop}%
\bibitem [{\citenamefont {Villazon}\ \emph {et~al.}(2021)\citenamefont {Villazon}, \citenamefont {Claeys}, \citenamefont {Polkovnikov},\ and\ \citenamefont {Chandran}}]{Villazon2021ShortcutsToDynamic}%
  \BibitemOpen
  \bibfield  {author} {\bibinfo {author} {\bibfnamefont {T.}~\bibnamefont {Villazon}}, \bibinfo {author} {\bibfnamefont {P.~W.}\ \bibnamefont {Claeys}}, \bibinfo {author} {\bibfnamefont {A.}~\bibnamefont {Polkovnikov}},\ and\ \bibinfo {author} {\bibfnamefont {A.}~\bibnamefont {Chandran}},\ }\bibfield  {title} {\bibinfo {title} {Shortcuts to dynamic polarization},\ }\href {https://doi.org/10.1103/physrevb.103.075118} {\bibfield  {journal} {\bibinfo  {journal} {Phys. Rev. B}\ }\textbf {\bibinfo {volume} {103}},\ \bibinfo {pages} {075118} (\bibinfo {year} {2021})}\BibitemShut {NoStop}%
\bibitem [{\citenamefont {Ji}\ \emph {et~al.}(2022)\citenamefont {Ji}, \citenamefont {Zhou}, \citenamefont {Chen}, \citenamefont {Liu}, \citenamefont {Li}, \citenamefont {Zhou},\ and\ \citenamefont {Peng}}]{Ji2022CounterdiabaticTransfer}%
  \BibitemOpen
  \bibfield  {author} {\bibinfo {author} {\bibfnamefont {Y.}~\bibnamefont {Ji}}, \bibinfo {author} {\bibfnamefont {F.}~\bibnamefont {Zhou}}, \bibinfo {author} {\bibfnamefont {X.}~\bibnamefont {Chen}}, \bibinfo {author} {\bibfnamefont {R.}~\bibnamefont {Liu}}, \bibinfo {author} {\bibfnamefont {Z.}~\bibnamefont {Li}}, \bibinfo {author} {\bibfnamefont {H.}~\bibnamefont {Zhou}},\ and\ \bibinfo {author} {\bibfnamefont {X.}~\bibnamefont {Peng}},\ }\bibfield  {title} {\bibinfo {title} {Counterdiabatic transfer of a quantum state in a tunable {H}eisenberg spin chain via the variational principle},\ }\href {https://doi.org/10.1103/physreva.105.052422} {\bibfield  {journal} {\bibinfo  {journal} {Phys. Rev. A}\ }\textbf {\bibinfo {volume} {105}},\ \bibinfo {pages} {052422} (\bibinfo {year} {2022})}\BibitemShut {NoStop}%
\bibitem [{\citenamefont {Passarelli}\ and\ \citenamefont {Lucignano}(2023)}]{Passarelli2023CounterdiabaticReverse}%
  \BibitemOpen
  \bibfield  {author} {\bibinfo {author} {\bibfnamefont {G.}~\bibnamefont {Passarelli}}\ and\ \bibinfo {author} {\bibfnamefont {P.}~\bibnamefont {Lucignano}},\ }\bibfield  {title} {\bibinfo {title} {Counterdiabatic reverse annealing},\ }\href {https://doi.org/10.1103/physreva.107.022607} {\bibfield  {journal} {\bibinfo  {journal} {Phys. Rev. A}\ }\textbf {\bibinfo {volume} {107}},\ \bibinfo {pages} {022607} (\bibinfo {year} {2023})}\BibitemShut {NoStop}%
\bibitem [{\citenamefont {Hegade}\ \emph {et~al.}(2021{\natexlab{b}})\citenamefont {Hegade}, \citenamefont {Paul}, \citenamefont {Albarrán-Arriagada}, \citenamefont {Chen},\ and\ \citenamefont {Solano}}]{Hegade2021DigitizedAdiabatic}%
  \BibitemOpen
  \bibfield  {author} {\bibinfo {author} {\bibfnamefont {N.~N.}\ \bibnamefont {Hegade}}, \bibinfo {author} {\bibfnamefont {K.}~\bibnamefont {Paul}}, \bibinfo {author} {\bibfnamefont {F.}~\bibnamefont {Albarrán-Arriagada}}, \bibinfo {author} {\bibfnamefont {X.}~\bibnamefont {Chen}},\ and\ \bibinfo {author} {\bibfnamefont {E.}~\bibnamefont {Solano}},\ }\bibfield  {title} {\bibinfo {title} {Digitized adiabatic quantum factorization},\ }\href {https://doi.org/10.1103/physreva.104.l050403} {\bibfield  {journal} {\bibinfo  {journal} {Phys. Rev. A}\ }\textbf {\bibinfo {volume} {104}},\ \bibinfo {pages} {L050403} (\bibinfo {year} {2021}{\natexlab{b}})}\BibitemShut {NoStop}%
\bibitem [{\citenamefont {Hegade}\ \emph {et~al.}(2022{\natexlab{a}})\citenamefont {Hegade}, \citenamefont {Chen},\ and\ \citenamefont {Solano}}]{Hegade2022DigitizedCounterdiabatic}%
  \BibitemOpen
  \bibfield  {author} {\bibinfo {author} {\bibfnamefont {N.~N.}\ \bibnamefont {Hegade}}, \bibinfo {author} {\bibfnamefont {X.}~\bibnamefont {Chen}},\ and\ \bibinfo {author} {\bibfnamefont {E.}~\bibnamefont {Solano}},\ }\bibfield  {title} {\bibinfo {title} {Digitized counterdiabatic quantum optimization},\ }\href {https://doi.org/10.1103/physrevresearch.4.l042030} {\bibfield  {journal} {\bibinfo  {journal} {Phys. Rev. Res.}\ }\textbf {\bibinfo {volume} {4}},\ \bibinfo {pages} {L042030} (\bibinfo {year} {2022}{\natexlab{a}})}\BibitemShut {NoStop}%
\bibitem [{\citenamefont {Hegade}\ \emph {et~al.}(2022{\natexlab{b}})\citenamefont {Hegade}, \citenamefont {Chandarana}, \citenamefont {Paul}, \citenamefont {Chen}, \citenamefont {Albarrán-Arriagada},\ and\ \citenamefont {Solano}}]{Hegade2022PortfolioOptimization}%
  \BibitemOpen
  \bibfield  {author} {\bibinfo {author} {\bibfnamefont {N.~N.}\ \bibnamefont {Hegade}}, \bibinfo {author} {\bibfnamefont {P.}~\bibnamefont {Chandarana}}, \bibinfo {author} {\bibfnamefont {K.}~\bibnamefont {Paul}}, \bibinfo {author} {\bibfnamefont {X.}~\bibnamefont {Chen}}, \bibinfo {author} {\bibfnamefont {F.}~\bibnamefont {Albarrán-Arriagada}},\ and\ \bibinfo {author} {\bibfnamefont {E.}~\bibnamefont {Solano}},\ }\bibfield  {title} {\bibinfo {title} {Portfolio optimization with digitized counterdiabatic quantum algorithms},\ }\href {https://doi.org/10.1103/physrevresearch.4.043204} {\bibfield  {journal} {\bibinfo  {journal} {Phys. Rev. Res.}\ }\textbf {\bibinfo {volume} {4}},\ \bibinfo {pages} {043204} (\bibinfo {year} {2022}{\natexlab{b}})}\BibitemShut {NoStop}%
\bibitem [{\citenamefont {Hegade}\ and\ \citenamefont {Solano}()}]{Hegade2023DigitizedCounterdiabatic}%
  \BibitemOpen
  \bibfield  {author} {\bibinfo {author} {\bibfnamefont {N.~N.}\ \bibnamefont {Hegade}}\ and\ \bibinfo {author} {\bibfnamefont {E.}~\bibnamefont {Solano}},\ }\bibfield  {title} {\bibinfo {title} {Digitized-counterdiabatic quantum factorization},\ }\href@noop {} {\bibinfo  {journal} {\href{https://arxiv.org/abs/2301.11005}{arXiv preprint arXiv:2301.11005 (2023)}}\ }\BibitemShut {NoStop}%
\bibitem [{\citenamefont {Guan}\ \emph {et~al.}(2024)\citenamefont {Guan}, \citenamefont {Zhou}, \citenamefont {Albarrán-Arriagada}, \citenamefont {Chen}, \citenamefont {Solano}, \citenamefont {Hegade},\ and\ \citenamefont {Huang}}]{Guan2024SingleLayer}%
  \BibitemOpen
\bibfield  {journal} {  }\bibfield  {author} {\bibinfo {author} {\bibfnamefont {H.}~\bibnamefont {Guan}}, \bibinfo {author} {\bibfnamefont {F.}~\bibnamefont {Zhou}}, \bibinfo {author} {\bibfnamefont {F.}~\bibnamefont {Albarrán-Arriagada}}, \bibinfo {author} {\bibfnamefont {X.}~\bibnamefont {Chen}}, \bibinfo {author} {\bibfnamefont {E.}~\bibnamefont {Solano}}, \bibinfo {author} {\bibfnamefont {N.~N.}\ \bibnamefont {Hegade}},\ and\ \bibinfo {author} {\bibfnamefont {H.-L.}\ \bibnamefont {Huang}},\ }\bibfield  {title} {\bibinfo {title} {Single-layer digitized-counterdiabatic quantum optimization for $p$-spin models},\ }\href {https://doi.org/10.1088/2058-9565/ad7880} {\bibfield  {journal} {\bibinfo  {journal} {Quantum Sci. Technol.}\ }\textbf {\bibinfo {volume} {10}},\ \bibinfo {pages} {015006} (\bibinfo {year} {2024})}\BibitemShut {NoStop}%
\bibitem [{\citenamefont {Romero}\ \emph {et~al.}(2025)\citenamefont {Romero}, \citenamefont {Visuri}, \citenamefont {Cadavid}, \citenamefont {Simen}, \citenamefont {Solano},\ and\ \citenamefont {Hegade}}]{Romero2025BiasField}%
  \BibitemOpen
  \bibfield  {author} {\bibinfo {author} {\bibfnamefont {S.~V.}\ \bibnamefont {Romero}}, \bibinfo {author} {\bibfnamefont {A.-M.}\ \bibnamefont {Visuri}}, \bibinfo {author} {\bibfnamefont {A.~G.}\ \bibnamefont {Cadavid}}, \bibinfo {author} {\bibfnamefont {A.}~\bibnamefont {Simen}}, \bibinfo {author} {\bibfnamefont {E.}~\bibnamefont {Solano}},\ and\ \bibinfo {author} {\bibfnamefont {N.~N.}\ \bibnamefont {Hegade}},\ }\bibfield  {title} {\bibinfo {title} {Bias-field digitized counterdiabatic quantum algorithm for higher-order binary optimization},\ }\href {https://doi.org/10.1038/s42005-025-02270-3} {\bibfield  {journal} {\bibinfo  {journal} {Commun. Phys.}\ }\textbf {\bibinfo {volume} {8}},\ \bibinfo {pages} {348} (\bibinfo {year} {2025})}\BibitemShut {NoStop}%
\bibitem [{\citenamefont {Petiziol}\ \emph {et~al.}(2024)\citenamefont {Petiziol}, \citenamefont {Mintert},\ and\ \citenamefont {Wimberger}}]{Petiziol2024QuantumControl}%
  \BibitemOpen
  \bibfield  {author} {\bibinfo {author} {\bibfnamefont {F.}~\bibnamefont {Petiziol}}, \bibinfo {author} {\bibfnamefont {F.}~\bibnamefont {Mintert}},\ and\ \bibinfo {author} {\bibfnamefont {S.}~\bibnamefont {Wimberger}},\ }\bibfield  {title} {\bibinfo {title} {Quantum control by effective counterdiabatic driving},\ }\href {https://doi.org/10.1209/0295-5075/ad19e3} {\bibfield  {journal} {\bibinfo  {journal} {EPL}\ }\textbf {\bibinfo {volume} {145}},\ \bibinfo {pages} {15001} (\bibinfo {year} {2024})}\BibitemShut {NoStop}%
\bibitem [{\citenamefont {Petiziol}\ \emph {et~al.}(2018)\citenamefont {Petiziol}, \citenamefont {Dive}, \citenamefont {Mintert},\ and\ \citenamefont {Wimberger}}]{Petiziol2018FastAdiabatic}%
  \BibitemOpen
  \bibfield  {author} {\bibinfo {author} {\bibfnamefont {F.}~\bibnamefont {Petiziol}}, \bibinfo {author} {\bibfnamefont {B.}~\bibnamefont {Dive}}, \bibinfo {author} {\bibfnamefont {F.}~\bibnamefont {Mintert}},\ and\ \bibinfo {author} {\bibfnamefont {S.}~\bibnamefont {Wimberger}},\ }\bibfield  {title} {\bibinfo {title} {Fast adiabatic evolution by oscillating initial {H}amiltonians},\ }\href {https://doi.org/10.1103/physreva.98.043436} {\bibfield  {journal} {\bibinfo  {journal} {Phys. Rev. A}\ }\textbf {\bibinfo {volume} {98}},\ \bibinfo {pages} {043436} (\bibinfo {year} {2018})}\BibitemShut {NoStop}%
\bibitem [{\citenamefont {Petiziol}\ \emph {et~al.}(2019)\citenamefont {Petiziol}, \citenamefont {Dive}, \citenamefont {Carretta}, \citenamefont {Mannella}, \citenamefont {Mintert},\ and\ \citenamefont {Wimberger}}]{Petiziol2019AcceleratingAdiabatic}%
  \BibitemOpen
  \bibfield  {author} {\bibinfo {author} {\bibfnamefont {F.}~\bibnamefont {Petiziol}}, \bibinfo {author} {\bibfnamefont {B.}~\bibnamefont {Dive}}, \bibinfo {author} {\bibfnamefont {S.}~\bibnamefont {Carretta}}, \bibinfo {author} {\bibfnamefont {R.}~\bibnamefont {Mannella}}, \bibinfo {author} {\bibfnamefont {F.}~\bibnamefont {Mintert}},\ and\ \bibinfo {author} {\bibfnamefont {S.}~\bibnamefont {Wimberger}},\ }\bibfield  {title} {\bibinfo {title} {Accelerating adiabatic protocols for entangling two qubits in circuit {QED}},\ }\href {https://doi.org/10.1103/physreva.99.042315} {\bibfield  {journal} {\bibinfo  {journal} {Phys. Rev. A}\ }\textbf {\bibinfo {volume} {99}},\ \bibinfo {pages} {042315} (\bibinfo {year} {2019})}\BibitemShut {NoStop}%
\bibitem [{\citenamefont {Claeys}\ \emph {et~al.}(2019)\citenamefont {Claeys}, \citenamefont {Pandey}, \citenamefont {Sels},\ and\ \citenamefont {Polkovnikov}}]{Claeys2019FloquetEngineering}%
  \BibitemOpen
  \bibfield  {author} {\bibinfo {author} {\bibfnamefont {P.~W.}\ \bibnamefont {Claeys}}, \bibinfo {author} {\bibfnamefont {M.}~\bibnamefont {Pandey}}, \bibinfo {author} {\bibfnamefont {D.}~\bibnamefont {Sels}},\ and\ \bibinfo {author} {\bibfnamefont {A.}~\bibnamefont {Polkovnikov}},\ }\bibfield  {title} {\bibinfo {title} {Floquet-engineering counterdiabatic protocols in quantum many-body systems},\ }\href {https://doi.org/10.1103/PhysRevLett.123.090602} {\bibfield  {journal} {\bibinfo  {journal} {Phys. Rev. Lett.}\ }\textbf {\bibinfo {volume} {123}},\ \bibinfo {pages} {090602} (\bibinfo {year} {2019})}\BibitemShut {NoStop}%
\bibitem [{\citenamefont {Schindler}\ and\ \citenamefont {Bukov}(2024)}]{Schindler2024CounterdiabaticDriving}%
  \BibitemOpen
  \bibfield  {author} {\bibinfo {author} {\bibfnamefont {P.~M.}\ \bibnamefont {Schindler}}\ and\ \bibinfo {author} {\bibfnamefont {M.}~\bibnamefont {Bukov}},\ }\bibfield  {title} {\bibinfo {title} {Counterdiabatic driving for periodically driven systems},\ }\href {https://doi.org/10.1103/physrevlett.133.123402} {\bibfield  {journal} {\bibinfo  {journal} {Phys. Rev. Lett.}\ }\textbf {\bibinfo {volume} {133}},\ \bibinfo {pages} {123402} (\bibinfo {year} {2024})}\BibitemShut {NoStop}%
\bibitem [{\citenamefont {Kim}\ \emph {et~al.}(2024)\citenamefont {Kim}, \citenamefont {Fishman},\ and\ \citenamefont {Sels}}]{Kim2024PRXQuantum}%
  \BibitemOpen
  \bibfield  {author} {\bibinfo {author} {\bibfnamefont {H.}~\bibnamefont {Kim}}, \bibinfo {author} {\bibfnamefont {M.}~\bibnamefont {Fishman}},\ and\ \bibinfo {author} {\bibfnamefont {D.}~\bibnamefont {Sels}},\ }\bibfield  {title} {\bibinfo {title} {Variational adiabatic transport of tensor networks},\ }\href {https://doi.org/10.1103/PRXQuantum.5.020361} {\bibfield  {journal} {\bibinfo  {journal} {PRX Quantum}\ }\textbf {\bibinfo {volume} {5}},\ \bibinfo {pages} {020361} (\bibinfo {year} {2024})}\BibitemShut {NoStop}%
\bibitem [{\citenamefont {Mc~Keever}\ and\ \citenamefont {Lubasch}(2024)}]{McKeever2024PRXQuantum}%
  \BibitemOpen
  \bibfield  {author} {\bibinfo {author} {\bibfnamefont {C.}~\bibnamefont {Mc~Keever}}\ and\ \bibinfo {author} {\bibfnamefont {M.}~\bibnamefont {Lubasch}},\ }\bibfield  {title} {\bibinfo {title} {Towards adiabatic quantum computing using compressed quantum circuits},\ }\href {https://doi.org/10.1103/PRXQuantum.5.020362} {\bibfield  {journal} {\bibinfo  {journal} {PRX Quantum}\ }\textbf {\bibinfo {volume} {5}},\ \bibinfo {pages} {020362} (\bibinfo {year} {2024})}\BibitemShut {NoStop}%
\bibitem [{\citenamefont {Hatomura}\ and\ \citenamefont {Takahashi}(2021)}]{Hatomura2021ControllingAndExploring}%
  \BibitemOpen
  \bibfield  {author} {\bibinfo {author} {\bibfnamefont {T.}~\bibnamefont {Hatomura}}\ and\ \bibinfo {author} {\bibfnamefont {K.}~\bibnamefont {Takahashi}},\ }\bibfield  {title} {\bibinfo {title} {Controlling and exploring quantum systems by algebraic expression of adiabatic gauge potential},\ }\href {https://doi.org/10.1103/PhysRevA.103.012220} {\bibfield  {journal} {\bibinfo  {journal} {Phys. Rev. A}\ }\textbf {\bibinfo {volume} {103}},\ \bibinfo {pages} {012220} (\bibinfo {year} {2021})}\BibitemShut {NoStop}%
\bibitem [{\citenamefont {Bhattacharjee}()}]{Bhattacharjee2023ALanczosApproach}%
  \BibitemOpen
  \bibfield  {author} {\bibinfo {author} {\bibfnamefont {B.}~\bibnamefont {Bhattacharjee}},\ }\bibfield  {title} {\bibinfo {title} {A {L}anczos approach to the adiabatic gauge potential},\ }\href@noop {} {\bibinfo  {journal} {\href{https://arxiv.org/abs/2302.07228}{arXiv preprint arXiv:2302.07228 (2023)}}\ }\BibitemShut {NoStop}%
\bibitem [{\citenamefont {Takahashi}\ and\ \citenamefont {del Campo}(2024)}]{Takahashi2024ShortcutsToAdiabaticity}%
  \BibitemOpen
\bibfield  {journal} {  }\bibfield  {author} {\bibinfo {author} {\bibfnamefont {K.}~\bibnamefont {Takahashi}}\ and\ \bibinfo {author} {\bibfnamefont {A.}~\bibnamefont {del Campo}},\ }\bibfield  {title} {\bibinfo {title} {Shortcuts to adiabaticity in {K}rylov space},\ }\href {https://doi.org/10.1103/physrevx.14.011032} {\bibfield  {journal} {\bibinfo  {journal} {Phys. Rev. X}\ }\textbf {\bibinfo {volume} {14}},\ \bibinfo {pages} {011032} (\bibinfo {year} {2024})}\BibitemShut {NoStop}%
\bibitem [{\citenamefont {Mbeng}\ and\ \citenamefont {Lechner}()}]{Mbeng2022RotatedAnsatz}%
  \BibitemOpen
  \bibfield  {author} {\bibinfo {author} {\bibfnamefont {G.~B.}\ \bibnamefont {Mbeng}}\ and\ \bibinfo {author} {\bibfnamefont {W.}~\bibnamefont {Lechner}},\ }\bibfield  {title} {\bibinfo {title} {Rotated ansatz for approximate counterdiabatic driving},\ }\href@noop {} {\bibinfo  {journal} {\href{https://arxiv.org/abs/2207.03553}{arXiv preprint arXiv:2207.03553 (2022)}}\ }\BibitemShut {NoStop}%
\bibitem [{\citenamefont {\v{C}epait\.{e}}\ \emph {et~al.}(2023)\citenamefont {\v{C}epait\.{e}}, \citenamefont {Polkovnikov}, \citenamefont {Daley},\ and\ \citenamefont {Duncan}}]{Cepaite2023CounterdiabaticOptimized}%
  \BibitemOpen
\bibfield  {journal} {  }\bibfield  {author} {\bibinfo {author} {\bibfnamefont {I.}~\bibnamefont {\v{C}epait\.{e}}}, \bibinfo {author} {\bibfnamefont {A.}~\bibnamefont {Polkovnikov}}, \bibinfo {author} {\bibfnamefont {A.~J.}\ \bibnamefont {Daley}},\ and\ \bibinfo {author} {\bibfnamefont {C.~W.}\ \bibnamefont {Duncan}},\ }\bibfield  {title} {\bibinfo {title} {Counterdiabatic optimized local driving},\ }\href {https://doi.org/10.1103/PRXQuantum.4.010312} {\bibfield  {journal} {\bibinfo  {journal} {PRX Quantum}\ }\textbf {\bibinfo {volume} {4}},\ \bibinfo {pages} {010312} (\bibinfo {year} {2023})}\BibitemShut {NoStop}%
\bibitem [{\citenamefont {Yao}\ \emph {et~al.}(2021)\citenamefont {Yao}, \citenamefont {Lin},\ and\ \citenamefont {Bukov}}]{Yao2021ReinforcementLearning}%
  \BibitemOpen
  \bibfield  {author} {\bibinfo {author} {\bibfnamefont {J.}~\bibnamefont {Yao}}, \bibinfo {author} {\bibfnamefont {L.}~\bibnamefont {Lin}},\ and\ \bibinfo {author} {\bibfnamefont {M.}~\bibnamefont {Bukov}},\ }\bibfield  {title} {\bibinfo {title} {Reinforcement learning for many-body ground-state preparation inspired by counterdiabatic driving},\ }\href {https://doi.org/10.1103/physrevx.11.031070} {\bibfield  {journal} {\bibinfo  {journal} {Phys. Rev. X}\ }\textbf {\bibinfo {volume} {11}},\ \bibinfo {pages} {031070} (\bibinfo {year} {2021})}\BibitemShut {NoStop}%
\bibitem [{\citenamefont {Wurtz}\ and\ \citenamefont {Love}(2022)}]{Wurtz2022CounterdiabaticityAndTheQuantum}%
  \BibitemOpen
  \bibfield  {author} {\bibinfo {author} {\bibfnamefont {J.}~\bibnamefont {Wurtz}}\ and\ \bibinfo {author} {\bibfnamefont {P.~J.}\ \bibnamefont {Love}},\ }\bibfield  {title} {\bibinfo {title} {Counterdiabaticity and the quantum approximate optimization algorithm},\ }\href {https://doi.org/10.22331/q-2022-01-27-635} {\bibfield  {journal} {\bibinfo  {journal} {Quantum}\ }\textbf {\bibinfo {volume} {6}},\ \bibinfo {pages} {635} (\bibinfo {year} {2022})}\BibitemShut {NoStop}%
\bibitem [{\citenamefont {Chandarana}\ \emph {et~al.}(2022)\citenamefont {Chandarana}, \citenamefont {Hegade}, \citenamefont {Paul}, \citenamefont {Albarrán-Arriagada}, \citenamefont {Solano}, \citenamefont {del Campo},\ and\ \citenamefont {Chen}}]{Chandarana2022DigitizedCounterdiabatic}%
  \BibitemOpen
  \bibfield  {author} {\bibinfo {author} {\bibfnamefont {P.}~\bibnamefont {Chandarana}}, \bibinfo {author} {\bibfnamefont {N.~N.}\ \bibnamefont {Hegade}}, \bibinfo {author} {\bibfnamefont {K.}~\bibnamefont {Paul}}, \bibinfo {author} {\bibfnamefont {F.}~\bibnamefont {Albarrán-Arriagada}}, \bibinfo {author} {\bibfnamefont {E.}~\bibnamefont {Solano}}, \bibinfo {author} {\bibfnamefont {A.}~\bibnamefont {del Campo}},\ and\ \bibinfo {author} {\bibfnamefont {X.}~\bibnamefont {Chen}},\ }\bibfield  {title} {\bibinfo {title} {Digitized-counterdiabatic quantum approximate optimization algorithm},\ }\href {https://doi.org/10.1103/physrevresearch.4.013141} {\bibfield  {journal} {\bibinfo  {journal} {Phys. Rev. Res.}\ }\textbf {\bibinfo {volume} {4}},\ \bibinfo {pages} {013141} (\bibinfo {year} {2022})}\BibitemShut {NoStop}%
\bibitem [{\citenamefont {Chandarana}\ \emph {et~al.}(2023)\citenamefont {Chandarana}, \citenamefont {Vieites}, \citenamefont {Hegade}, \citenamefont {Solano}, \citenamefont {Ban},\ and\ \citenamefont {Chen}}]{Chandarana2023MetaLearning}%
  \BibitemOpen
  \bibfield  {author} {\bibinfo {author} {\bibfnamefont {P.}~\bibnamefont {Chandarana}}, \bibinfo {author} {\bibfnamefont {P.~S.}\ \bibnamefont {Vieites}}, \bibinfo {author} {\bibfnamefont {N.~N.}\ \bibnamefont {Hegade}}, \bibinfo {author} {\bibfnamefont {E.}~\bibnamefont {Solano}}, \bibinfo {author} {\bibfnamefont {Y.}~\bibnamefont {Ban}},\ and\ \bibinfo {author} {\bibfnamefont {X.}~\bibnamefont {Chen}},\ }\bibfield  {title} {\bibinfo {title} {Meta-learning digitized-counterdiabatic quantum optimization},\ }\href {https://doi.org/10.1088/2058-9565/ace54a} {\bibfield  {journal} {\bibinfo  {journal} {Quantum Sci. Technol.}\ }\textbf {\bibinfo {volume} {8}},\ \bibinfo {pages} {045007} (\bibinfo {year} {2023})}\BibitemShut {NoStop}%
\bibitem [{\citenamefont {Wurtz}\ \emph {et~al.}(2020)\citenamefont {Wurtz}, \citenamefont {Claeys},\ and\ \citenamefont {Polkovnikov}}]{Wurtz2020VariationalSchrieffer}%
  \BibitemOpen
  \bibfield  {author} {\bibinfo {author} {\bibfnamefont {J.}~\bibnamefont {Wurtz}}, \bibinfo {author} {\bibfnamefont {P.~W.}\ \bibnamefont {Claeys}},\ and\ \bibinfo {author} {\bibfnamefont {A.}~\bibnamefont {Polkovnikov}},\ }\bibfield  {title} {\bibinfo {title} {Variational {S}chrieffer-{W}olff transformations for quantum many-body dynamics},\ }\href {https://doi.org/10.1103/physrevb.101.014302} {\bibfield  {journal} {\bibinfo  {journal} {Phys. Rev. B}\ }\textbf {\bibinfo {volume} {101}},\ \bibinfo {pages} {014302} (\bibinfo {year} {2020})}\BibitemShut {NoStop}%
\bibitem [{\citenamefont {Wurtz}\ and\ \citenamefont {Polkovnikov}(2020)}]{Wurtz2020EmergentConservation}%
  \BibitemOpen
  \bibfield  {author} {\bibinfo {author} {\bibfnamefont {J.}~\bibnamefont {Wurtz}}\ and\ \bibinfo {author} {\bibfnamefont {A.}~\bibnamefont {Polkovnikov}},\ }\bibfield  {title} {\bibinfo {title} {Emergent conservation laws and nonthermal states in the mixed-field {I}sing model},\ }\href {https://doi.org/10.1103/physrevb.101.195138} {\bibfield  {journal} {\bibinfo  {journal} {Phys. Rev. B}\ }\textbf {\bibinfo {volume} {101}},\ \bibinfo {pages} {195138} (\bibinfo {year} {2020})}\BibitemShut {NoStop}%
\bibitem [{\citenamefont {Santos}\ and\ \citenamefont {Sarandy}(2021)}]{Santos2021GeneralizedTransitionless}%
  \BibitemOpen
  \bibfield  {author} {\bibinfo {author} {\bibfnamefont {A.~C.}\ \bibnamefont {Santos}}\ and\ \bibinfo {author} {\bibfnamefont {M.~S.}\ \bibnamefont {Sarandy}},\ }\bibfield  {title} {\bibinfo {title} {Generalized transitionless quantum driving for open quantum systems},\ }\href {https://doi.org/10.1103/physreva.104.062421} {\bibfield  {journal} {\bibinfo  {journal} {Phys. Rev. A}\ }\textbf {\bibinfo {volume} {104}},\ \bibinfo {pages} {062421} (\bibinfo {year} {2021})}\BibitemShut {NoStop}%
\bibitem [{\citenamefont {Passarelli}\ \emph {et~al.}(2022)\citenamefont {Passarelli}, \citenamefont {Fazio},\ and\ \citenamefont {Lucignano}}]{Passarelli2022OptimalQuantum}%
  \BibitemOpen
  \bibfield  {author} {\bibinfo {author} {\bibfnamefont {G.}~\bibnamefont {Passarelli}}, \bibinfo {author} {\bibfnamefont {R.}~\bibnamefont {Fazio}},\ and\ \bibinfo {author} {\bibfnamefont {P.}~\bibnamefont {Lucignano}},\ }\bibfield  {title} {\bibinfo {title} {Optimal quantum annealing: A variational shortcut-to-adiabaticity approach},\ }\href {https://doi.org/10.1103/physreva.105.022618} {\bibfield  {journal} {\bibinfo  {journal} {Phys. Rev. A}\ }\textbf {\bibinfo {volume} {105}},\ \bibinfo {pages} {022618} (\bibinfo {year} {2022})}\BibitemShut {NoStop}%
\bibitem [{\citenamefont {Gjonbalaj}\ \emph {et~al.}(2022)\citenamefont {Gjonbalaj}, \citenamefont {Campbell},\ and\ \citenamefont {Polkovnikov}}]{Gjonbalaj2022CounterdiabaticDriving}%
  \BibitemOpen
  \bibfield  {author} {\bibinfo {author} {\bibfnamefont {N.~O.}\ \bibnamefont {Gjonbalaj}}, \bibinfo {author} {\bibfnamefont {D.~K.}\ \bibnamefont {Campbell}},\ and\ \bibinfo {author} {\bibfnamefont {A.}~\bibnamefont {Polkovnikov}},\ }\bibfield  {title} {\bibinfo {title} {Counterdiabatic driving in the classical $\beta$-{F}ermi-{P}asta-{U}lam-{T}singou chain},\ }\href {https://doi.org/10.1103/physreve.106.014131} {\bibfield  {journal} {\bibinfo  {journal} {Phys. Rev. E}\ }\textbf {\bibinfo {volume} {106}},\ \bibinfo {pages} {014131} (\bibinfo {year} {2022})}\BibitemShut {NoStop}%
\bibitem [{\citenamefont {Sugiura}\ \emph {et~al.}(2021)\citenamefont {Sugiura}, \citenamefont {Claeys}, \citenamefont {Dymarsky},\ and\ \citenamefont {Polkovnikov}}]{Sugiura2021AdiabaticLandscape}%
  \BibitemOpen
  \bibfield  {author} {\bibinfo {author} {\bibfnamefont {S.}~\bibnamefont {Sugiura}}, \bibinfo {author} {\bibfnamefont {P.~W.}\ \bibnamefont {Claeys}}, \bibinfo {author} {\bibfnamefont {A.}~\bibnamefont {Dymarsky}},\ and\ \bibinfo {author} {\bibfnamefont {A.}~\bibnamefont {Polkovnikov}},\ }\bibfield  {title} {\bibinfo {title} {Adiabatic landscape and optimal paths in ergodic systems},\ }\href {https://doi.org/10.1103/physrevresearch.3.013102} {\bibfield  {journal} {\bibinfo  {journal} {Phys. Rev. Res.}\ }\textbf {\bibinfo {volume} {3}},\ \bibinfo {pages} {013102} (\bibinfo {year} {2021})}\BibitemShut {NoStop}%
\bibitem [{\citenamefont {Steeb}\ and\ \citenamefont {Hardy}(2010)}]{Steeb2010QuantumMechanics}%
  \BibitemOpen
  \bibfield  {author} {\bibinfo {author} {\bibfnamefont {W.-H.}\ \bibnamefont {Steeb}}\ and\ \bibinfo {author} {\bibfnamefont {Y.}~\bibnamefont {Hardy}},\ }\href@noop {} {\emph {\bibinfo {title} {Quantum mechanics using computer algebra}}}\ (\bibinfo  {publisher} {World Scientific},\ \bibinfo {address} {Singapore},\ \bibinfo {year} {2010})\BibitemShut {NoStop}%
\bibitem [{\citenamefont {Ohga}\ and\ \citenamefont {Hatomura}()}]{GitHub}%
  \BibitemOpen
  \bibfield  {author} {\bibinfo {author} {\bibfnamefont {N.}~\bibnamefont {Ohga}}\ and\ \bibinfo {author} {\bibfnamefont {T.}~\bibnamefont {Hatomura}},\ }\href@noop {} {}\bibinfo {note} {{I}mplementation and data of weighted variational method for counterdiabatic driving, \href{https://doi.org/10.5281/zenodo.17106271}{doi:10.5281/zenodo.17106271 (2025)}}\BibitemShut {NoStop}%
\bibitem [{\citenamefont {Kato}(1950)}]{Kato1950OnTheAdiabaticTheorem}%
  \BibitemOpen
  \bibfield  {author} {\bibinfo {author} {\bibfnamefont {T.}~\bibnamefont {Kato}},\ }\bibfield  {title} {\bibinfo {title} {On the adiabatic theorem of quantum mechanics},\ }\href {https://doi.org/10.1143/jpsj.5.435} {\bibfield  {journal} {\bibinfo  {journal} {J. Phys. Soc. Jpn.}\ }\textbf {\bibinfo {volume} {5}},\ \bibinfo {pages} {435} (\bibinfo {year} {1950})}\BibitemShut {NoStop}%
\bibitem [{\citenamefont {Lychkovskiy}\ \emph {et~al.}(2017)\citenamefont {Lychkovskiy}, \citenamefont {Gamayun},\ and\ \citenamefont {Cheianov}}]{Lychkovskiy2017TimeScale}%
  \BibitemOpen
  \bibfield  {author} {\bibinfo {author} {\bibfnamefont {O.}~\bibnamefont {Lychkovskiy}}, \bibinfo {author} {\bibfnamefont {O.}~\bibnamefont {Gamayun}},\ and\ \bibinfo {author} {\bibfnamefont {V.}~\bibnamefont {Cheianov}},\ }\bibfield  {title} {\bibinfo {title} {Time scale for adiabaticity breakdown in driven many-body systems and orthogonality catastrophe},\ }\href {https://doi.org/10.1103/physrevlett.119.200401} {\bibfield  {journal} {\bibinfo  {journal} {Phys. Rev. Lett.}\ }\textbf {\bibinfo {volume} {119}},\ \bibinfo {pages} {200401} (\bibinfo {year} {2017})}\BibitemShut {NoStop}%
\bibitem [{\citenamefont {Baylis}(1996)}]{Baylis1996PauliAlgebra}%
  \BibitemOpen
  \bibfield  {author} {\bibinfo {author} {\bibfnamefont {W.~E.}\ \bibnamefont {Baylis}},\ }\bibinfo {title} {Pauli-algebra calculations in {MAPLE} {V}},\ in\ \href@noop {} {\emph {\bibinfo {booktitle} {Clifford Algebras with Numeric and Symbolic Computations}}},\ \bibinfo {editor} {edited by\ \bibinfo {editor} {\bibfnamefont {R.}~\bibnamefont {Ab{\l}amowicz}}, \bibinfo {editor} {\bibfnamefont {J.~M.}\ \bibnamefont {Parra}},\ and\ \bibinfo {editor} {\bibfnamefont {P.}~\bibnamefont {Lounesto}}}\ (\bibinfo  {publisher} {Birkh{\"a}user Boston},\ \bibinfo {year} {1996})\ pp.\ \bibinfo {pages} {69--82}\BibitemShut {NoStop}%
\bibitem [{\citenamefont {Filip}\ and\ \citenamefont {Filip}(2010)}]{Filip2010SdCasSpinDynamics}%
  \BibitemOpen
  \bibfield  {author} {\bibinfo {author} {\bibfnamefont {X.}~\bibnamefont {Filip}}\ and\ \bibinfo {author} {\bibfnamefont {C.}~\bibnamefont {Filip}},\ }\bibfield  {title} {\bibinfo {title} {{SD-CAS}: Spin dynamics by computer algebra system},\ }\href {https://doi.org/10.1016/j.jmr.2010.08.014} {\bibfield  {journal} {\bibinfo  {journal} {J. Magn. Reson.}\ }\textbf {\bibinfo {volume} {207}},\ \bibinfo {pages} {95} (\bibinfo {year} {2010})}\BibitemShut {NoStop}%
\bibitem [{YiZ()}]{YiZhuangMathematicaPackages}%
  \BibitemOpen
  \href@noop {} {}\bibinfo {note} {Y.-Z. You, Mathematica packages for physicists, \urlstyle{same}\url{https://github.com/EverettYou/Mathematica-for-physics}.}\BibitemShut {Stop}%
\bibitem [{\citenamefont {Loizeau}\ \emph {et~al.}(2025)\citenamefont {Loizeau}, \citenamefont {Peacock},\ and\ \citenamefont {Sels}}]{Loizeau2025QuantumMany}%
  \BibitemOpen
  \bibfield  {author} {\bibinfo {author} {\bibfnamefont {N.}~\bibnamefont {Loizeau}}, \bibinfo {author} {\bibfnamefont {J.~C.}\ \bibnamefont {Peacock}},\ and\ \bibinfo {author} {\bibfnamefont {D.}~\bibnamefont {Sels}},\ }\bibfield  {title} {\bibinfo {title} {Quantum many-body simulations with {P}auli{S}trings.jl},\ }\href {https://doi.org/10.21468/scipostphyscodeb.54} {\bibfield  {journal} {\bibinfo  {journal} {SciPost Phys. Codebases}\ }\textbf {\bibinfo {volume} {\hspace{-.37em}}},\ \bibinfo {pages} {54} (\bibinfo {year} {2025})}\BibitemShut {NoStop}%
\bibitem [{\citenamefont {Trefethen}\ and\ \citenamefont {Bau~III}(1997)}]{Trefethen1997NumericalLinear}%
  \BibitemOpen
  \bibfield  {author} {\bibinfo {author} {\bibfnamefont {L.~N.}\ \bibnamefont {Trefethen}}\ and\ \bibinfo {author} {\bibfnamefont {D.}~\bibnamefont {Bau~III}},\ }\href@noop {} {\emph {\bibinfo {title} {Numerical Linear Algebra}}}\ (\bibinfo  {publisher} {Society for Industrial and Applied Mathematics},\ \bibinfo {address} {Philadelphia},\ \bibinfo {year} {1997})\BibitemShut {NoStop}%
\bibitem [{\citenamefont {Yarkoni}\ \emph {et~al.}(2022)\citenamefont {Yarkoni}, \citenamefont {Raponi}, \citenamefont {Bäck},\ and\ \citenamefont {Schmitt}}]{Yarkoni2022QuantumAnnealing}%
  \BibitemOpen
  \bibfield  {author} {\bibinfo {author} {\bibfnamefont {S.}~\bibnamefont {Yarkoni}}, \bibinfo {author} {\bibfnamefont {E.}~\bibnamefont {Raponi}}, \bibinfo {author} {\bibfnamefont {T.}~\bibnamefont {Bäck}},\ and\ \bibinfo {author} {\bibfnamefont {S.}~\bibnamefont {Schmitt}},\ }\bibfield  {title} {\bibinfo {title} {Quantum annealing for industry applications: introduction and review},\ }\href {https://doi.org/10.1088/1361-6633/ac8c54} {\bibfield  {journal} {\bibinfo  {journal} {Rep. Prog. Phys.}\ }\textbf {\bibinfo {volume} {85}},\ \bibinfo {pages} {104001} (\bibinfo {year} {2022})}\BibitemShut {NoStop}%
\bibitem [{Sup()}]{Suppli}%
  \BibitemOpen
  \href@noop {} {}\bibinfo {note} {See Supplemental Material at [URL] for supplemental figures.}\BibitemShut {Stop}%
\bibitem [{\citenamefont {Johansson}\ \emph {et~al.}(2012)\citenamefont {Johansson}, \citenamefont {Nation},\ and\ \citenamefont {Nori}}]{Johansson2012QuTiP}%
  \BibitemOpen
  \bibfield  {author} {\bibinfo {author} {\bibfnamefont {J.}~\bibnamefont {Johansson}}, \bibinfo {author} {\bibfnamefont {P.}~\bibnamefont {Nation}},\ and\ \bibinfo {author} {\bibfnamefont {F.}~\bibnamefont {Nori}},\ }\bibfield  {title} {\bibinfo {title} {{QuTiP}: An open-source {P}ython framework for the dynamics of open quantum systems},\ }\href {https://doi.org/10.1016/j.cpc.2012.02.021} {\bibfield  {journal} {\bibinfo  {journal} {Comput. Phys. Commun.}\ }\textbf {\bibinfo {volume} {183}},\ \bibinfo {pages} {1760} (\bibinfo {year} {2012})}\BibitemShut {NoStop}%
\bibitem [{\citenamefont {Johansson}\ \emph {et~al.}(2013)\citenamefont {Johansson}, \citenamefont {Nation},\ and\ \citenamefont {Nori}}]{Johansson2013QuTiP}%
  \BibitemOpen
  \bibfield  {author} {\bibinfo {author} {\bibfnamefont {J.}~\bibnamefont {Johansson}}, \bibinfo {author} {\bibfnamefont {P.}~\bibnamefont {Nation}},\ and\ \bibinfo {author} {\bibfnamefont {F.}~\bibnamefont {Nori}},\ }\bibfield  {title} {\bibinfo {title} {{QuTiP} 2: A {P}ython framework for the dynamics of open quantum systems},\ }\href {https://doi.org/10.1016/j.cpc.2012.11.019} {\bibfield  {journal} {\bibinfo  {journal} {Comput. Phys. Commun.}\ }\textbf {\bibinfo {volume} {184}},\ \bibinfo {pages} {1234} (\bibinfo {year} {2013})}\BibitemShut {NoStop}%
\bibitem [{\citenamefont {Domb}\ and\ \citenamefont {Green}(1974)}]{Domb1972ModernQuantumMechanics}%
  \BibitemOpen
  \bibfield  {author} {\bibinfo {author} {\bibfnamefont {C.}~\bibnamefont {Domb}}\ and\ \bibinfo {author} {\bibfnamefont {M.~S.}\ \bibnamefont {Green}},\ }\href@noop {} {\emph {\bibinfo {title} {Phase Transitions and Critical Phenomena. Vol. 3: Series Expansions for Lattice Models}}}\ (\bibinfo  {publisher} {Academic Press},\ \bibinfo {address} {London},\ \bibinfo {year} {1974})\BibitemShut {NoStop}%
\bibitem [{\citenamefont {Schmidt}\ \emph {et~al.}(2011)\citenamefont {Schmidt}, \citenamefont {Lohmann},\ and\ \citenamefont {Richter}}]{Schmidt2011EighthOrder}%
  \BibitemOpen
  \bibfield  {author} {\bibinfo {author} {\bibfnamefont {H.-J.}\ \bibnamefont {Schmidt}}, \bibinfo {author} {\bibfnamefont {A.}~\bibnamefont {Lohmann}},\ and\ \bibinfo {author} {\bibfnamefont {J.}~\bibnamefont {Richter}},\ }\bibfield  {title} {\bibinfo {title} {Eighth-order high-temperature expansion for general {H}eisenberg {H}amiltonians},\ }\href {https://doi.org/10.1103/physrevb.84.104443} {\bibfield  {journal} {\bibinfo  {journal} {Phys. Rev. B}\ }\textbf {\bibinfo {volume} {84}},\ \bibinfo {pages} {104443} (\bibinfo {year} {2011})}\BibitemShut {NoStop}%
\bibitem [{\citenamefont {Shingu}\ and\ \citenamefont {Hatomura}(2025)}]{Shingu2025GeometricalScheduling}%
  \BibitemOpen
  \bibfield  {author} {\bibinfo {author} {\bibfnamefont {Y.}~\bibnamefont {Shingu}}\ and\ \bibinfo {author} {\bibfnamefont {T.}~\bibnamefont {Hatomura}},\ }\bibfield  {title} {\bibinfo {title} {Geometrical scheduling of adiabatic control without information of energy spectra},\ }\href {https://doi.org/10.1103/n5xj-phwv} {\bibfield  {journal} {\bibinfo  {journal} {Phys. Rev. A}\ }\textbf {\bibinfo {volume} {112}},\ \bibinfo {pages} {022410} (\bibinfo {year} {2025})}\BibitemShut {NoStop}%
\bibitem [{\citenamefont {Dengis}\ \emph {et~al.}(2025)\citenamefont {Dengis}, \citenamefont {Wimberger},\ and\ \citenamefont {Schlagheck}}]{Dengis2025MultimodeNoon}%
  \BibitemOpen
  \bibfield  {author} {\bibinfo {author} {\bibfnamefont {S.}~\bibnamefont {Dengis}}, \bibinfo {author} {\bibfnamefont {S.}~\bibnamefont {Wimberger}},\ and\ \bibinfo {author} {\bibfnamefont {P.}~\bibnamefont {Schlagheck}},\ }\bibfield  {title} {\bibinfo {title} {Multimode {NOON}-state generation with ultracold atoms via geodesic counterdiabatic driving},\ }\href {https://doi.org/10.1103/qxqg-qnnq} {\bibfield  {journal} {\bibinfo  {journal} {Phys. Rev. A}\ }\textbf {\bibinfo {volume} {112}},\ \bibinfo {pages} {042610} (\bibinfo {year} {2025})}\BibitemShut {NoStop}%
\bibitem [{\citenamefont {Pandey}\ \emph {et~al.}(2020)\citenamefont {Pandey}, \citenamefont {Claeys}, \citenamefont {Campbell}, \citenamefont {Polkovnikov},\ and\ \citenamefont {Sels}}]{Pandey2020AdiabaticEigenstate}%
  \BibitemOpen
  \bibfield  {author} {\bibinfo {author} {\bibfnamefont {M.}~\bibnamefont {Pandey}}, \bibinfo {author} {\bibfnamefont {P.~W.}\ \bibnamefont {Claeys}}, \bibinfo {author} {\bibfnamefont {D.~K.}\ \bibnamefont {Campbell}}, \bibinfo {author} {\bibfnamefont {A.}~\bibnamefont {Polkovnikov}},\ and\ \bibinfo {author} {\bibfnamefont {D.}~\bibnamefont {Sels}},\ }\bibfield  {title} {\bibinfo {title} {Adiabatic eigenstate deformations as a sensitive probe for quantum chaos},\ }\href {https://doi.org/10.1103/physrevx.10.041017} {\bibfield  {journal} {\bibinfo  {journal} {Phys. Rev. X}\ }\textbf {\bibinfo {volume} {10}},\ \bibinfo {pages} {041017} (\bibinfo {year} {2020})}\BibitemShut {NoStop}%
\bibitem [{\citenamefont {Zanardi}\ \emph {et~al.}(2007)\citenamefont {Zanardi}, \citenamefont {Giorda},\ and\ \citenamefont {Cozzini}}]{Zanardi2007InformationTheoretic}%
  \BibitemOpen
  \bibfield  {author} {\bibinfo {author} {\bibfnamefont {P.}~\bibnamefont {Zanardi}}, \bibinfo {author} {\bibfnamefont {P.}~\bibnamefont {Giorda}},\ and\ \bibinfo {author} {\bibfnamefont {M.}~\bibnamefont {Cozzini}},\ }\bibfield  {title} {\bibinfo {title} {Information-theoretic differential geometry of quantum phase transitions},\ }\href {https://doi.org/10.1103/physrevlett.99.100603} {\bibfield  {journal} {\bibinfo  {journal} {Phys. Rev. Lett.}\ }\textbf {\bibinfo {volume} {99}},\ \bibinfo {pages} {100603} (\bibinfo {year} {2007})}\BibitemShut {NoStop}%
\bibitem [{\citenamefont {Cormen}\ \emph {et~al.}(2009)\citenamefont {Cormen}, \citenamefont {Leiserson}, \citenamefont {Rivest},\ and\ \citenamefont {Stein}}]{Cormen2009IntroductionToAlgorithms}%
  \BibitemOpen
  \bibfield  {author} {\bibinfo {author} {\bibfnamefont {T.~H.}\ \bibnamefont {Cormen}}, \bibinfo {author} {\bibfnamefont {C.~E.}\ \bibnamefont {Leiserson}}, \bibinfo {author} {\bibfnamefont {R.~L.}\ \bibnamefont {Rivest}},\ and\ \bibinfo {author} {\bibfnamefont {C.}~\bibnamefont {Stein}},\ }\href@noop {} {\emph {\bibinfo {title} {Introduction to Algorithms}}},\ \bibinfo {edition} {3rd}\ ed.\ (\bibinfo  {publisher} {MIT Press},\ \bibinfo {address} {Cambridge},\ \bibinfo {year} {2009})\BibitemShut {NoStop}%
\end{thebibliography}
%

\onecolumngrid

\appendix 
\counterwithout{equation}{section} 

\clearpage 

\onecolumngrid 

\setcounter{equation}{0} 
\setcounter{figure}{0} 
\setcounter{table}{0} 
\setcounter{page}{1} 

\renewcommand{\theequation}{S\arabic{equation}} 
\renewcommand{\thefigure}{S\arabic{figure}} 
\renewcommand{\thesection}{\Alph{section}} 
\renewcommand{\thesubsection}{\arabic{subsection}}

\begin{center} 
\textbf{\large Supplemental Material for\\[0.1em]
Improving variational counterdiabatic driving with weighted actions and computer algebra}

\vspace*{1em}
Naruo\hspace{0.35em}Ohga\hspace{0.35em}and\hspace{0.35em}Takuya\hspace{0.35em}Hatomura

\vspace*{1em}
\end{center}

\noindent This Supplemental Material contains two figures referenced in the main text.

\vspace{1em}

\begin{figure*}[ht]
\centering
\includegraphics{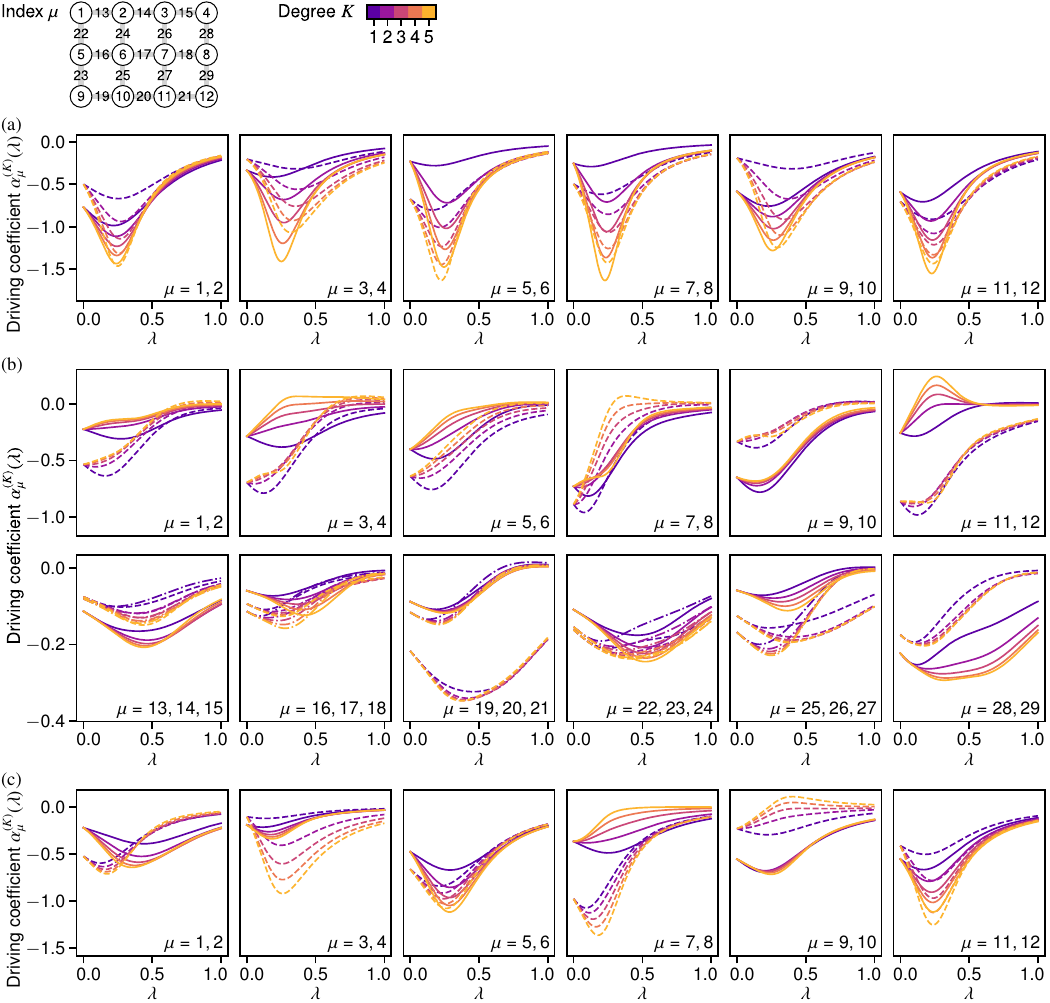}
\caption{Driving coefficients $\alpha^{(K)}_{\mu}(\lambda)$ obtained from the weighted variational method for representative systems with $N=12$. The index $\mu$ of driving terms is arranged as shown in the top, with $1\leq\mu\leq12$ representing one-body terms and $13\leq\mu\leq29$ representing two-body terms. Each panel shows $\alpha^{(K)}_{\mu}(\lambda)$ for two or three terms $\mu$. Solid curves are for the smallest $\mu$, and dashed curves are for the second smallest $\mu$. Dash-dotted curves are for the third, if it exists. (a) Ferromagnetic system with the one-body driving, for which some of the curves have been shown in Fig.~\ref{fig:ferromagnetic_weight_coeff}(b) in the main text. (b) Antiferromagnetic system with the two-body driving. (c) Spin-glass system with the one-body driving.
\label{fig:supplemental-driving}}
\end{figure*}

\begin{figure*}[ht]
\centering
\includegraphics{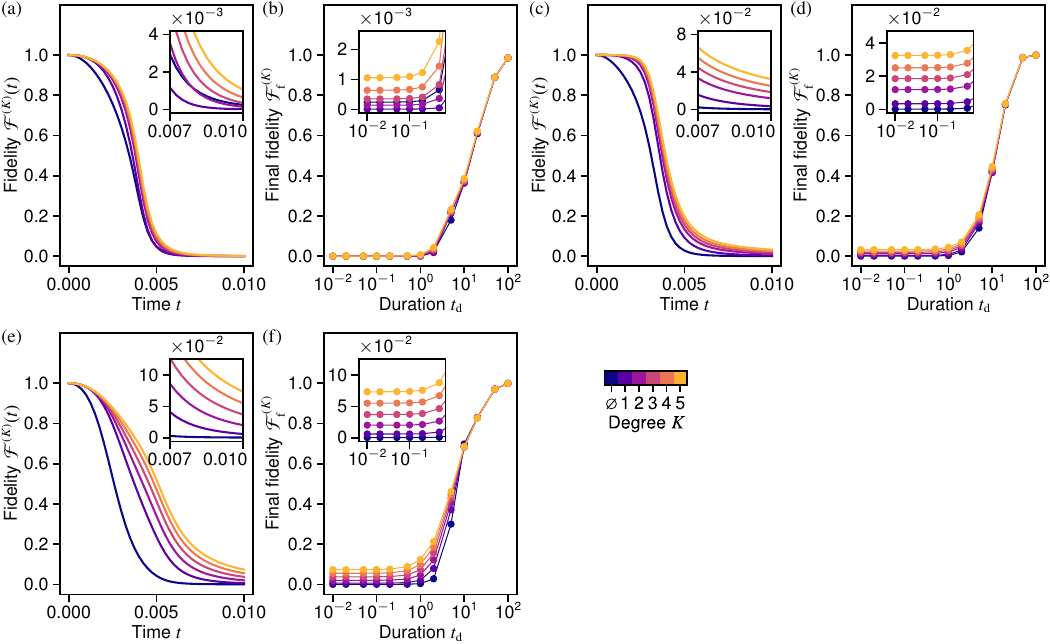}
\caption{Performance of the weighted variational method in representative antiferromagnetic and spin-glass systems of $N=12$. Similar plots for ferromagnetic systems have been shown in Fig.~\ref{fig:ferromagnetic_single} in the main text. Panels (a) and (b) show an antiferromagnetic system with the one-body driving, panels (c) and (d) show the same antiferromagnetic system with the two-body driving, and panels (e) and (f) are for a spin-glass system with the one-body driving. In panels (a), (c), and (e), we plot the time evolution of the fidelity to the ground state $\mathcal{F}^{(K)}(t)$, with the insets showing the close-up of the curves at later times. In panels (b), (d), and (f), we plot the final fidelity $\mathcal{F}_{\mathrm{f}}^{(K)}$ over varied protocol durations $\td$, where the insets are the zoom-up for short durations. In the antiferromagnetic systems with the one-body driving (panels (a) and (b)), the time evolution without CD driving ($K=\varnothing$) performs better than the time evolution with variational CD driving for small $K$, as discussed in the main text.
\label{fig:supplemental-single}}
\end{figure*}

\end{document}